\documentclass[preprint]{ptephy_v2}

\preprintnumber{} 
\usepackage{hyperref}
\usepackage{amsmath} 
\usepackage{graphics} 
\usepackage{bm} 
\usepackage{subfigure} 
\usepackage[normalem]{ulem}
\usepackage{xcolor}

\newcommand{\ee}{$e^+e^- $} 
\newcommand{\KK}{$K^+K^- $} 
\newcommand{\pipi}{$\pi^+\pi^- $}

\newcommand{\omegameson}{$\omega$ meson }

\newcommand{\degree}{$^{\circ}$}
\newcommand{\um}{$\mu$m} 
\newcommand{\MEV}{MeV/$c^2$}
\newcommand{\GEV}{GeV/$c^2$}

\newcommand{\cherenkov}{\v Cerenkov } 
\newcommand{\bg}{$\beta\gamma$}

\newcommand{\expo}[2]{$ #1 \times 10^{#2} $}
\newcommand{\Xp}{X$^{\prime}$}
\newcommand{\Vp}{V$^{\prime}$}

\newcommand{\FIGciteNaruki}{Adapted from \cite{naruki_D}.}
\newcommand{\FIGciteMuto}{Adapted from \cite{muto_D}.}
\newcommand{\FIGciteSakuma}{Adapted from \cite{sakuma_D}.}

\newcommand{\TobeChecked}[1]{\textcolor{red} {#1} }

\definecolor{green2}{cmyk}{ 1.0, 0.0, 0.5, 0.4}

\newcommand{\Gtot}{\Gamma_\mathrm{tot}}
\newcommand{\Gee}{\Gamma_{ee}}

\begin{document}

\title{
Velocity dependence of the mass modifications of $\rho$ and $\omega$ mesons in 12 GeV $p+A$ reactions
}



\newcommand\KEK {Present address: IPNS, KEK, Tsukuba 305-0801, Japan}
\newcommand\KEKacc {Present address: Accelerator Laboratory, KEK,  Tsukuba 305-0801, Japan}
\newcommand\KYOTOF {Present address: Institute for Liberal Arts and Sciences, Kyoto University,  Kyoto 606-8501, Japan}
\newcommand\NIAS {Present address: Institute for Innovative Science and Technology, Nagasaki Institute of Applied Science, Nagasaki, 851-0121, Japan}
\newcommand\ICEPP {Present address: International Center for the Elementary Particle Physics (ICEPP), University of Tokyo, Tokyo, 113-0033, Japan}
\newcommand\RCNP {Present address: Research Center for Nuclear Physics (RCNP), Osaka University, Ibaraki, 567-0047, Japan}
\newcommand\NAGOYA {Present address: Kobayashi-Maskawa Institute for the Origin of Particles and the Universe (KMI), Nagoya University, Nagoya 464-8602, Japan}
\newcommand\TOHOKU {Present address: Department of Physics, Tohoku University, Sendai 980-8578, Japan}
\newcommand\RIKENkaitaku {Present address: RIKEN Cluster for Pioneering Research, RIKEN, Wako 351-0198, Japan}
\newcommand\RIKENnishina {Present address: RIKEN Nishina Center, RIKEN, Wako 351-0198, Japan}
\newcommand\Deceased {Deceased}

\makeatletter
\def\@fnsymbol#1{\ensuremath{\ifcase#1\or \hbox{*}\or *\or
            \dagger\or a\or b\or c\or d\or e\or f\or g\or h\or i\or j\or k\or l
            \else\@ctrerr\fi}}

\makeatother

\author{Wataru~Nakai$^1$\thanks{E-mail: wnakai@post.kek.jp},
Kazuya~Aoki$^1$,
Junsei~Chiba$^1$\thanks{\Deceased},
Hideto~En'yo$^2$,
Yoshinori~Fukao$^{3,}$\thanks{\KEK},          
Haruhiko~Funahashi$^{3,}$\thanks{\KYOTOF},
Hideki~Hamagaki$^{4,}$\thanks{\NIAS},
Masaharu~Ieiri$^1$,
Masaya~Ishino$^{3,}$\thanks{\ICEPP},
Hiroki~Kanda$^{3,}$\thanks{\RCNP},
Koki~Kanno$^2$,
Masaaki~Kitaguchi$^{3,}$\thanks{\NAGOYA},
Satoshi~Mihara$^{3,\mathrm{a}}$, 
Koji~Miwa$^{3,}$\thanks{\TOHOKU}$^{,\mathrm{a}}$,
Takuya~Miyashita$^3$, 
Tetsuya~Murakami$^3$, 
Ryotaro~Muto$^{3,}$\thanks{\KEKacc}, 
Terunao~Nakura$^3$,  
Megumi~Naruki$^3$,   
Kyoichiro~Ozawa$^1$, 
Fuminori~Sakuma$^{3,}$\thanks{\RIKENkaitaku},
Osamu~Sasaki$^1$,       
Michiko~Sekimoto$^1$,  
Tsuguchika~Tabaru$^2$,   
Kazuhiro~Tanaka$^1$,    
Manabu~Togawa$^{3,\mathrm{a}}$,   
Satoru~Yamada$^{3,\mathrm{a}}$,      
Satoshi~Yokkaichi$^2$,   
and Yoshio~Yoshimura$^3$
\authorcr
{\bf (KEK-PS E325 collaboration)}
}

\affil{
$^1${Institute of Particle and Nuclear Studies (IPNS), High Energy Accelerator Research Organization (KEK), Tsukuba, 305-0801, Japan}\\
$^2${RIKEN Nishina Center, RIKEN, Wako, 351-0198, Japan}\\
$^3${Department of Physics, Kyoto University, Kyoto 606-8502, Japan}\\
$^4${Center for Nuclear Study, Graduate School of Science, University of Tokyo, Tokyo 113-0033, Japan}\\
}






\begin{abstract}%
This study measured the invariant mass spectra of $\rho$ and $\omega$ mesons
in the \ee~decay channel for 12 GeV (12.9 GeV/$c$) $p+\mathrm{C}$
and $p+\mathrm{Cu}$ reactions ($\sqrt{s}_{NN}=5.1$ GeV) at the KEK 12-GeV Proton Synchrotron.
The measured spectra were divided into three \bg\ regions to examine their
velocity dependence.
Across all regions, significant excesses were
observed on the low-mass side of the \omegameson peak,
beyond the contributions of known hadronic sources, in the data of the C and Cu targets.
Model calculations were subsequently performed to evaluate 
the magnitudes of the mass modifications of $\rho$ and $\omega$ mesons. 
\end{abstract}

\subjectindex{xxxx, xxx}

\maketitle
\vspace{-1mm}
    \section{Introduction}
Quantum chromodynamics (QCD) research in the nonperturbative domain has garnered widespread attention in hadron physics. 
A prominent topic within this field is the dynamic generation of hadron mass,
believed to be predominantly caused by the spontaneous breaking of chiral symmetry within
``QCD vacuum''~\cite{bib:njl}.
Based on this theory, the amount of the generated hadron mass can be assumed to depend on the properties of the surrounding QCD medium.
Consequently, studies on hadron mass within a QCD medium can offer crucial insights into both the QCD medium itself 
and its chiral symmetry. 
To date, numerous theoretical and experimental 
studies have been focused on modified mass spectra of vector mesons, as such modifications can be directly related to the restoration of chiral symmetry within a given medium.
Historically, such studies relied on high-energy heavy-ion collisions to observe a quark-gluon-plasma (QGP), 
a high-temperature substance, as modifications in the mass spectra of vector mesons were considered indicative of QGP formation.
Numerous measurements were performed at the Super Proton Synchrotron and Relativistic Heavy Ion Collider~\cite{bib:CERES:2005, bib:na60_2008, bib:PHENIX:2015}.



Theoretically, the mass spectra of vector mesons can be modified in finite-density matter.
For instance,  Brown and Rho proposed the in-medium scaling law,
which states an approximate 20\% reduction in the masses of $\sigma$, $\rho$ and $\omega$ mesons and nucleons
at the normal nuclear density, $\rho_0$, using an effective
chiral Lagrangian~\cite{bib:br_scaling}.
Meanwhile, Hatsuda and Lee developed the in-medium QCD sum rule, 
predicting a decrease in vector meson masses
proportional to the baryon density, $\rho$,
in the range of $0 \sim 2 \rho_0$.
According to their predictions, the 
expected mass decrease at $\rho_0$ was 
16 $\pm$ 6\% for $\rho$ and $\omega$ mesons 
and approximately 1--4\% for $\phi$ mesons~\cite{bib:HL92, bib:HL95}.
The results with finite three-momentum were also calculated, although the momentum region is less than 1 GeV/$c$~\cite{bib:shlee98}.
In addition to vector mesons, numerous theoretical studies have also investigated hadrons in dense and/or hot matter using the QCD sum rule,
as summarized by Gubler and Satow~\cite{GS19}.

Furthermore, several theoretical studies have attempted to compute
spectral functions, mass and width changes of vector mesons in dense matter, using effective models.
For instance, Klingl \etal~\cite{KKW97,KWW98}
computed
the spectral shapes of $\rho$, $\omega$ and $\phi$ mesons
within the range of $0 \sim 1 \rho_0$,
using their
effective Lagrangian based on $KN$ interactions, ultimately claiming that the obtained shapes were consistent
with the predictions of the in-medium QCD sum rule.
Meanwhile, the authors of \cite{PM02, COV02-rho} performed similar calculations for $\rho$ mesons,
those of \cite{Ramos13, CR14} focused on $\omega$ mesons, and those of \cite{OR01, GW16} considered $\phi$ mesons.

Additionally, numerous experiments have been conducted to examine the properties of vector mesons in finite-density matter, utilizing nucleus targets and low-energy photon, hadron, and heavy-ion beams. 
While these experiments have reported modifications of the mass spectra of vector mesons, a comprehensive understanding still remains elusive.
Additional details pertaining to these experiments are outlined in Sec. \ref{sec:intro_e325} and \ref{sec:previous_experiments}.
The subsequent sections report the results of the updated analysis of the KEK-PS E325 experiment, performed to obtain additional information using existing data. 

\section{Previous experimental results}
\subsection{KEK-PS E325 experiment}\label{sec:intro_e325}
Previously, an experiment, known as the KEK-PS E325 experiment, was performed to study the effects of nuclear matter on the mass spectra of vector mesons~\cite{bib:sekimoto_nim, bib:tabaru, bib:naruki_prl, bib:muto_prl, bib:sakuma_prl}.
In this context, the nucleus was regarded as finite-density QCD matter.  
Specifically, the KEK-PS E325 experiment measured the mass spectra of vector mesons within nuclei through 
12~GeV $pA$ reactions, focusing on the $e^+e^-$ decay mode of the mesons. 
In particular, spectra were obtained for C and Cu targets.
The obtained spectra were subsequently fitted against known hadronic sources, including $\omega\rightarrow e^+e^-$, $\rho\rightarrow e^+e$, $\phi\rightarrow e^+e^-$, $\eta\rightarrow e^+e^-\gamma$, and $\omega\rightarrow e^+e^-\pi^0$, and a combinatorial background, with its shape
generated by the event mixing method.
Significant excesses were observed on the low-mass side of the $\omega$ peak~\cite{bib:naruki_prl}.
To explain these excesses, model calculations assuming mass modifications of $\rho$ and $\omega$ mesons were performed.
These calculations indicated a mass decrease of $9.2 \%$ for both $\rho$ and $\omega$ mesons.
Conversely, another model, assuming width broadening of $\rho$ and $\omega$ mesons, failed to accurately fit the data.

The E325 experiment included another study focusing on the $\phi$ meson mass region~\cite{bib:muto_prl}.
Here, the mass spectra of the $\phi$ meson mass region were divided into three $\beta\gamma$ regions ($\beta\gamma < 1.25$, $1.25 < \beta\gamma < 1.75$, and $1.75 < \beta\gamma$) for C and Cu targets.
Significant excess was observed on the low-mass side of the $\phi$ peak, exclusively in the spectrum of the low-$\beta\gamma$ region and the Cu target.
Similar to the $\rho$ and $\omega$ mesons, model fitting was performed, assuming a decrease in the common mass and broadening of the width
for all \bg\ regions and targets. In this case, a density-dependent width broadening was assumed.
The fitting result indicated a mass decrease of $3.4 \%$ and a 3.6-fold width broadening at the normal nuclear matter density for the $\phi$ meson.

\subsection{
Other experimental results
}\label{sec:previous_experiments}

In heavy-ion collisions at lower incident energies, 
the HADES experiment at GSI measured \ee\ invariant mass spectra, 
reporting enhancements over their models
within the range of 0.15 $<M_{ee}<$ 0.5 \GEV\, 
in 1--2 A$\cdot$GeV $\mathrm{C}+\mathrm{C}$ 
reactions~\cite{HADES-2AGeV,HADES-1AGeV}. 
The obtained data aligned with 
the results of the 1.0 A GeV $\mathrm{C}+\mathrm{C}$ reaction
performed in the DLS experiment at Bevalac~\cite{DLS}.
However, new data obtained from 1.25 A GeV $p+p$ and $d+p$ 
reactions~\cite{HADES-ppnp} demonstrated unexpectedly
large enhancement of \ee\ spectra in the $n+p$ channel 
compared to those in the $p+p$ channel. By considering these data, 
they successfully reproduced the excess observed in the above $\mathrm{C}+\mathrm{C}$ data by  
superposing $p+p$ and $n+p$ collisions.

Subsequent studies reported a low-mass enhancement over the
superposition of p+p and n+p collisions within larger collision systems,
such as those of the 1.76 A GeV $\mathrm{Ar}+\mathrm{KCl}$ reaction~\cite{HADES-ArKCl}, 
and 1.23 A GeV $\mathrm{Au}+\mathrm{Au}$ reaction~\cite{HADES-AuAu}, with a larger enhancement factor in the latter.
Additionally, the HADES experiment measured 
\ee\ spectra in 3.5 GeV $p+\mathrm{Nb}$ reactions~\cite{HADES-pNb}.
By selecting pairs with lower momenta ($P_{ee} < 0.8~\mathrm{GeV}/c$), 
an enhancement was observed below the $\omega$-meson peak.

In addition to the above experiments, several others using photon-induced reactions 
at lower energies have also contributed to this field.
For instance, the TAGX~\cite{TAGX} experiment reported modified $\rho$ mesons in the \pipi\
decay channel 
using 0.6--1.12 GeV $\gamma + ^{3}\mathrm{He},~^{12}\mathrm{C}$ reactions.
Meanwhile, the LEPS~\cite{LEPS} experiment detected $\phi$ mesons in the \KK\ decay channel
through 1.5--2.4  GeV $\gamma + A$ reactions. Notably, the momentum
range of measured $\phi$ mesons was 1.0--2.2 GeV/$c$,
with an average of 1.8 GeV/$c$.
Although no spectral modifications were
observed, an anomalous nuclear dependence of
the production cross-section was reported. 
Furthermore, model calculations based on $\phi N$ interactions in nuclei
were performed to reproduce the dependency, resulting in an 
interaction cross-section of 35$^{+17}_{-11}$ mb,
approximately four times the value reported by previous experiments~\cite{old-phiN}.

The CLAS experiment measured $e^+e^-$ invariant mass spectra in $\gamma A$ collisions~\cite{bib:clas_g7} using D$_2$, C, Fe, and Ti target data.
The measured momentum range for the Fe and Ti target data was 0.8 $<p_{\rho}<$ 3.0 GeV/$c$.
By subtracting the contributions of $\omega$ and $\phi$ mesons,
they obtained a $\rho$ meson mass spectrum.
On comparing the data of C and Fe-Ti targets with those of the D$_2$ target, they observed significant mass broadening of the $\rho$ meson, by a factor of 1.45 approximately, 
for the Fe-Ti targets. They claimed that it is consistent with a theoretical estimation using Boltzmann--Uehling--Uhlenbeck (BUU) transport model.


Furthermore, the CLAS experiment studied in-medium modifications of $\omega$ mesons through transparency ratio measurements~\cite{bib:clas_trans}.
The transparency ratio, $T$, is defined as the ratio of the production cross-sections of vector mesons in $\gamma$ nucleus collisions 
to those in $\gamma$ proton collisions multiplied by the mass number of the nucleus.  
They discussed the in-medium width of the $\omega$ meson with a Glauber analysis of the measured transparency ratio, and compared with other theoretical calculations.
The results indicated that the widths of the $\omega$ mesons within nuclei exceeded 200~MeV/$c^2$.

The CB/TAPS detector at ELSA in Bonn explored the properties of $\omega$ mesons in nuclear matter via $\gamma A \rightarrow \pi^0 \gamma X$ reactions.
They detected three gammas using CB photon calorimeters comprised of CsI(Tl) crystals and the TAPS comprised of BaF$_2$ detectors.
The study concluded that no significant mass shifts could be identified for the $\omega$ meson~\cite{bib:taps_shape_2010}.
However, this lack of observed mass shift for $\omega$ mesons within nuclei does not necessarily imply its complete absence.
Instead, it could be attributed to the low detection probability for $\omega$ meson decays within nuclei in this experiment, thus resulting in poor sensitivity to mass modifications.
Furthermore, no significant differences were observed between the obtained line shapes for the LH$_2$ target and Nb targets, even under model calculations assuming a 16\% lower pole mass of the $\omega$ meson compared to its value in vacuum. 

A similar experiment using the CB/TAPS detector was performed at the Mainz Microtron MAMI-C electron accelerator~\cite{bib:taps_shape_2013}.
Specifically, they analyzed $\omega$ meson line shapes with greater statistics and identical resolutions using photon beams with energies ranging from 900 to 1300~MeV.
However, the conclusion remained unchanged: No significant enhancement was observed, and the line shapes were consistent with the modified spectra resulting from collisional broadenings.

They also performed the transparency ratio measurements to study the in-medium modifications of $\omega$ mesons using the CBELSA/TAPS detector~\cite{bib:taps_transparency_2008, bib:taps_transparency_2016}.
The results suggested a significant increase in the vector meson widths within nuclei. 
Specifically, the width of the $\omega$ meson within nuclei were found to vary from approximately 60~MeV to about 200~MeV depending on the meson momentum.

Table~\ref{tab:previous_results_summary} summarizes the results of previous related experiments,
indicating significant variations among them.
Specifically, the results of the KEK experiment appear to contradict those of the CLAS and TAPS experiments.
\begin{table}[hbpt]
   \caption{Summary of the existing experiments.
   Each experiment varied in terms of the production reaction, the decay channel being measured, and the method employed to estimate modifications.
   The details are described in Sec.~\ref{sec:previous_experiments}.
   }
   \label{tab:previous_results_summary}
   \begin{center}
      \begin{tabular}{ccccc}
         \hline

          & Reaction & Channel & Method & Conclusion \\
          \hline\hline
          CLAS~\cite{bib:clas_g7} & $\gamma A$ & $\rho       \rightarrow e^+e^-$       & Mass shape & No mass shift, \\
          & & & & Broadening \\ \hline
          CLAS~\cite{bib:clas_trans} & $\gamma A$ & $\omega     \rightarrow e^+e^-$       & Transparency ratio & Broadening \\
          & & & & ($>200$~MeV/$c^2$) \\ \hline
          CB/TAPS~\cite{bib:taps_shape_2010,bib:taps_shape_2013}  & $\gamma A$ & $\omega     \rightarrow \pi^0\gamma$  & Mass shape & Low sensitivity \\ \hline
          CB/TAPS~\cite{bib:taps_transparency_2008, bib:taps_transparency_2016}  & $\gamma A$ & $\omega     \rightarrow \pi^0\gamma$  & Transparency ratio & Broadening \\
          & & & & (60-200 MeV/$c^2$) \\ \hline
          KEK E325~\cite{bib:naruki_prl} & $pA$       & $\rho/\omega\rightarrow e^+e^-$       & Mass shape & Mass shift ($\sim9$\%),\\
          & & & & No broadening \\
          \hline


      \end{tabular}
   \end{center}

\end{table}

The KEK experiment concluded a finite mass shift but no broadening for $\rho$ and $\omega$ mesons.
Conversely, the CLAS experiment concluded broadening for the $\rho$ meson, attributed to collisional broadening, without any discernible mass shifts. Furthermore, its upper limit was lower than that of the KEK result.
Meanwhile, the TAPS results indicated broadening of the $\omega$ meson, despite differences in the decay channels and measurement methods.

\subsection{Updated analysis of the E325}
To better understand these discrepancies, we conducted a thorough examination of E325 data.
Initially, we examined the $\beta\gamma$-dependence of the mass spectra as our $\phi$ meson analysis~\cite{bib:muto_prl}.
The mean flight length before the decay,
is expressed by $\beta\gamma c \tau$.
Here, $\tau$ is the mean life of the particle.
When we classify the meson spectra data by \bg,
for the lower \bg\ data, 
the larger fraction of inside-nucleus decay is
expected in the spectrum and the shape modification
could be also larger.
To this end, we increased the statistics of the analyzed data.
Although the analysis of $\phi$ meson~\cite{bib:muto_prl} used all statistics we have, the previous analysis for $\rho$ and $\omega$ meson~\cite{bib:naruki_prl} used only two-thirds of the full statistics. In this time, 
the remaining one-third of data are newly used.
In addition, we refined our analysis by optimizing  
kinematical cuts to widen the acceptance coverage of the detector and improving fiducial cut parameters to ensure 
uniform acceptance across all target positions. 
As a result, the number of obtained $\omega$ meson decays were significantly increased from 6900 in the previous study to 14860 in the current study.

New model calculations were performed incorporating the following effects.
We considered a non-Breit-Wigner distribution to model the mass distributions in a medium. While our original analysis relied on relativistic Breit-Wigner distributions, the CLAS analysis employed a different, nonsymmetric distribution. We adopted the same distribution as in the CLAS analysis.
Additionally, based on recent theoretical developments, we updated the form factors of the Dalitz decays of $\omega$ and $\eta$ mesons. We utilized the event generator Pluto~\cite{bib:pluto} to include them.
Moreover, internal radiative corrections were applied to each mass spectra using the code PHOTOS~\cite{bib:photos}, before the detector simulation was applied
to take account into the experimental effects.


\section{Experimental apparatus of the E325 experiment}	
\label{experiment}
\subsection{Overview}
\label{experiment-overview}

We performed the E325 experiment at the 
KEK 12~GeV Proton Synchrotron (KEK-PS).
This experiment was proposed in 1993, and spectrometer construction began in 1996.
Moving ahead, physics data acquisition commenced in 1997 and continued until 2002. In this manuscript, the last two data-collection runs, one performed in November and December 2001, and the second performed in February and March 2002, were analyzed.
The subsequent sections provide information on the detectors adopted in the 2001 and 2002 experiments.
More detailed descriptions of the spectrometer can be found in Ref.~\cite{bib:sekimoto_nim}.

\begin{figure*}[tb]
\begin{center}
\includegraphics[width=12cm]{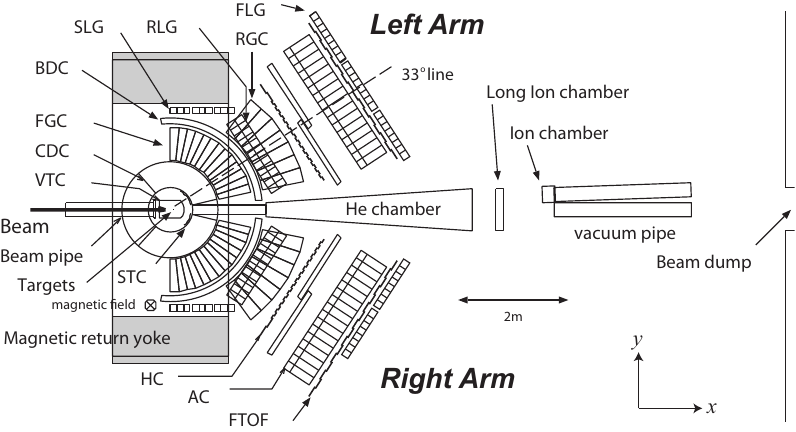}
 \caption{ Schematic plan view of the spectrometer. The magnetic field is aligned perpendicular to the paper. The $x$ axis is aligned along the beam axis, whereas the $y$ axis is horizontal and points toward the left when viewed from the upstream direction.}
 \label{fig-spectrometer1}
 \end{center}
\end{figure*}

\begin{figure}[tb]
\begin{center}
\includegraphics[width=10cm]{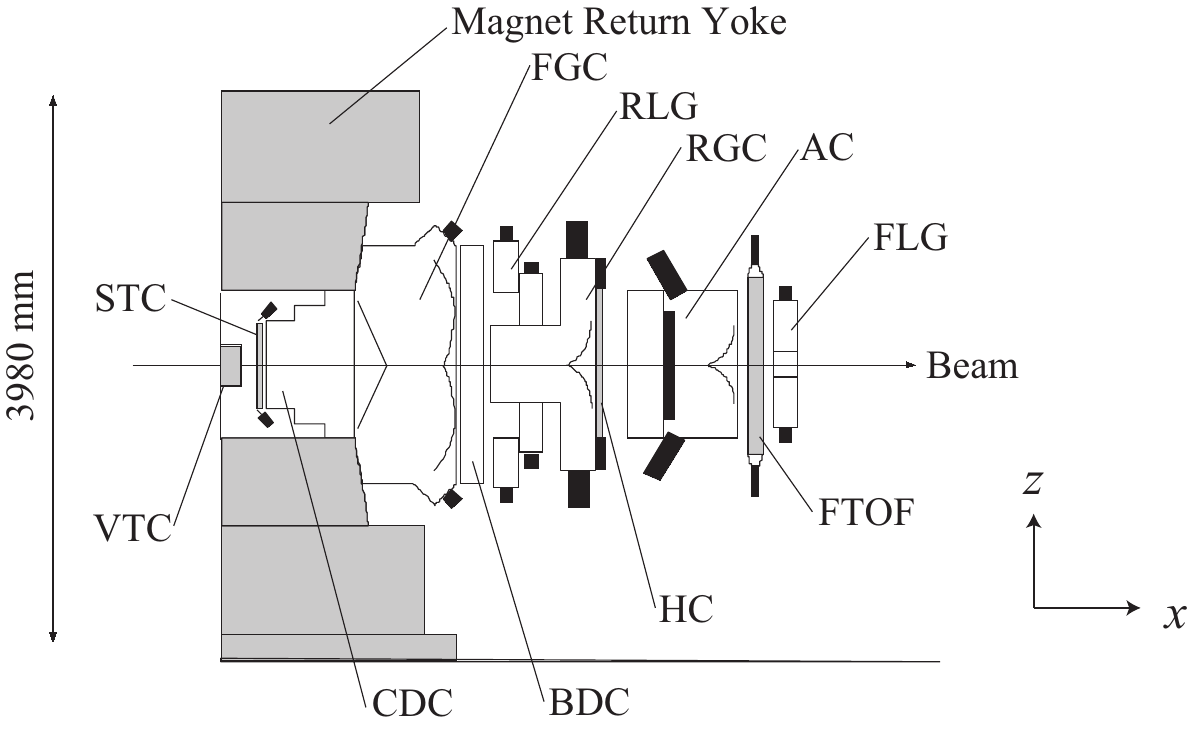}
 \caption{ 
Vertical cross-section of the spectrometer along the 
line 33\degree \ from the beamline 
depicted in Fig.\ref{fig-spectrometer1}. The $x$ axis coincides with the beam direction, while the $z$ axis points upward.
}
 \label{fig-spectrometer2}
 \end{center}
 \end{figure}

The E325 experiment utilized the magnetic spectrometer, designed to detect  \ee\ pairs and \KK\ pairs
from vector meson decays, situated at the 
EP1--B primary beamline~\cite{bib:takasaki}
in the North counter hall of KEK--PS. 
As depicted in the schematic views presented in Figs.~\ref{fig-spectrometer1} and 
\ref{fig-spectrometer2},
the spectrometer featured two arms, left and right.
During the experiment, a 12~GeV primary proton beam was delivered
onto nuclear targets positioned at the center of the 
spectrometer magnet, producing vector mesons such as $\rho$, $\omega$, and $\phi$ mesons.
Within the right-handed coordinate system of the setup, the $x$ axis was directed along the beam axis, the $z$ axis pointed upward, and the $y$ axis was directed to the left when viewed from the upstream direction.
%
The dipole magnet was designed and placed to allow the beam injection along the horizontal center line
of the magnet in the absence of a magnetic field.
Conversely, under the normal operation of the magnet, the magnetic field was directed 
from top to bottom, horizontally deflecting the 12 GeV proton beam by 2.2\degree.
Furthermore, the steering magnet installed on the beamline ensured precise alignment of the beam onto the center of the targets.

Targets were aligned along the center line of the spectrometer
magnet. 
The experimental setup for electron identification included two types of gas \cherenkov counters and three types of lead glass electromagnetic calorimeters.
The angle coverages of the electron identification counters were $\pm$23\degree\ in the vertical direction and 
$\pm$\,12\degree\ to $\pm$\,90\degree\ in the horizontal direction, as
measured from the center line of the spectrometer magnet.
Drift chambers were used as trackers.
Under the normal operating magnetic field conditions of the spectrometer magnet, particles produced at the target with a momentum below approximately 0.1 GeV/$c$
were unable to reach most outside trackers.
The total radiation length of detector materials
integrated from the center of the spectrometer
to the outermost tracker was 4.6\% $X_0$.
Table~\ref{table-material} lists the utilized detector materials.
The total number of read-out channels of the spectrometer was approximately 3400,
including the number of photomultiplier tubes (PMTs) and signal wires.

\begin{table}[tb]
\caption{ Detector materials and their respective radiation length within the tracking region, extending from the center of the spectrometer to radius of 1680~mm, where the Barrel Drift Chambers (BDCs) were located. Targets, described in Table~\ref{table-target}, are not included here.
}
\label{table-material}
\begin{center}
{\normalsize
\begin{tabular}{l|l|r|c|r|r}
\hline
Counter & Material & 
\multicolumn{1}{l|}{Radial position}  & 
\multicolumn{1}{c|}{Thickness} & 
\multicolumn{1}{c|}{$X_{0}$} & 
\multicolumn{1}{c}{$X/X_{0}$}\\
& & 
\multicolumn{1}{c|}{[mm]} & 
\multicolumn{1}{c|}{$X$ [mm]} &
\multicolumn{1}{c|}{[mm]} & 
\multicolumn{1}{c}{[\%]} \\
\hline
\hline
Vertex drift
  & Ar-C$_2$H$_6$(50:50) &   0.00 & 245.00   & 177608.0 & 0.138 \\
chamber   & Wire(Cu-Be)    &  -     &   0.0173 &     15.8 & 0.109 \\
(VTC) & Wire(W)        &  -     &   0.0004 &      3.5 & 0.011 \\
        & Mylar          & 245.00 &   0.05   &    287.4 & 0.017 \\
\hline
        & Air            & 245.05 & 134.95   & 305225.0 & 0.044 \\
\hline                                     
Start timing & Scintillator   &  380.00 &   5.00 & 424.0     & 1.179 \\
counter (STC)        & Wrapping        &  385.00 &   0.20 & 287.0     & 0.070 \\
\hline
        & Air            &  385.20 &  14.80 & 305225.0  & 0.005 \\
\hline                                                  
Cylindrical & Mylar          &  400.00 &   0.05   &    287.4 & 0.017 \\
drift chamber & Ar-C$_2$H$_6$(50:50) &  400.05 & 479.95   & 177608.0 & 0.270 \\
(CDC)       & Wire (Cu-Be)   &    -    &   0.031  &     15.8 & 0.196 \\
        & Wire (W)       &       - &   0.0005 &      3.5 & 0.014 \\
        & Mylar          &  880.00 &   0.05   &    287.4 & 0.017 \\
\hline
        & Air            &  880.05 &  19.90   & 305225.0 & 0.007 \\
\hline                                                  
Front gas     & Mylar          &  899.95 &   0.05 & 287.4     & 0.017 \\
\cherenkov        & Isobutane      &  900.00 & 660.00 & 181801.0  & 0.363 \\
counter        & Acrylic mirror & 1560.00 &   3.00 & 340.8     & 0.880 \\
(FGC)        & Aluminum wall  & 1563.00 &   1.00 & 88.9      & 1.125 \\
\hline
        & Air            & 1564.00 &   5.95 & 305225.0  & 0.002 \\
\hline                                                  
BDC & Mylar          & 1569.95 &   0.05 & 287.4     & 0.017 \\
  &Ar-C$_2$H$_6$(50:50) & 1570.00 & 110.00 & 177608.0  & 0.062 \\
     & Wire (Cu-Be)   & -       & 0.010  & 15.8      & 0.063 \\
        & Wire (W)       & -       & 0.0001 &  3.5      & 0.003 \\
\hline
\hline
Total     &                &         &        &            & 4.628 \\
\hline

\end{tabular}
}

\end{center}
\end{table}


\subsection{Beamline and targets}
 \label{section-beamline}
\begin{figure*}[tb]

\begin{center}
\includegraphics[width=15cm]{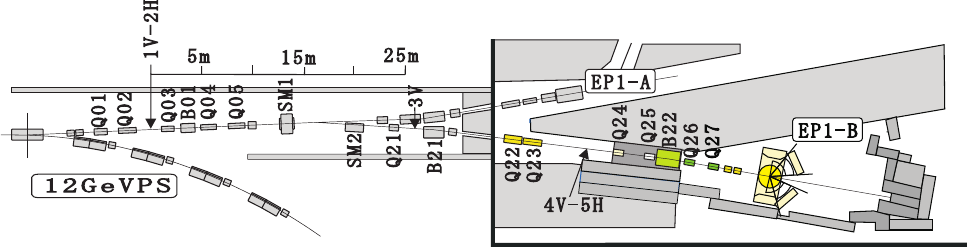}

 \caption{ Schematic view of the beamlines around the extraction
point from the 12 GeV PS to EP1 primary beamline. The spectrometer
is also depicted in the right-hand side of the figure. The positions
of five beam collimators (1V--2H, 3V, 4V--5H) are indicated by arrows.
\FIGciteNaruki
}
 \label{fig-beamline}
 \end{center}
 \end{figure*}

The KEK--PS typically operated with a beam intensity of \expo{3}{12} protons per spill (pps).
Injection and acceleration processes required 2 sec, with a beam spill duration of 2 sec, resulting in a repetition cycle of 4 sec.
KEK--PS was equipped with two slow--extraction beamlines, EP1 and EP2.
Figure~\ref{fig-beamline} depicts the extraction point from the 12 GeV PS to the EP1 beamline. 
EP1--B, commonly used for primary-beam experiments, was a branch line from the main EP1--A line, dedicated to secondary-beam production.

In the aforementioned data-collection periods, 
an intensity-asymmetric double-slow extraction
was performed~\cite{EP1B}. 
Almost the entire beam was directed toward EP2 and utilized for
the secondary beam production in other experiments.
Meanwhile, the beam intensity directed toward EP1 was 1/100 that directed toward EP2, 
that is, \expo{1\text{--}2}{10} pps.
For beam delivery to the EP1--B line,
the intensity of the extracted beam was adjusted to approximately \expo{1}{9} pps using beam collimator 1V, while the beam halo was reduced using collimators 2H--5H,  indicated by arrows in Fig.~\ref{fig-beamline}.
Radiation protection constraints limited the maximum intensity of the beamline to \expo{4}{9} pps.

The beam was delivered through a vacuum pipe into the spectrometer setup.
Figure~\ref{fig-spectrometer1} illustrates the plan view of the vacuum pipe and the spectrometer.
The vacuum pipe was terminated by the 200 \um\ of mylar placed just upstream
of the VTC. 
This VTC, wherein
the targets were installed,
was positioned at the center of the magnet. 
As depicted in  Fig.~\ref{fig-spectrometer1}, 
a helium-filled pipe and a tapered-shape 
chamber, resulting in a total length of 5.5 m, were installed
just downstream of the VTC to reduce the room background
from the beam-air interactions surrounding the detectors.
The tapered chamber was designed to handle both scenarios with and without magnetic field deflected.

The beam protons were not counted particle by particle. Instead, 
the beam intensity was monitored using two ionization chambers~\cite{IonChamber}
positioned downstream of the helium chamber.  
The trajectories of the beam differed significantly in the presence and absence of the spectrometer magnetic field.
Among the two ionization chambers, the larger ionization chamber covered both trajectories, while the smaller chamber covered only the deflected trajectory.
To mitigate room background, 2.5-m long vacuum pipes were
placed between the ion chambers and the beam dump.
The beam pipes, helium pipe and chamber, and ion chambers are not depicted in 
Fig.~\ref{fig-spectrometer2}.

The targets were installed along the beam path
using rotatable target holders set within the VTC.
Each target foil was supported by a 1-mm-thick and 10-mm-wide polyethylene stay.
Table~\ref{table-target} lists the targets adopted in the 2001 and 2002 experimental runs, along with their dimensions
and positions. In the 2001 experimental run, the C target was positioned at the center, with two Cu targets located upstream and downstream, each spaced 48~mm from the C target. In the 2002 experimental run, the C target was positioned at the center with four Cu targets located upstream and downstream, each spaced 24~mm.
\begin{table}
    \caption{
      Targets utilized in the 2001 and 2002 experimental runs.
    }
      \label{table-target}
  \begin{center}
    \begin{tabular}{ c| c c c c c c c}
      \hline\hline
     Year & Target & Relative position &Width&Height&Thickness& $\lambda_I$ &
     $X_0$   \\
          & material& (mm)    &(mm)&(mm)  & (mg/cm$^2$)&(\%)  &(\%)  \\
      \hline
      2001 & C  & 0                   &25&25&92 & 0.11 & 0.21 \\
           & Cu ($\times2$)& $\pm$48  &25&25&73 & 0.054& 0.57 \\
           & (in total)    &          &  &  &   & 0.22 &   \\
      \hline
      2002 & C  & 0                   &10&25&184& 0.21 & 0.43 \\
         & Cu ($\times4$) & $\pm$24, $\pm$48 &10&25&73 & 0.054 & 0.57 \\
         & (in total)    &                  &  &  &   &  0.43 &  \\
                      \hline \hline
    \end{tabular}
  \end{center}
\end{table}

\begin{figure}[ptb]
\begin{center}
\includegraphics[width=8cm]{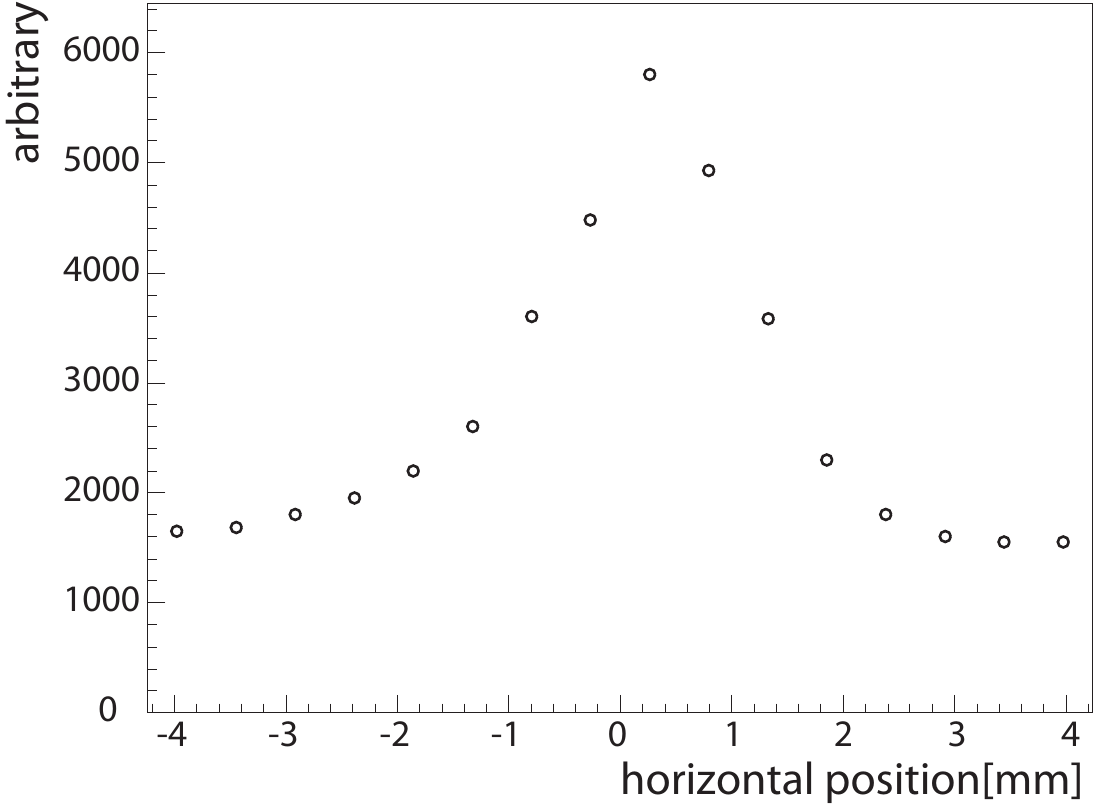}

 \caption{ Horizontal beam profile of the 2002 experimental run, recorded as the trigger rate by varying 
the current of the bending magnet B22. The current was converted into the horizontal 
distance, with the origin of the position being arbitrary.
}
 \label{fig-beamprofile}
 \end{center}
 \end{figure}

The beam profile at the center of the spectrometer was recorded using
the C target, rotated by 90\degree, as a probe with a width of
0.5--1 mm. This probe remained fixed while the beam position
was adjusted horizontally by the bending magnet B22, 
which is depicted in Fig.~\ref{fig-beamline}. 
The counting rate of the particles from the target 
as a function of the beam position was adopted as 
a measure of the beam profile.
The recorded profiles are shown in Fig.~\ref{fig-beamprofile}. 
The horizontal widths of the beam in standard deviations were 1.59 mm and 0.83 mm in
the 2001 and 2002 runs, respectively. Each deviation
was evaluated from the profiles considering the probe width.

In the 2002 experimental run, the typical beam intensity was \expo{7}{8} pps, 
with a total target thickness of 0.4\% the nuclear 
interaction length. This resulted in a value of 1.4~MHz for the
typical interaction rate on targets, considering the spill duration of 2 sec.
In practice, 1.8 sec of the flat region of the spill was used for the data
taking. Table~\ref{table-protons-target} summarizes the total number of protons on the target.
\begin{table}
    \caption{
      Number of protons on the targets
   in the 2001 and 2002 runs.
    }
 \label{table-protons-target}
  \begin{center}
    \begin{tabular}{ c| c|c c c| c}
      \hline\hline
     Year & Days &Number & Spill  & Average protons & Trig \\
          & & of protons  & & per spill & \\
      \hline
      2001 & 34& \expo{3.8}{14} & 435 k &\expo{8.7}{8}&310M\\
      \hline
      2002 & 32&  \expo{3.2}{14} &508 k &\expo{6.4}{8}&510M\\
                      \hline \hline
    \end{tabular}
  \end{center}
\end{table}

\subsection{Spectrometer magnet}
\label{section-magnet}
The spectrometer magnet utilized in the E325 experiment was a dipole-type magnet, weighing approximately 300 tons,
with dimensions
of 5655 mm (W) $\times$ 3980 mm (H) $\times$ 2120 mm (D).
The diameter of the pole pieces
was 1760 mm, while the pole gap space was 907 mm. 
The return yoke and pole pieces
were recycled from the INS \footnote{Institute for Nuclear Study,
University of Tokyo} 160~cm FM/FF cyclotron magnet~\cite{FMmagnet}.
The pole pieces were then remodeled, and their height was extended.
The upper and lower coils, each with 168 turns, were newly fabricated.

Under a normal operating current of 1560 A
and voltage of 300 V,
the magnetic field strength 
at the center of the pole pieces was 0.71~T,
while the field integral within the tracking region was 0.81 T$\cdot$m.
The tracking region is extended from the center of the spectrometer
to a radius of 1680 mm, where the outer trackers
were located.
Field mapping was performed using Hall probes, and the recorded
map was compared with a map derived from a three-dimensional 
calculation performed using the finite element solver, TOSCA~\cite{TOSCA}.
The comparison revealed good agreement between the two. 
Following a 0.2\% of scale correction of the field strength,
the determined momentum values agree within $\pm$ 0.2\% for particles with momenta in the range of 0.5 to 2.0 GeV/$c$. Owing to this high level of accuracy, the computed map was used for further analysis.

Throughout the experiment, the field strength was monitored spill-by-spill
using a nuclear magnetic resonance probe located at the center of the bottom pole piece.
The drift of the magnetic field was within 0.07\% throughout the 
data-collection period. This value
was corrected run-by-run by tuning the scale of the field map.

\subsection{Tracking chambers}
The experiment adopted three types of tracking chambers, namely the VTC, CDC, and BDC, arranged sequentially from the inside to the outside of the spectrometer, as depicted in Fig.~\ref{fig-spectrometer1}.
All drift chambers were filled with a gas mixture comprising argon(50\%) and ethane(50\%) at 1~atm.
The gas of the VTC was supplied through
an ethanol bubbler at room temperature
to prevent discharge through
hydro-carbon deposition on the anode wire.
For the final momentum
determination of the particles, only the CDC and BDC were used. 
The typical
position resolutions of the CDC and BDC were both below 300 $\mu$m.

The CDC comprised 10 layers of drift cells, grouped
into three super-layers: the inner, middle, and outer super-layers.
Each super-layer comprised three or four layers, labeled
\{X, \Xp, U\}, \{X, \Xp, V, \Vp\}, \{U, X, \Xp\},
arranged from the inside to the outside.
Three layers demonstrated angle coverages of 
$\pm$12\degree \ to $\pm$132\degree\ in the horizontal direction and 
$\pm$23\degree \ in the vertical direction.
Meanwhile, the BDC featured four layers, labeled \{X, \Xp, V, U\}, arranged from the inside to the outside.
These layers demonstrated angle coverages of
$\pm$6\degree \ to $\pm$96\degree\ in the horizontal direction and 
$\pm$23\degree \ in the vertical direction.
The "X" and "\Xp" layers were equipped with vertical wires, while the "U" and "V" layers were equipped with tilted wires to measure the vertical position.
The tilt angle was approximately 0.11 rad
for both the CDC and BDC.

The VTC comprised two super-layers, \{X, \Xp, X\} and \{X, \Xp, X\},
arranged from the inside to the outside,
referred to as the backward and forward super-layers, 
respectively~\cite{bib:sekimoto_nim}. 
The forward (backward) super-layer exhibited angle overages of
$\pm$6\degree (18\degree) \ to $\pm$24\degree (141\degree)\ 
at a radius of 200 (100) mm in the horizontal direction
and 
$\pm$29\degree (48\degree) \ in the vertical direction.
The VTC was not equipped with any tilted-wire layers.
While the VTC was used for calibrating and checking the tracking system,
it was not involved in the final momentum determination
owing to its limited acceptance and efficiencies, as outlined in Table~\ref{table-chambereff}.

\subsection{Electron identification counters\label{eid-counters}}
The experimental setup included two or three stages of electron identification (eID) counters.
Among these, FGC 
constituted the first-stage eID counters, situated between the CDC and BDC, covering a horizontal angle of $\pm$12\degree \ to $\pm$90\degree .
Meanwhile, three types of 
eID counters positioned just behind the BDC constituted the second-stage counters. Among these,
the side lead glass EM calorimeters (SLG) covered a horizontal angle range of  $\pm$57\degree\  to $\pm$90\degree, with nine segments in each arm.
The rear gas \cherenkov\ counters (RGC) covered angles of 
$\pm$12\degree\ to $\pm$54\degree\ horizontally and $\pm$6\degree\
vertically. Rear lead glass EM calorimeters (RLG) covered angles of
$\pm$12\degree\  to $\pm$56\degree\ horizontally, with 12 segments, and  $\pm$8\degree\  
to $\pm$23\degree\ vertically, with two segments.
Forward lead glass EM calorimeters (FLG), positioned
just behind the forward time-of-flight (FTOF) counters, constituted the third-stage counters, covering RGC acceptances with 24 segments in
the left arm. In the right arm, only 16 segments were located, by the limitation
due to the layout of the experimental area.

Both the FGC and RGC were threshold-type gas \cherenkov counters 
using isobutane as their radiator gas.
The refractive index of isobutane is 1.00127 (1.0019), and its
momentum threshold for pions is 2.7 (2.3) GeV/$c$ at room temperature
(0\degree C). For each arm, the FGC was divided into 13 segments, while
the RGC was divided into seven segments horizontally, with
each separated by aluminized mylar, featuring a horizontal coverage of 6\degree.

Lead glass SF6W, recycled from the TOPAZ~\cite{TOPAZLG} detector at 
TRISTAN, was utilized for the RLG, SLG, and FLG. This lead glass featured a refractive
index of 1.8 and a radiation length of 17 mm.
After being reshaped to rectangular forms, 
five (three) lead glass blocks were 
vertically stacked and glued together using epoxy resin for the SLG (FLG).
Each stack was read out by a PMT from the top and bottom.  
The RLG utilized the original tapered-shaped lead glass counters from TOPAZ,
and these were read out from only one end.
Unlike typical calorimeters, these counters covered their own acceptances as hodoscopes, to cover large acceptances at low costs. Consequently, their depths, measured in
the radiation length (6.5$-$7.9 $X_0$), were smaller than that in TOPAZ, 20~$X_0$.
Although the achieved energy resolution,
15\%/$\sqrt{E(\mathrm{GeV})}$, was lower than the TOPAZ value of 8\%/$\sqrt{E(\mathrm{GeV})}$~\cite{TOPAZLG}, it was sufficient to distinguish electrons from pions in this experiment.

\subsection{Trigger and data acquisition}
\label{section-trigger} 
To trigger the vector mesons in the \ee\ channel, 
an electron (or positron) candidate was required
in each spectrometer arm (left and
right). Furthermore, a single hit on the STC in
each arm was required. The charge of each
track was not identified by the trigger.
An electron candidate was defined by the coincidence of the first-stage 
eID counters (FGC) and the geometrically matched
second-stage eID counters (RGC, RLG, and SLG).
The abovementioned geometrical
matching was determined using the matrix-coincidence logic module. 
The matching window size corresponded to a
track momentum of approximately 0.4 GeV/$c$.
The typical trigger-request (accept) rates per spill were
1.2k (0.7k) and 1.6k (0.9k)
for the 2001 and 2002 runs, respectively.

To gather various control data, the following triggers were used.
Minimum bias trigger, requiring only one hit on the STC, and one or two charged-particle trigger. 
The beam time for the control data, including
pedestal runs and no-magnetic field runs, was 
approximately 15\% of the total
data-collection duration for the 2001 and 2002 runs.

The online data acquisition system was built on the VME, TKO~\cite{TKO}
and CAMAC-based front-end electronics.
The data from the TKO front-end modules (the analog-to-digital converter (ADC) and time-to-digital converter (TDC)) were transferred in an event-by-event manner to 
memory modules on a VME crate using flat cables.
A CPU board (HP model 743) on the VME bus,
operated by HP-RT OS~\cite{HPRTDAQ}, was used to read the data during
the beam off time of 2 s in the beam cycle
from the memory modules and CAMAC/VME front-end modules (scalers)
via the VME bus. This board was also utilized to record the data
on a digital data storage tape (DDS-4) directly connected to the CPU board via a SCSI cable.
The data monitoring
system was constructed on a network-distributed computing
environment using the UNIX OS and Ethernet.
The software for data collection, recording, and monitoring 
was developed using the programming language C/C++, while
the graphical user interfaces of these systems were developed using the script language Ruby/Tk. 

The typical data-collection rate was 1.8 MB/spill in a
2-sec beam spill, corresponding to approximately
1 k events/spill. The typical event-data size
was 1.8 KB after the zero suppression and
pedestal subtraction by the front-end modules, 
but not compressed.
The collected data amounted to
approximately 1.2 TB during the 2001 run and 1.1 TB during the 2002 run.
The recorded data were also 
read out from the DDS tape in the counting house and 
transferred in semi real-time from KEK to RIKEN via WAN,
and recorded on the mass storage system HPSS~\cite{HPSS}.
The typical transfer rate in this process was 2 MB/sec.

     \section{Analysis}
\subsection{Outline of the data analysis scheme}
This section outlines the procedure adopted for analyzing the data collected during the 2001 and 2002 experimental runs.
Sec.~\ref{sec:calib} focuses on detector calibration.
For tracking detectors, the calibration process includes establishing the $x$-$t$ relation
and aligning the positions of the CDC and BDC.
For EM calorimeters, the emphasis is on energy calibration.
Sec.~\ref{sec:reconstruction} details the process of track reconstruction and
vertex position determination using the hits in the CDC and BDC.
During the track reconstruction process, once track candidates were identified, they were fitted using the Runge-Kutta method to deduce their individual momenta.
An event vertex position was determined by identifying the closest point of approach from the fitted tracks in the event,
and then the interaction target was selected based on its proximity to the vertex.
Finally, two oppositely charged tracks were fitted simultaneously with a constraint requiring the pair to have their vertex point on the target foil. 

Sec.~\ref{sec:eid} covers the electron identification process.
Here, associated hits in the counters were required to have ADC values exceeding specific threshold,
position matching with track candidates,
and timing matching with time-of-flights derived from the timing information of the electron identification counters and the STCs.
For the EM calorimeters, an $E/p$ cut was also applied.
Additional event selection criteria are summarized in Sec.~\ref{sec:selection}.

In order to evaluate all the experimental effects, we performed a detector simulation using Geant4~\cite{bib:geant4}.
Details of the simulation are presented in Sec.~\ref{sec:simulation}.
Based on this simulation, we subsequently evaluated misalignment parameters of the BDCs, mass scale factors, and the effects of additional mass smearing to reproduce the obtained mass spectra.
The acceptance cut was also improved to ensure uniform acceptances across all target positions.

\subsection{Calibration\label{sec:calib}}
\subsubsection{Drift time measurement and drift length determination}
The drift times of the CDC and BDC were measured using TKO 32-channel drift-chamber TDC modules (Dr.~T~II).
The full scale of the TDC modules was 1000~ns, with a typical time resolution of 350~ps.
The time gain was calibrated using a CAMAC TDC tester (RPC070),
and the measured gain was 0.7~ns/channel.

\begin{figure}[htbp]
 \begin{center}
  \includegraphics[width=10cm]{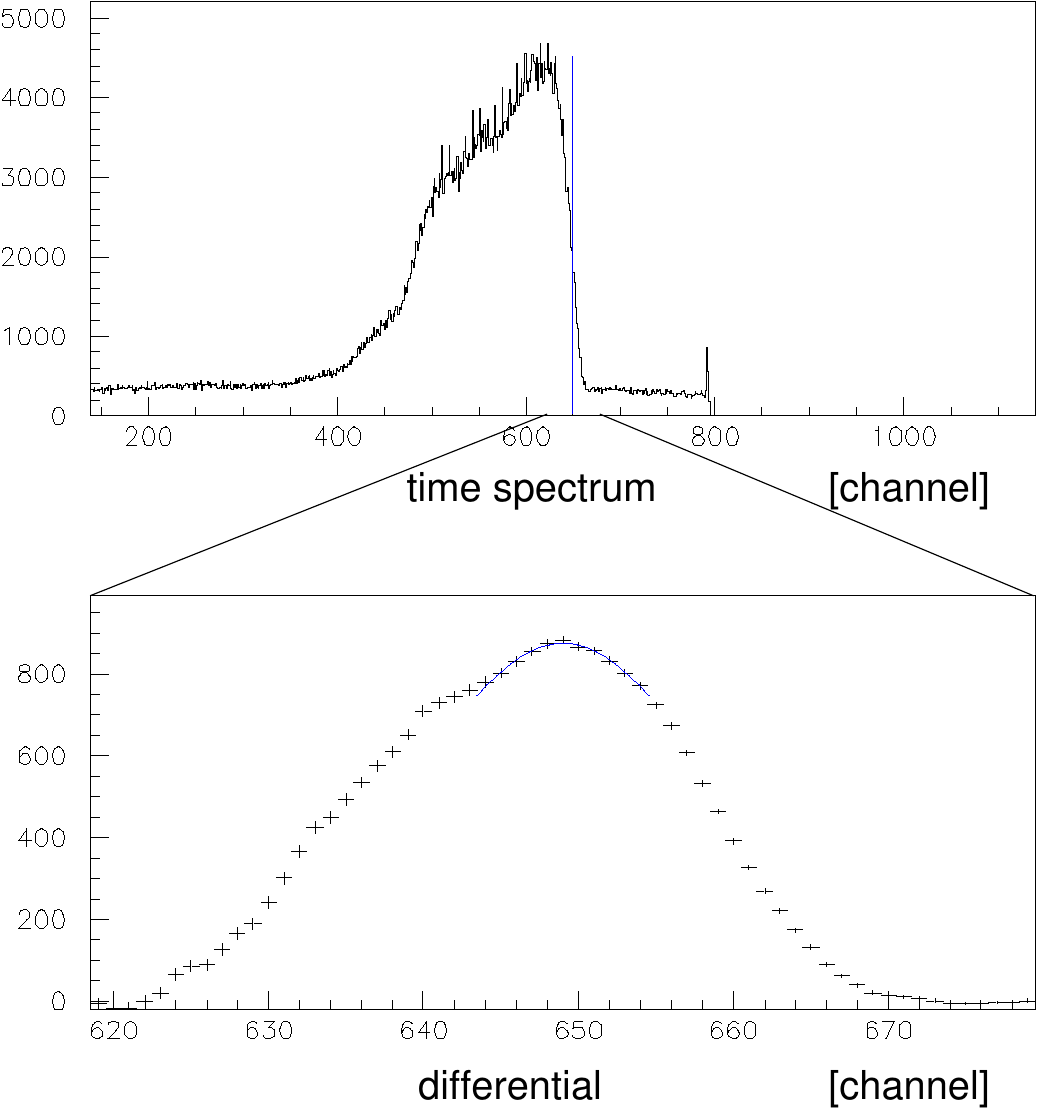}
  \caption{Typical drift-time distribution of the CDC (upper panel) and its differential distribution (lower panel).
  The blue solid line in the upper panel corresponds to the peak position indicated in the lower panel.
  \FIGciteMuto
    }
  \label{fig:t0_calib}
 \end{center}
\end{figure}

The relative time offset for each channel was determined based on the obtained drift time distribution.
As depicted in Fig.~\ref{fig:t0_calib},
the distribution exhibited an edge structure corresponding to signals generated near a wire.
The peak position of the differentiated spectrum served as the relative time offset of the channel.

\begin{figure}[htbp]
 \begin{center}
  \subfigure[2001]{
  \includegraphics[width=6cm]{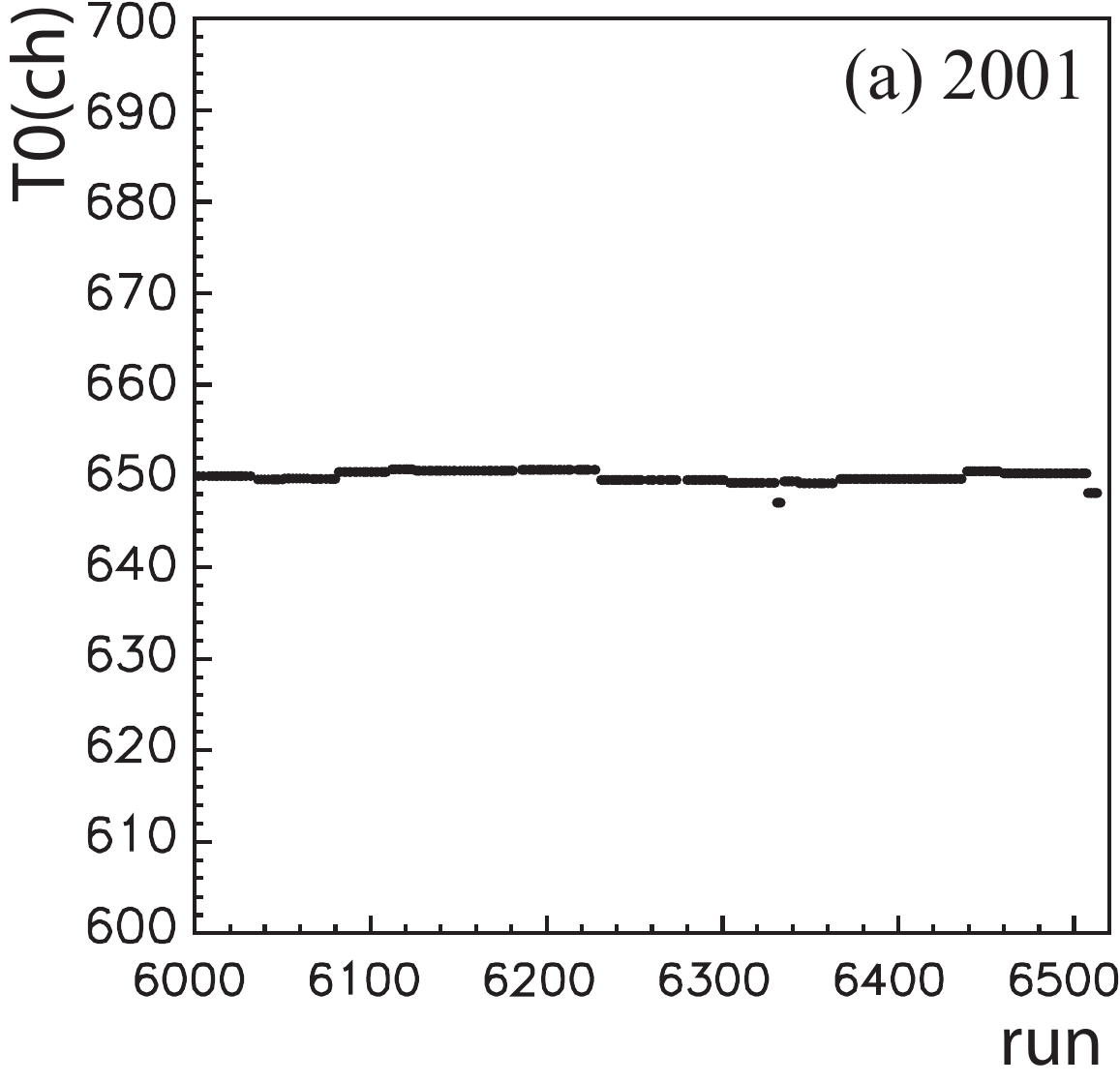}
  \label{fig:t0_rundep01}}
  \subfigure[2002]{
  \includegraphics[width=6cm]{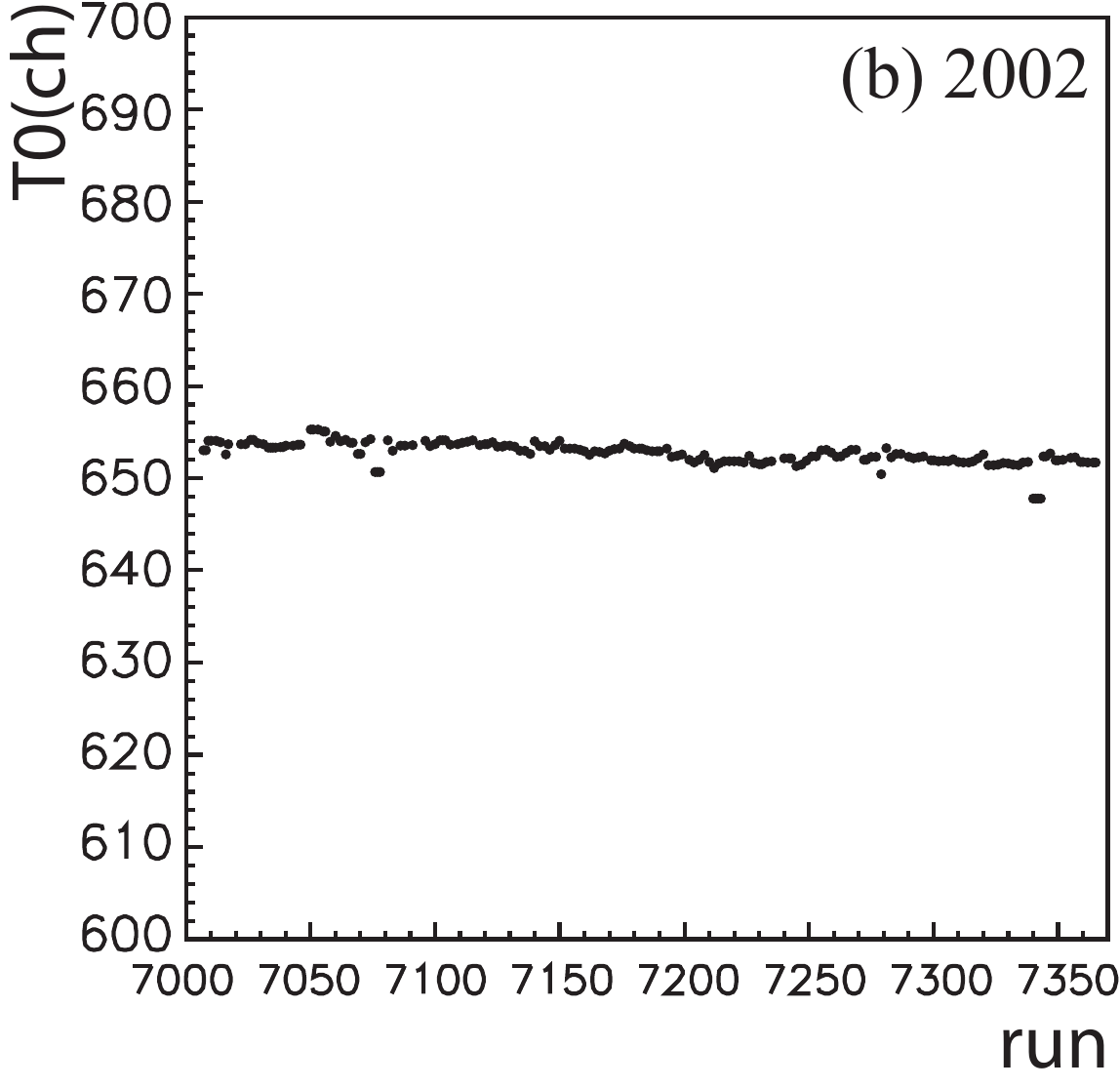}
  \label{fig:t0_rundep02}}
  \caption{Run dependence of the time offset of a typical wire in the CDC, recorded (a) in the 2001 and (b) 2002 experimental runs.
  The typical drift was within 4~ns.
  \FIGciteSakuma
       }
  \label{fig:t0_rundep}
 \end{center}
\end{figure}

Calibration procedures were conducted in a run-by-run manner to compensate for timing drifts in cables and a common stop logic resulting from temperature changes.
Figure~\ref{fig:t0_rundep} illustrates the run dependence of the time offset. The typical drift was within 4~ns.

\begin{figure}[htbp]
 \begin{center}
  \includegraphics[width=15cm]{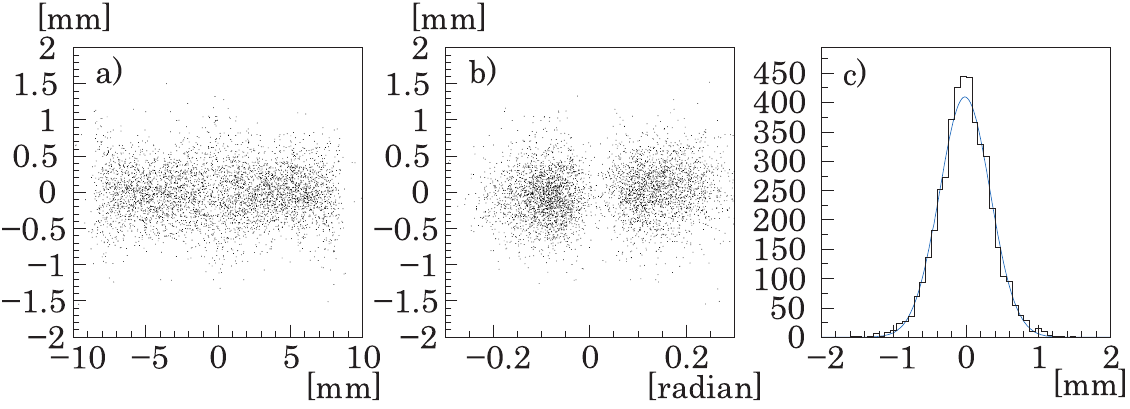}
  \caption{Typical residual distribution of the CDC after calibration.
  Residual as a function of the drift length (a), as a function of the incident angle (b),
  and the overall residual distribution (c).
  \FIGciteSakuma
    }
  \label{fig:cdc_residual}
 \end{center}
\end{figure}

To convert measured drift times into hit positions, the relation between drift length ($x$) and drift time ($t$) was evaluated for each layer of the CDC and BDC.
First, the $x$-$t$ relation derived by GARFIELD~\cite{bib:garfield}
served as the basis for the following iterative adjustments.
Residuals, as functions of the drift length, between the hit positions and track positions determined by the Runge-Kutta fitting method, were computed based on the initial $x$-$t$ relation.
The layer that to be calibrated was excluded from this fitting.
By minimizing systematic shifts in the residual distributions, a new $x$-$t$ relation was calculated.
Here, the common time zero and scaling factor of the drift velocity were adopted as parameters.
This procedure was repeated until convergence.
Because the incident angle dependence was negligibly small, it was disregarded.
Figure~\ref{fig:cdc_residual} illustrates a typical residual distribution of the CDC after calibration.
The typical position resolution was 350~\textmu m.
Additional details regarding the resolution of the tracking chambers for the detector simulation are presented in Sec.~\ref{sec:simulation}.

\subsubsection{Chamber efficiency and resolution}
\label{section-residual}
The efficiency of each tracking chamber layer 
was evaluated based on the \ee\ triggered events.
The efficiency was defined using the numbers of
two track categories: $N_{\rm{all}}$ for tracks with hits in all layers and $N_{\rm{off}}(layer)$ for tracks
with hits in all layers except the relevant $layer$.
Accordingly, the efficiency ${\cal{E}}(layer)$ was defined as follows:
\begin{equation}
  {\cal{E}}(layer) = N_{\rm{all}}/(N_{\rm{all}}+N_{\rm{off}}(layer)).
 \end{equation}
Table~\ref{table-chambereff} presents the evaluated efficiency values.

\begin{table}[hbpt]
 \caption{Evaluated efficiency for each chamber layer.}
 \label{table-chambereff}
 \begin{center}
  \begin{tabular}{l|l|l}
   \hline
   Super-layer & Layer         & Efficiency [\%]\\
   \hline
   VTC backward & (X,\Xp,X)     & (73.5, 64.7, 52.4) \\
   VTC forward  & (X,\Xp,X)     & (92.4, 100.0, 93.3) \\
   CDC inner    & (X,\Xp,U)     & (96.1, 96.8, 90.2) \\
   CDC middle   & (X,\Xp,V,\Vp) & (94.6, 95.7, 94.6, 93.1) \\
   CDC outer    & (U,X,\Xp)      & (97.2, 95.7, 94.2) \\
   BDC          & (X,\Xp,U,V)   & (99.0, 99.1, 99.0, 96.8) \\
   \hline
  \end{tabular}
 \end{center}
\end{table}

\begin{figure}[htbp]
 \begin{center}
  \subfigure[Residual]{
  \includegraphics[width=7cm,keepaspectratio]{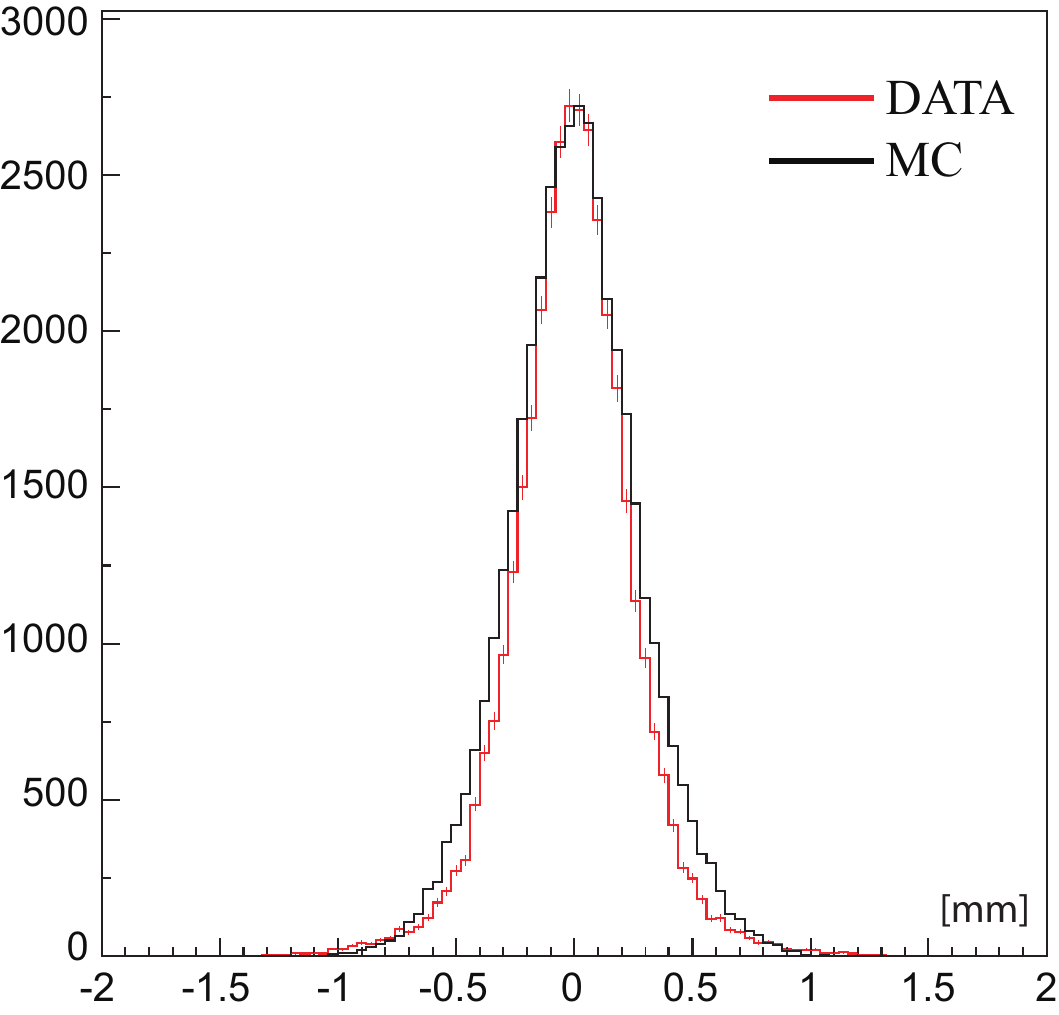}}
  \subfigure[Reduced $\chi^2$]{
  \includegraphics[width=7cm,keepaspectratio]{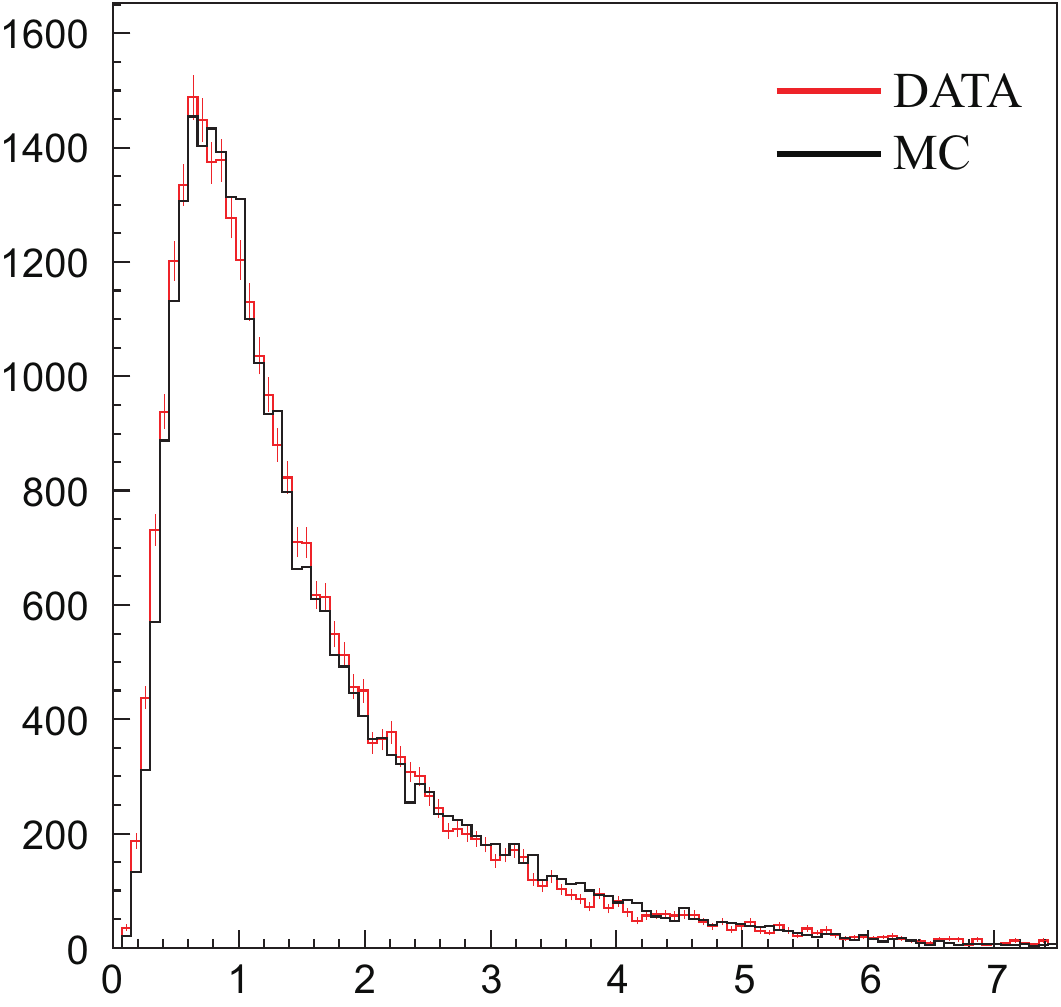}}
  \caption{Typical residual distribution of the tracking chambers and reduced $\chi^2$
  distributions derived from the real data (red) and tuned detector simulation (black).
  \FIGciteSakuma 
  }
   \label{fig-residual}
 \end{center}
\end{figure}

The track residual in each layer
was approximated assuming a double-Gaussian shape,
obtained by summing the main Gaussian component with
the broader Gaussian component to reproduce the
broad tail in the residual distribution.
The standard deviation 
of the main component, $\sigma_1$, was obtained 
for each layer, while the $\sigma_2$ value for the broader
component and the amplitude ratio of the broader
component to the main component
were obtained for each super-layer. Furthermore, shape parameters were
determined iteratively 
through the detector simulation described in Sec.~\ref{section-sim}
to reproduce the residual
shape obtained for each layer and the track $\chi^2/\mathrm{ndf}$ distribution
of the final \ee\ sample.
Table~\ref{table-residual} outlines the determined parameters,
and Fig.~\ref{fig-residual} illustrates the typical residual distribution and the track $\chi^2/\mathrm{ndf}$
distribution.

\begin{table*}
 \caption{Evaluated residual parameters for each tracking chamber layer.
 Here, $\sigma_1$ denotes the widths of the main component of the double Gaussian distribution, $\sigma_2$ denotes those of the second
 component, and ``ratio'' denotes the amplitude ratios of the second component
to the main component for each super-layer.}
 \label{table-residual}
 \begin{center}
  \small
  \begin{tabular}{l|l|l|l|l}
   \hline
   Super-layer & Layer & $\sigma_1$ [mm] & $\sigma_2$ [mm] & Ratio \\
   \hline
   VTC backward & (X,\Xp,X)     & (0.389, 0.394, 0.429)        & 0.75  & 0.37
   \\
   VTC forward  & (X,\Xp,X)     & (0.174, 0.212, 0.183)        & 0.75  & 0.37
   \\
   CDC inner    & (X,\Xp,U)     & (0.197, 0.203, 0.286)        & 0.595 & 0.436
   \\
   CDC middle   & (X,\Xp,V,\Vp) & (0.238, 0.217, 0.183, 0.181) & 0.644 & 0.367
   \\
   CDC outer    & (U,X,\Xp)     & (0.285, 0.212, 0.261)        & 0.718 & 0.395
   \\
   BDC          & (X,\Xp,U,V)   & (0.292, 0.289, 0.340, 0.261) & 0.661 & 0.488
   \\
   \hline
  \end{tabular}
 \end{center}
\end{table*}

\subsubsection{Detector position calibration\label{ssec:detector_alignment}}

The coordinate system for the present analysis was established with its origin at the center of the CDC.
Other tracking chambers were positioned and surveyed within this coordinate system.
The surveys were conducted before and after the beam times.
The magnet position was also calibrated based on these surveys.

\begin{figure}[htbp]
 \begin{center}
  \includegraphics[width=12cm]{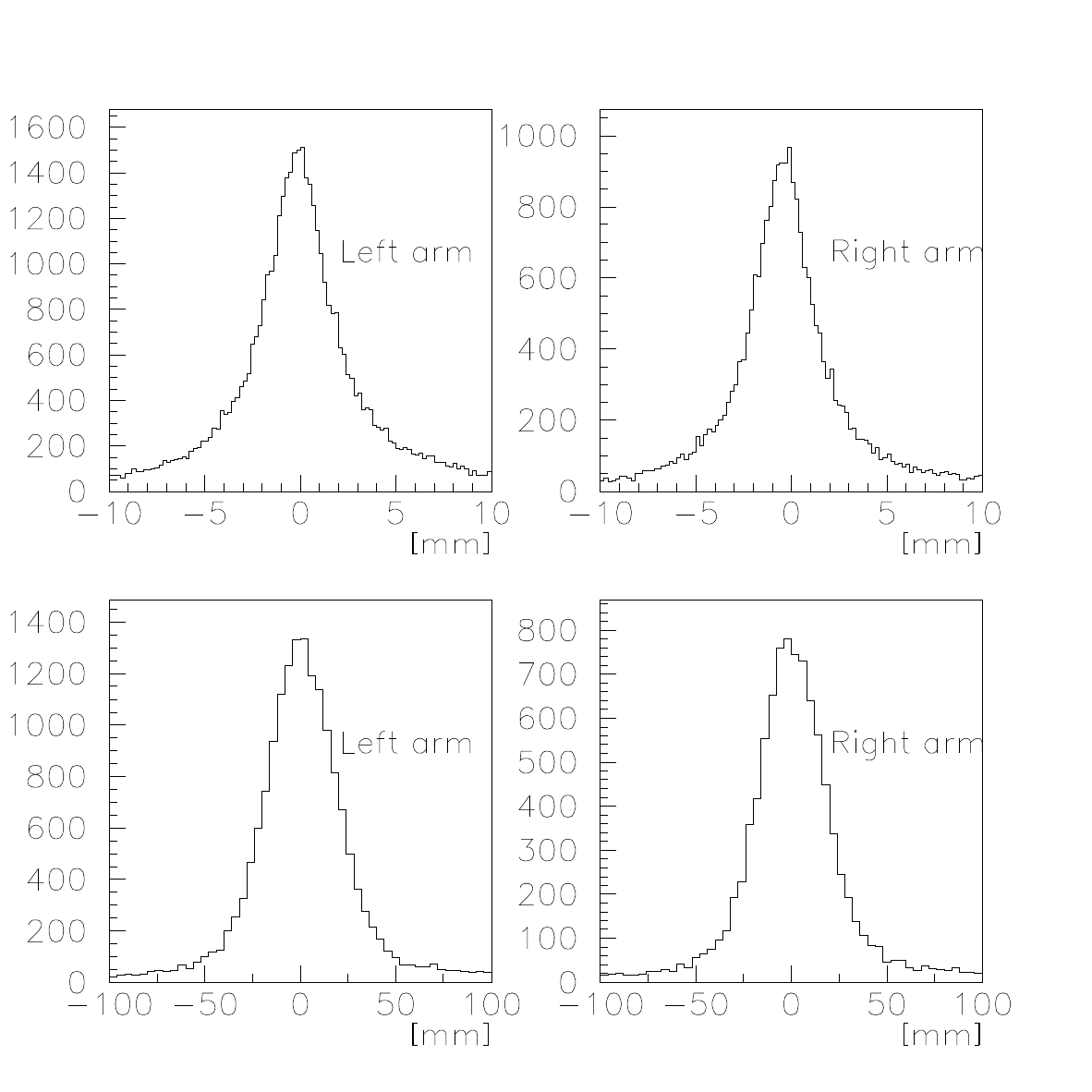}
  \caption{Residual distributions between the hit positions of the BDC and 
   extrapolated CDC tracks for the 2002 data.
   The upper panels present the residuals along the horizontal direction at
   the BDC location, while the lower panels present the residual along the vertical direction at the BDC position.
   \FIGciteNaruki
   }
   
   \label{fig:bdcalignres}
 \end{center}
\end{figure}

In addition to the magnet position, the position of the BDC was also calibrated based on the results of the surveys
and refined using the data recorded under a zero magnetic field.
In these zero field data, the CDC hits were fitted with straight tracks and extrapolated to the BDC's.
By minimizing the systematic shift in the residual distributions between the extrapolation of the CDC track and BDC hits,
we obtained six parameters determining the translation and rotation of each BDC.
The typical residual distributions after this alignment are depicted in Fig.~\ref{fig:bdcalignres}.
The parameters of the translation ($\varDelta x$, $\varDelta y$, $\varDelta z$) and rotation ($\tau_{x}$, $\tau_{y}$, $\tau_{z}$) are summarized in Table~\ref{tab:bdc_alignment}.

\begin{table}[htbp]
\caption{
Translation and rotation parameters of the BDC with respect to the
nominal position.
Two sets of parameters are available for the 2001 data as an
earthquake occurred during this data-collection period. 
To obtain the position of BDC,
it must be first translated by $(\varDelta x, \varDelta y, \varDelta z)$ and subsequently
rotated by $(\tau_x, \tau_y, \tau_z)$ around the three axes crossing the origin of the global coordinate system.
\label{tab:bdc_alignment}
}
\begin{center}
{\small
\begin{tabular}[tb]{|c|c|cccccc|}\hline
Year& \hspace{-3pt}Arm\hspace{-3pt} & $\varDelta x$ [mm] & $\varDelta y$ [mm] & $\varDelta z$ [mm] & $\tau_{x}$ [mrad] & $\tau_{y}$ [mrad] & $\tau_{z}$ [mrad]\\
\hline
2001&L& $+2.40 \pm 0.90$ & $-0.85 \pm 0.65$ & $+2.54 \pm 0.33$  &$-0.33 \pm 0.55$ & $+1.59 \pm 0.59$ & $+2.30 \pm 0.39$\\
(a)&R & $+2.35 \pm 0.99$ & $+0.47 \pm 0.70$ & $-0.35 \pm 0.44$  & $-0.19 \pm 0.58$ & $+0.02 \pm 0.61$ & $-1.69 \pm 0.37$ \\ \hline
2001&L & $+2.30 \pm 0.92 $ & $-0.62 \pm 0.67$ & $+2.15 \pm 0.14$  & $-0.13 \pm 0.37$ & $+1.32 \pm 0.35$ & $+2.12 \pm 0.29$\\
(b)&R & $+1.94 \pm 0.79$ & $+0.47 \pm 0.58$ & $-1.06 \pm 0.13$ & $-0.24 \pm 0.16$ & $-0.22 \pm 0.17$ & $-1.49 \pm 0.25$ \\ \hline
2002 &L& $+2.44 \pm 0.54$ & $-1.11 \pm 0.42$ & $+2.60 \pm 0.15$  & $-0.20 \pm 0.43$ & $+1.42 \pm 0.37$ & $+1.08 \pm 0.20$\\
&R & $+2.20 \pm 0.73$ & $-0.05 \pm 0.52$ & $-0.39 \pm 0.21$ & $-0.38 \pm 0.46$ & $-0.24 \pm 0.45$ & $-2.79 \pm 0.24$ \\ \hline
\end{tabular}
}

\end{center}
\end{table}

\subsubsection{Deposited energy calibration of the EM calorimeters}
The energy deposit of the EM calorimeters was calibrated using pure electron samples, as outlined in Sec.~\ref{ssec:eid_efficiency}.
We assume that the energy deposits of electrons within the calorimeters were proportional to their momenta.
We calibrated the gain such that the deposited energy divided by the momentum ($E/p$) resulted in a value of one.

\begin{figure}[htbp]
 \begin{center}
  \includegraphics[width=15cm]{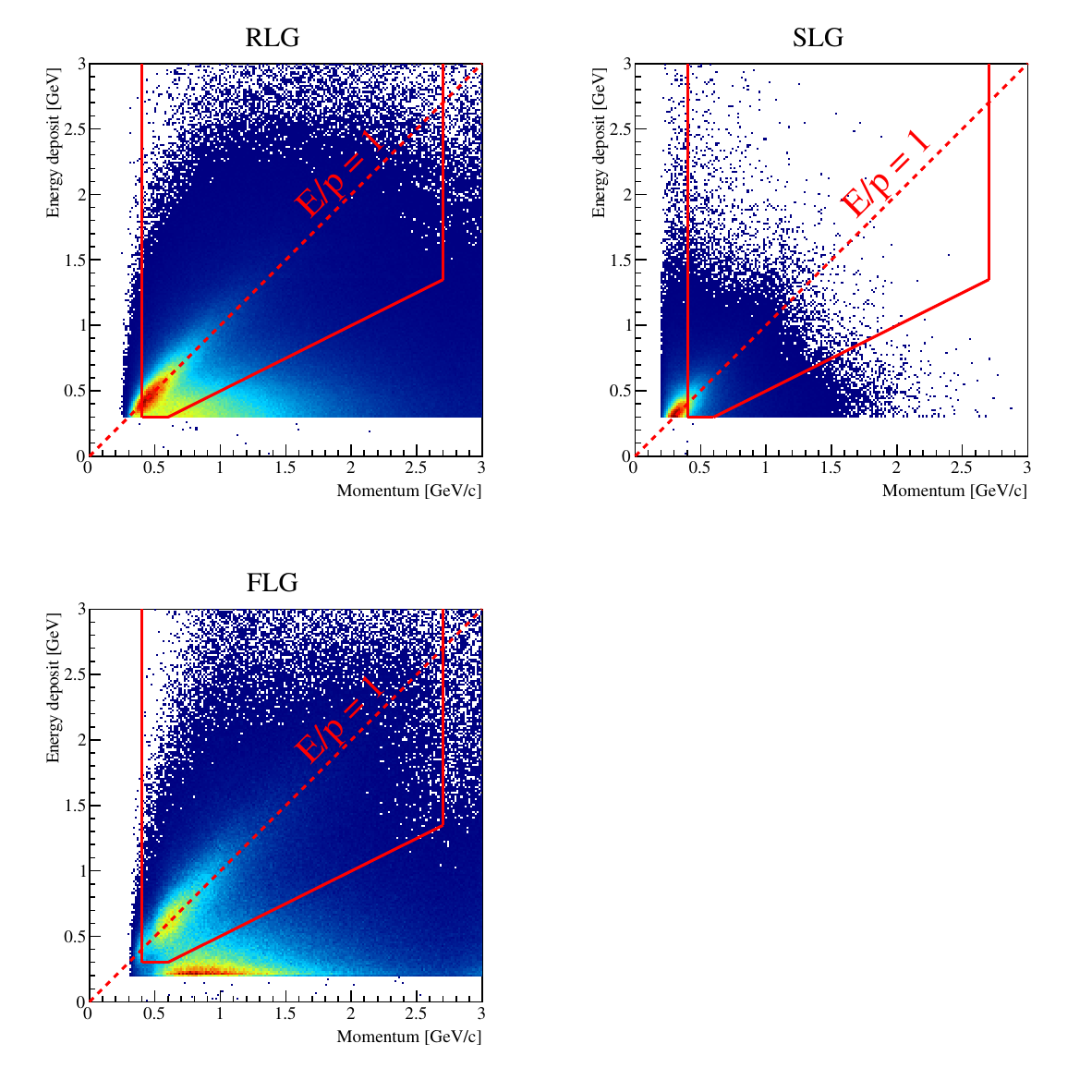}
  \caption{Scatter plot of the deposited energy versus momentum
  for the RLG (upper-left), SLG (upper-right), and FLG (lower-left).
  The dotted lines depict the $E/p = 1$ condition, serving as an eye guide. 
  The solid lines correspond to cut conditions, as described in Sec.
  \ref{ssec:eid_cut}}.
  \label{fig:lg_ep_cut}
 \end{center}
\end{figure}

\begin{figure}[htbp]
 \begin{center}
  \includegraphics[width=12cm,keepaspectratio]{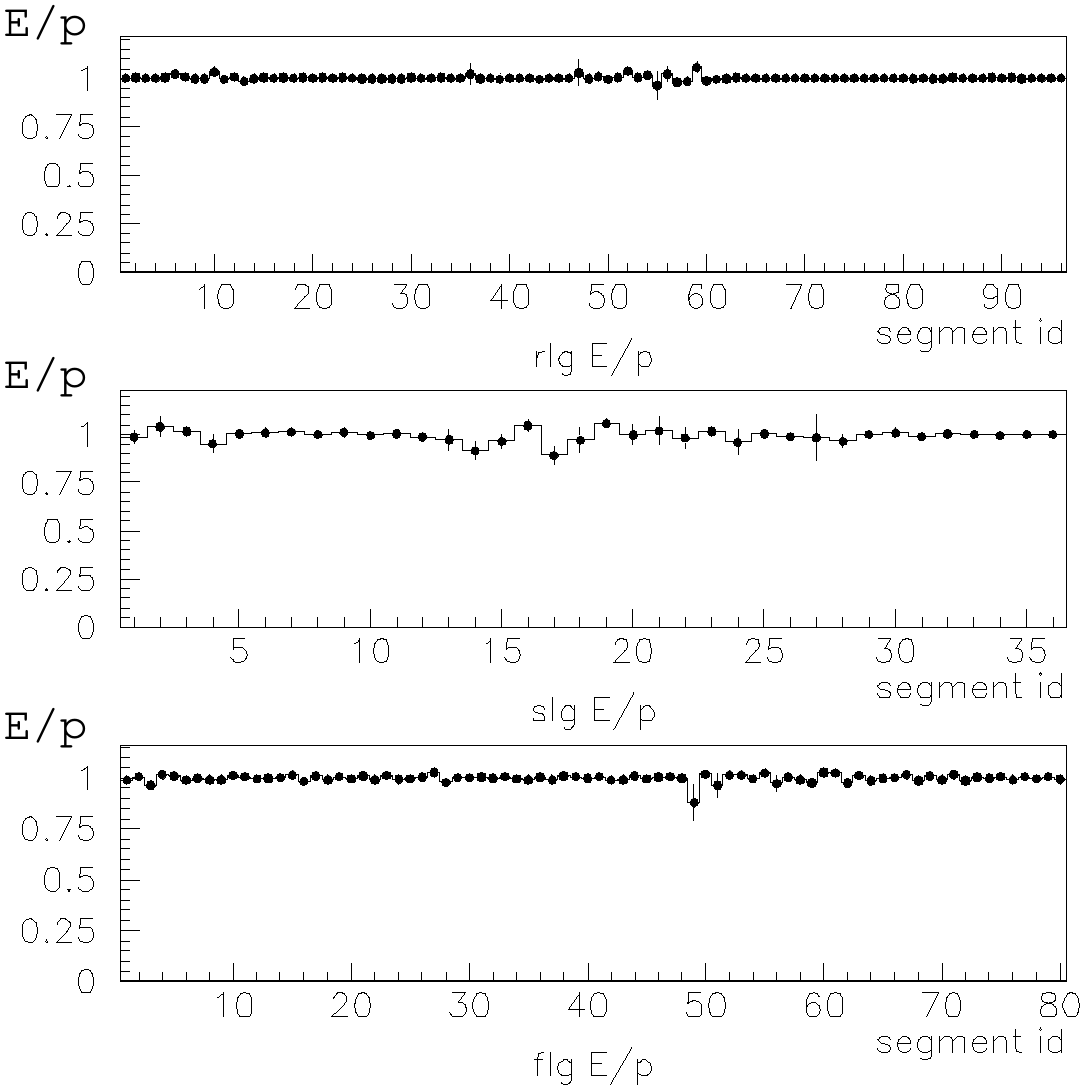}
  \caption{$E/p$ ratios in all segments of RLG, SLG, and FLG following calibration.
     The segment ID is numbered from the top PMT to the bottom PMT, from the forward direction to the backward direction, and from the left arm to the right arm.
     The error bars include only statistical errors.
     \FIGciteNaruki
     }
  \label{fig:lgcalib}
 \end{center}
\end{figure}

Following track reconstruction, the hit segments of the EM calorimeters were assigned individual tracks.
The matching windows were $\pm2.2$, $\pm2.0$, and $\pm2.5$ times the segment sizes for the RLG, SLG, and FLG, respectively.
Considering the spread of the deposited energy over several segments, $E$ was defined as the summation of the corrected ADC values of the associated segments.
For the SLG and FLG, the ADC values of the top and bottom PMTs were also summed.
Subsequently, the correction factor was determined iteratively.
Figure~\ref{fig:lg_ep_cut} depicts scatter plots of the energy deposited ($E$) within the EM calorimeters versus momentum ($p$).
After calibration, the peak positions of the $E/p$ distributions for all segments were adjusted to unity with an accuracy of 4\% as illustrated in Fig.~\ref{fig:lgcalib}.

\subsection{Event reconstruction\label{sec:reconstruction}}
\subsubsection{Track reconstruction\label{sec:tracking}}

Track reconstruction was performed as follows.
First, hit combinations were selected from the $X$ and $X'$ layers of each super-layer within the CDC.
Subsequently, two hits within adjacent cells of the $X$ and $X'$ layers were combined to form a hit pair, whose hit position was defined as the center of the two hits.

Possible combinations of three hit pairs in the three super-layers were established as track candidates, with each track candidate processing one hit pair within each super-layer.
A track trajectory was approximated by a circular line passing through the three hit pairs. A rough momentum estimate was obtained from this approximated trajectory.
This momentum estimate was required to be greater than 0.35~GeV/$c$. 
Further consistency checks were performed using the circular trajectory 
and local hit information. 
Next, a quadratic function was fitted to all hit positions within the $X$ and $X'$ layers. 
Furthermore, the reduced $\chi^2$ was required to be less than 5.5, and fit residuals for all layers were below six standard deviations.

Next, hits were searched for in the four tilted layers (inner $U$, middle $V$ and $V'$, and outer $U$ layers) that could be associated with each track candidate.
Each track candidate was fitted against a three-dimensional curve, represented by a quadratic function in the $x$-$y$ plane and a straight line in the $r$-$z$ plane.
Track candidates whose fit residuals were lower than three standard deviations for all tilted layers were classified as CDC track candidates.

Next, track candidates within the BDC were searched for independently of the CDC tracks.
Here, if the distance between two hits within the $X$ and $X'$ layers was less than a quarter of the cell size, the two hits were considered to belong to the same BDC track candidate;
otherwise, they were defined as two independent track candidates.
Furthermore, associated hits in the $U$ or $V$ layers were required to be located within a distance of nine times the cell size from the $X$ or $X'$ hit, as determined by the wire tilt angle. 

To connect a CDC track candidate with a BDC track candidate, the fitted trajectory of a CDC track candidate was extrapolated to the BDC.
Position agreements within 4.3~mm in the horizontal direction and within 60~mm in the vertical direction were required to establish a CDC--BDC track candidate.

All track candidates surviving the above process were fitted using the Runge--Kutta method to determine their precise trajectories.
The trajectories were traced in 50~mm steps under the application of a magnetic field, as described in Sec.~\ref{section-magnet}.
This was accomplished by solving the equation of motion using the fourth-order Runge--Kutta method.
Next, a $\chi^2$ value was calculated from the residuals between a trajectory and the hit positions of the chambers.
All possible combinations of the CDC--BDC track candidates were subjected to the Runge--Kutta fitting.
In cases wherein the track candidates shared the same hits, those with the best $\chi^2$ values were selected as the final track candidates.

\subsubsection{Event vertex reconstruction}
\begin{figure}[htbp]
 \begin{center}
  \includegraphics[width=15cm]{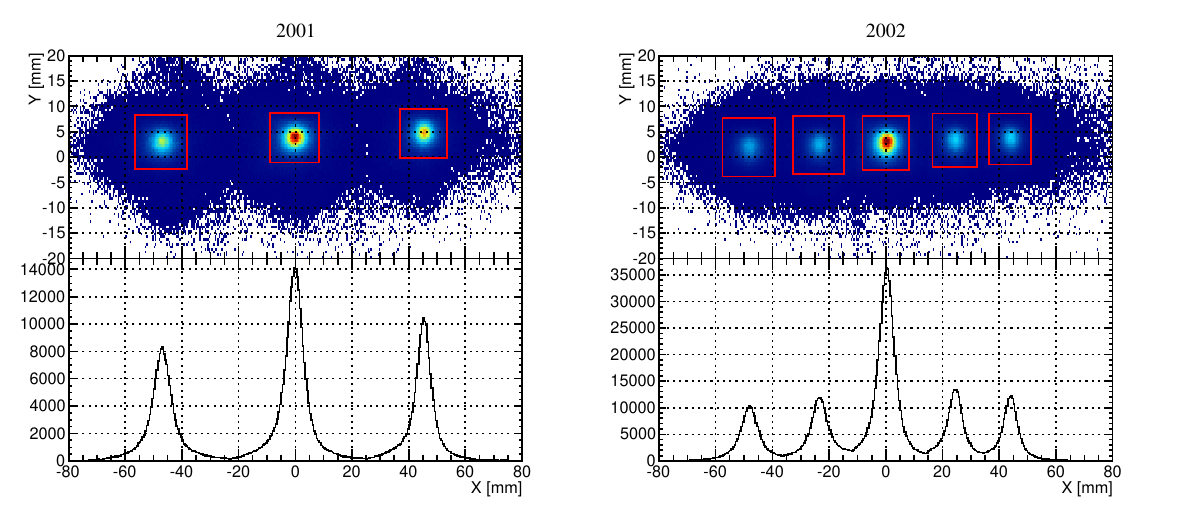}
  \caption{Vertex distributions corresponding to the double-arm events of 2001 (left) and 2002 (right) experimental runs.
  The cut regions for each target are indicated by the red rectangles in the upper panels.}
  \label{fig:vtx_cut}
 \end{center}
\end{figure}

The event vertex point was determined by minimizing $S$, which is defined as follows:
\begin{eqnarray}
   S = \frac{1}{3N_{\mathrm t}-4}\sum^{N_{\mathrm t}}_{j=1}\left[
        \left( \frac{d^j_x}{\sigma_x} \right)^2
      + \left( \frac{d^j_y}{\sigma_y} \right)^2
      + \left( \frac{d^j_z}{\sigma_z} \right)^2
   \right]
\end{eqnarray}
where $N_{\mathrm t}$ denotes the number of tracks in each event;
$\boldsymbol{d}^j = (d^j_x, d^j_y, d^j_z)$ represents the distance vector between the vertex and the $j$-th track;
and $\sigma_x, \sigma_y, \sigma_z$ denote the weight parameters for each axis, with values of $\sigma_x = 1.44~\mathrm{mm}$, $\sigma_y = 1.80~\mathrm{mm}$, and $\sigma_z = 7.15~\mathrm{mm}$.
Once the event vertex position was determined, track candidates not satisfying the condition
\begin{eqnarray}
   \left( \frac{d^j_x}{\sigma_x} \right)^2
 + \left( \frac{d^j_y}{\sigma_y} \right)^2
 + \left( \frac{d^j_z}{\sigma_z} \right)^2
 < 10.0
\end{eqnarray}
were discarded.
This procedure was repeated until all remaining tracks satisfied the above condition.
In case the number of the remaining tracks was less than two, the event was deemed not to have an appropriate vertex point.
The distributions of the reconstructed vertex positions for ``double-arm events'' are depicted in Fig.~\ref{fig:vtx_cut}. Here,  ``double-arm events'' are defined as those wherein both arms have at least one track.

The positions of the targets were determined by fitting the distributions against a Gaussian function.
The obtained target positions and the standard deviations are summarized in Table~\ref{tab:vtx}.
Events for which the minimized $S$ values was less than $5.0$ and the vertex position was 
located within $\pm3.0~\sigma$ from the closest target position for each axis were used 
in the current analysis.
Figure~\ref{fig:vtx_cut} also presents these cut regions.

\begin{table}[htbp]
 \caption{Peak positions of the vertex distributions (target positions) and the standard deviations derived from the double-arm events.}
  \label{tab:vtx}
\begin{center}
\begin{tabular}{c|ccc|ccccc} \hline
year & \multicolumn{3}{c}{2001\phantom{.}\phantom{00}} &
 \multicolumn{5}{c}{2002\phantom{.}\phantom{00}} \\ \hline
target & Cu1 & C & Cu2 & Cu1 & Cu2 & C & Cu3 & Cu4 \\ \hline\hline
$x$~[mm]        &    $-47.37$ & $-0.40$ & $45.18$ & $-48.36$ & $-23.66$ & $0.02$ & $24.27$ & $43.82$ \\
$y$~[mm]        &    $2.93$ & $3.81$ & $4.66$ & $1.87$ & $2.33$ & $2.75$ & $3.27$ & $3.58$           \\
$z$~[mm]        &    $0.59$ & $0.66$ & $0.61$ & $0.43$ & $0.51$ & $0.48$ & $0.49$ & $0.49$           \\
$\sigma_x$~[mm] &    $3.06$ & $2.88$ & $2.74$ & $3.09$ & $3.00$ & $2.73$ & $2.64$ & $2.45$           \\
$\sigma_y$~[mm] &    $1.78$ & $1.62$ & $1.62$ & $1.91$ & $1.91$ & $1.76$ & $1.76$ & $1.69$           \\
$\sigma_z$~[mm] &    $3.21$ & $2.99$ & $2.74$ & $3.47$ & $3.42$ & $3.23$ & $3.20$ & $3.11$           \\ \hline
\end{tabular}
 
 \end{center}
\end{table}

\subsubsection{Common vertex track fitting}
To improve the mass resolution, we adopted a procedure called common vertex track fitting (CVTF).
According to this procedure, two oppositely charged tracks were selected from the final candidates utilized to determine the vertex.
These two tracks were refitted with the Runge--Kutta method, imposing an additional constraint: both tracks must be generated from the same vertex point on the target plane.
The $\chi^2$ value to be minimized in the CVTF process was defined as
\begin{eqnarray}
   \chi^2_\mathrm{CVTF} = \sum_{q=\pm1}\sum_l\left( \frac{x_\mathrm{hit}^{ql}-x_\mathrm{track}^{ql}}{\sigma_l} \right)^2
   + \sum_{i=y, z} \left( \frac{v_i-c_i}{\sigma'_i} \right)^2
\end{eqnarray}
where $x_\mathrm{hit}^{ql}$ and $x_\mathrm{track}^{ql}$ denote the hit and track positions with a charge of $q$ and the $l$-th layer of the chambers, $\sigma_l$ represents the position resolution of the $l$-th layer,
$\boldsymbol{v}$ denotes the vertex position on the target plane,
$\boldsymbol{c}$ indicates the center of the target position,
and $\sigma'_y$ $\sigma'_z$ represent the beam size.
The beam size was only measured along the $y$ direction, as described in Sec.~\ref{section-beamline}, however, an identical value was also assumed in the $z$ direction
as the beam was tuned to create a disk-shaped profile on the luminescence board.
Thus, values of $\sigma'_y = \sigma'_z = 1.59~\mathrm{mm}$ and $\sigma'_y = \sigma'_z = 0.83~\mathrm{mm}$ were used for the 2001 and 2002 runs.

The reduced $\chi^2$ value was obtained as $\chi^2_\mathrm{CVTF}/\mathrm{ndf}$, where ndf denotes the number of degrees of freedom.
For the final samples, the track pairs were required to have a reduced $\chi^2_\mathrm{CVTF}$ value below 7.5.

\subsection{Electron identification\label{sec:eid}}
\subsubsection{Cut conditions\label{ssec:eid_cut}}
The eID process was accomplished using five types of the eID counters described in Sec.~\ref{eid-counters}.
For each counter, hit information associated with the track candidates, such as ADC values (or energy deposited in calorimeters), hit positions, and TOFs was determined.
By extrapolating (interpolating) the trajectories determined by fitting the hit positions in the CDC and BDC,
cross points of the counters were calculated.
Furthermore, ADC and TDC values of counter segments within a specific range from the extrapolated points were employed to identify electron tracks.
The ranges for the FGC, RGC, RLG, SLG, and FLG were $\pm1.5$, $\pm1.5$, $\pm2.2$, $\pm2.0$, and $\pm2.5$ segment sizes, respectively.
Finally, the ADC values of the segments were summed, hit positions were defined as the weighted means of the positions of the segment centers, and the TOFs were determined by computing $T_\mathrm{counter}-T_\mathrm{STC}$.

\begin{figure}[htbp]
 \begin{center}
  \includegraphics[width=15cm]{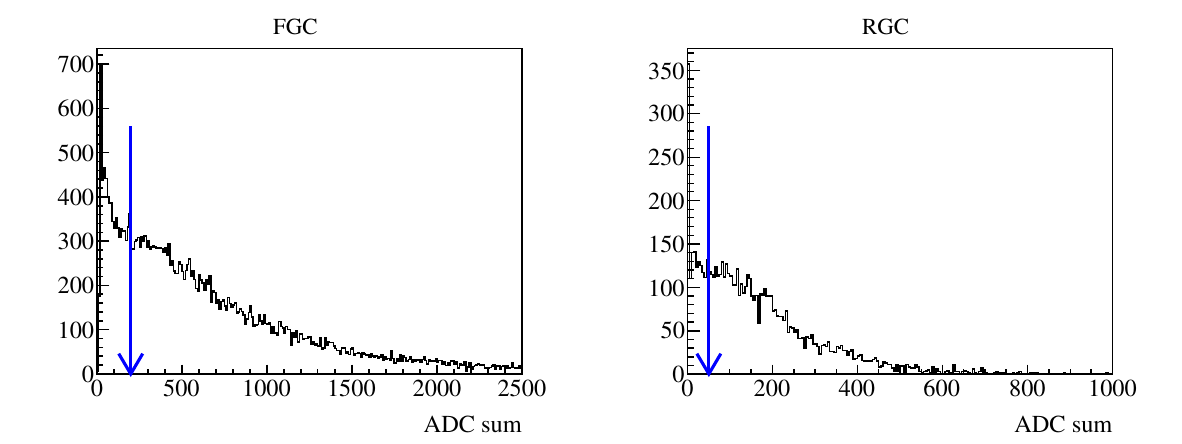}
  \caption{ADC distributions of the FGC (left) and RGC (right).
  Thresholds are indicated by blue arrows.}
  \label{fig:gc_adc_cut}
 \end{center}
\end{figure}

\begin{figure}[htbp]
 \begin{center}
  \includegraphics[width=15cm]{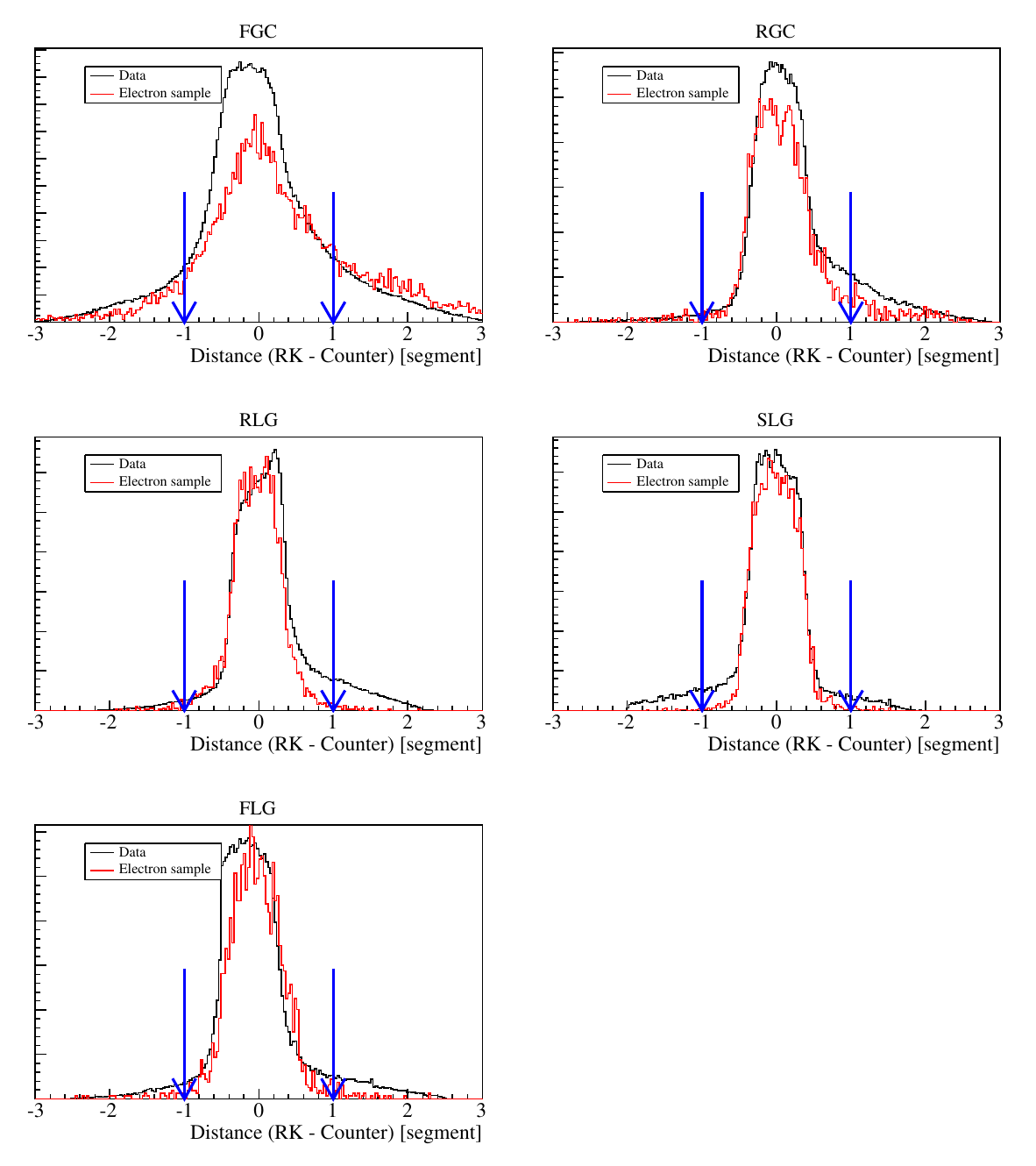}
  \caption{Distributions of the position differences of each eID counter.
  Cut regions are indicated by blue arrows.
  Black lines depict the data distributions, while red lines depict the distributions of the pure-electron sample described in Sec.~\ref{ssec:eid_efficiency}.}
  \label{fig:eid_cut_pdiff}
 \end{center}
\end{figure}

\begin{figure}[htbp]
 \begin{center}
  \includegraphics[width=15cm]{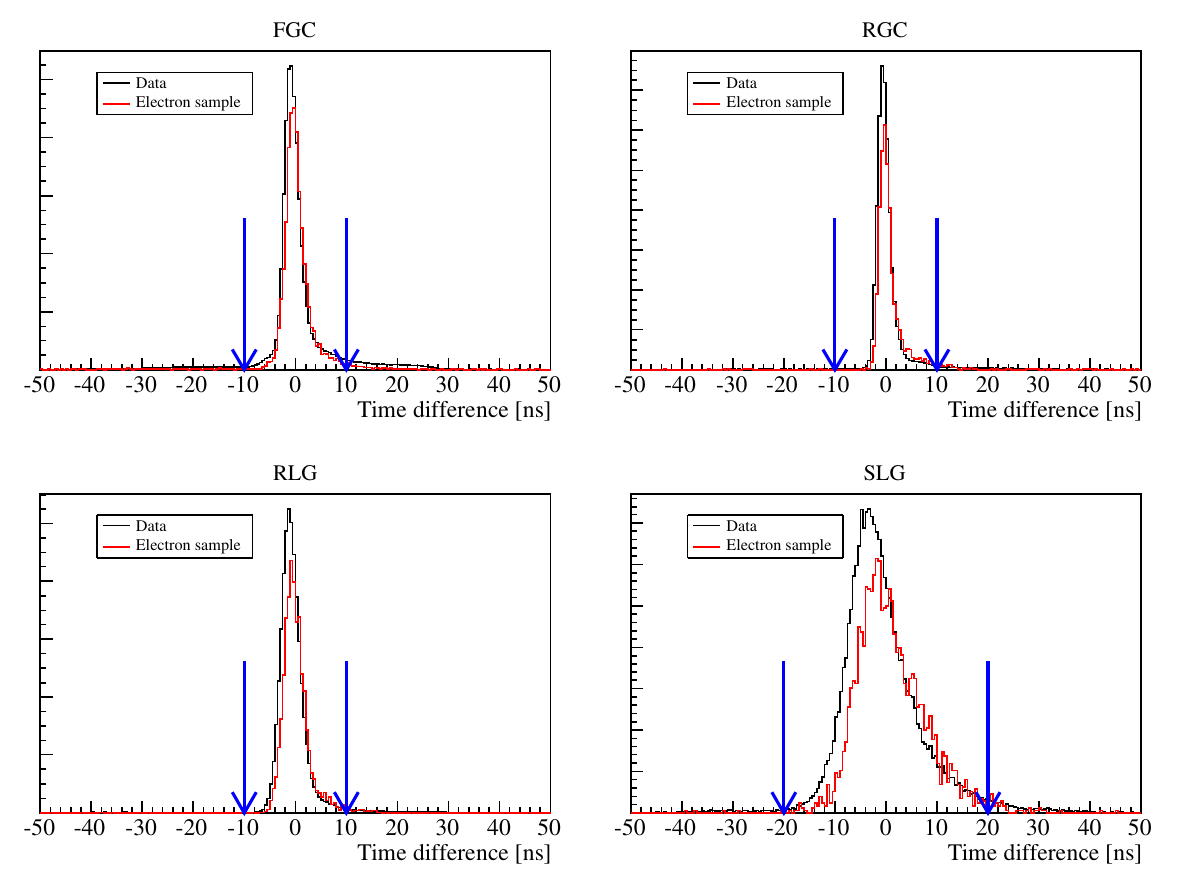}
  \caption{Distributions of the TOF difference of each eID counter.
  Cut regions are indicated by blue arrows.
  Black lines depict the data distributions, while red lines depict the distributions of the pure-electron sample described in Sec.~\ref{ssec:eid_efficiency}.}
  \label{fig:eid_cut_tdiff}
 \end{center}
\end{figure}

Electron candidates were required to satisfy the following conditions. The criteria of the cut parameters follow the published analysis. 
\begin{enumerate}
   \item The ADC sum must exceed 200 channels for the FGC and 50 channels for the RGC.
         The ADC distributions and the thresholds for the FGC and RGC are illustrated in Fig.~\ref{fig:gc_adc_cut}.
   \item The hit positions must match the trajectories of the track candidates.
         The match windows must have a range of $\pm1.0$ segment size for all counters.
         The distributions of the position difference and cut regions are depicted in Fig.~\ref{fig:eid_cut_pdiff}.
   \item The measured TOFs and the expected values calculated using the Runge--Kutta method must agree within 10~ns for the FGC, RGC, and RLG and 20~ns for the SLG.
         The distributions of the TOF difference and cut regions are presented in Fig.~\ref{fig:eid_cut_tdiff}.
         It is noted that the FLG presented no TDC data owing to the limitation of the R/O circuit.
   \item Energy deposits within LG calorimeters must exceed 0.3~GeV, and energy deposits divided by the momentum ($E/p$) must be greater than 0.5.
         The two-dimensional distributions of energy deposits versus momentum and cut regions are depicted in Fig.~\ref{fig:lg_ep_cut}.
\end{enumerate}

\subsubsection{Efficiency evaluation using a pure-electron sample\label{ssec:eid_efficiency}}
\begin{figure}[htbp]
 \begin{center}
  \includegraphics[width=15cm]{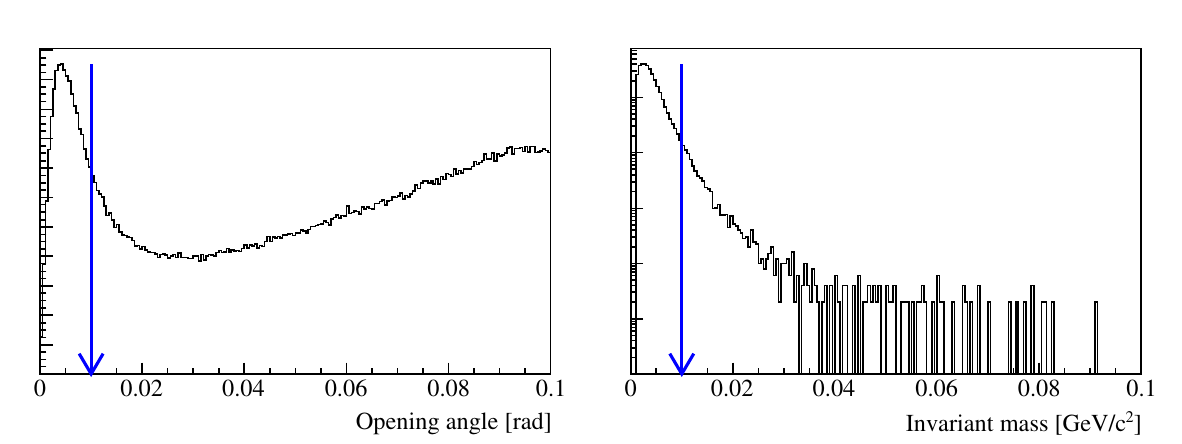}
  \caption{Distributions of the opening angle (left) and invariant mass (right) for 
  less-biased tracks.
  Cut regions are indicated by blue arrows.}
  \label{fig:opang_mass_unbias}
 \end{center}
\end{figure}

We evaluated the efficiencies of the eID counters using electrons and positrons originating from $\gamma$ conversions or $\pi^0$ Dalitz decays.
To avoid trigger biases,
one of the tracks satisfying the trigger condition was randomly excluded from each arm.
From among the remaining tracks, $e^+e^-$ pairs with opening angles less than 0.01~rad and invariant masses less than 0.01~GeV/$c^2$ were selected.
The opening angle and invariant mass distributions are depicted in Fig.~\ref{fig:opang_mass_unbias}.
Such tracks were considered as pure electron samples.

\begin{figure}[htbp]
 \begin{center}
  \includegraphics[width=15cm]{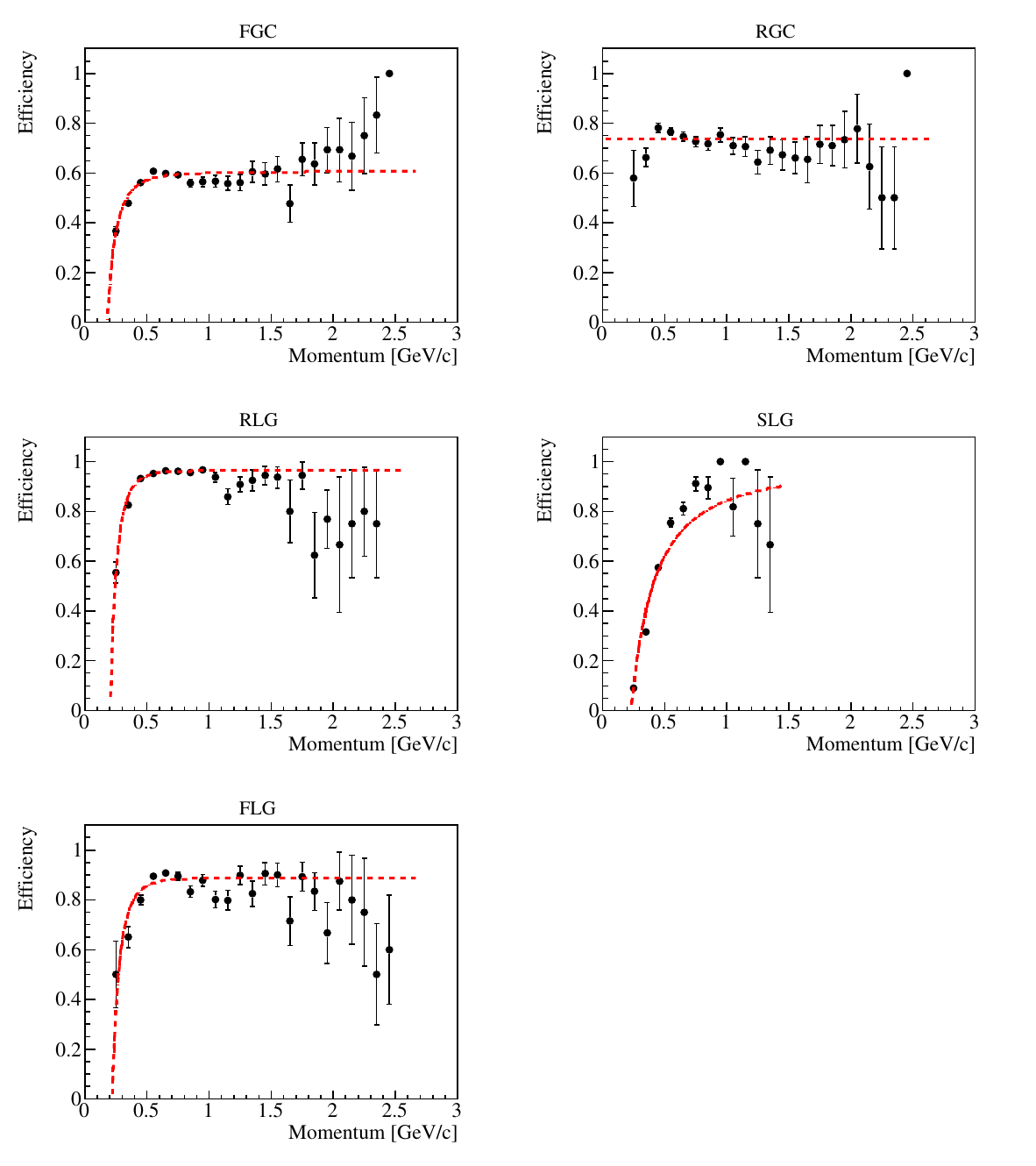}
  \caption{Efficiency of each eID counter as a function of the track momentum.
     Red dashed lines represent the fitting results obtained using Eq.~\ref{eqn:eff_curve}. The corresponding parameters are listed in Table~\ref{tab:eid_eff}}
  \label{fig:eid_eff}
 \end{center}
\end{figure}

To evaluate the efficiency of a counter, the tracks used were identified by the other counters.
For instance, to evaluate the efficiency of the FLG, we used tracks identified by the FGC and RGC.
By implementing the electron cuts described in Sec.~\ref{ssec:eid_cut} on the pure electron samples,
the efficiencies as functions of the momentum were evaluated.
Notably, the effects of position matching were also taken into account in the efficiency evaluation.  
The corresponding results are illustrated in Fig.~\ref{fig:eid_eff}.

The efficiency curves were fitted against empirical functions defined by
\begin{eqnarray}
   f(p) = a-b/p^c
   \label{eqn:eff_curve}
\end{eqnarray}
where $a$, $b$, and $c$ denote fitting parameters and $p$ represents the momentum.
The fitting results are depicted in Fig.~\ref{fig:eid_eff}, while the obtained parameters are summarized in Table~\ref{tab:eid_eff}.
The efficiency functions were employed for the detector simulation explained in Sec.~\ref{sec:simulation}. 
The remaining ambiguity in evaluations of electron efficiencies were 
taken into account as a source of systematic errors. 

\begin{table}[htbp]
\caption{Parameters of the efficiency curves depicted in Fig.~\ref{fig:eid_eff},  defined in Eq.~\ref{eqn:eff_curve}.}
\label{tab:eid_eff}
   \begin{center}
\begin{tabular}[tb]{cccc}\hline
 counter & $a$ & $b$ & $c$ \\ \hline\hline
 FGC & $0.603\pm0.008$ & $0.005\pm0.002$ & $2.8\pm0.3$ \\
 RGC & $0.737\pm0.007$ & 0 & 0 \\
 RLG & $0.963\pm0.005$ & $0.0013\pm0.0006$ & $4.2\pm0.4$ \\
 SLG & $1.000\pm0.003$ & $0.161\pm0.006$ & $1.26\pm0.03$ \\
 FLG & $0.886\pm0.009$ & $0.002\pm0.001$ & $4.0\pm0.5$ \\ \hline
\end{tabular}
\end{center}
\end{table}

\subsubsection{Contamination through misidentification}
Although electron cuts were applied to the final $e^+e^-$ samples, some contamination remained owing to the misidentification of pions.
To evaluate this contamination, we examined the signal-to-background ratio ($S/B$) in the mass region extending from 0.76 to 0.86~GeV/$c^2$ for various eID cut conditions.
The values of $S$ and $B$ were obtained by fitting the mass spectra as described in Sec.~\ref{sec:deconvolution}.
Given that $B$ is thought to originate from uncorrelated $e^+e^-$ pairs and $e^\pm\pi^\mp$ or $\pi^+\pi^-$ pairs,
we have $B=B_{ee}+B_\mathrm{cont}$.

\begin{figure}[htbp]
  \begin{center}
    \begin{minipage}{9.0cm}
    \includegraphics[width=9.0cm]{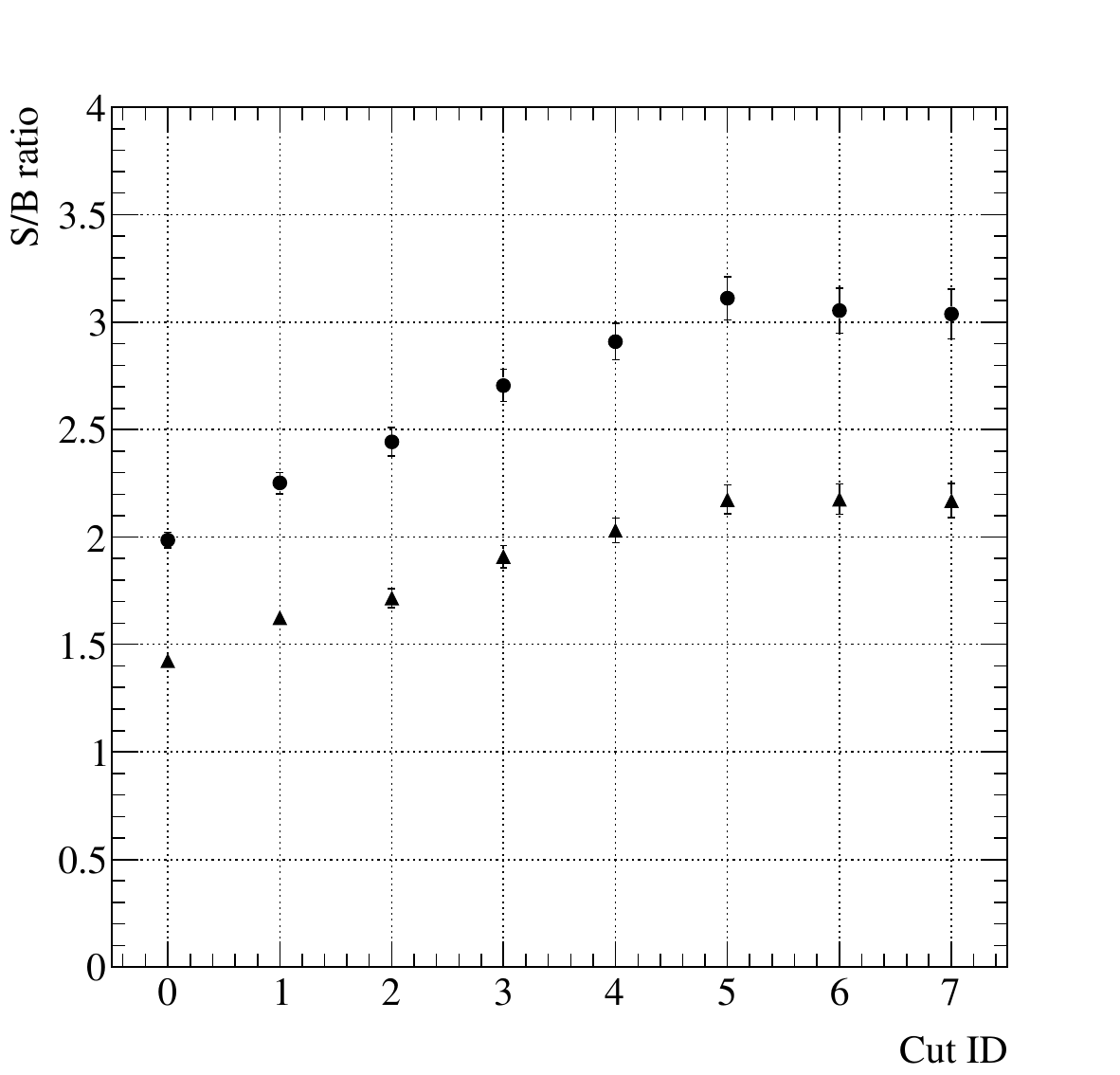}
    \end{minipage}
    \begin{minipage}{5.5cm}
    \begin{tabular}{cccc}\hline
    & \multicolumn{3}{c}{threshold}\\ \cline{2-4}
       &{\footnotesize FGC}&{\footnotesize RGC}& {\footnotesize LG}\\
    {\footnotesize ID} &{\footnotesize ADC}&{\footnotesize ADC}& {\footnotesize $E/p$}\\ \hline\hline
    0 &  200 &  50 & 0.5\\
    1 &  400 &  50 & 0.5\\
    2 &  600 &  50 & 0.5\\
    3 &  600 &  50 & 0.6\\
    4 &  600 &  50 & 0.7\\
    5 &  600 &  50 & 0.8\\
    6 &  600 & 100 & 0.8\\
    7 &  600 & 200 & 0.8\\ \hline
    \end{tabular}
    \end{minipage}
    \caption{
      \label{fig:sb_plot}
      $S/B$ in the mass region extending 
      from 0.76 to 0.86~GeV/$c^2$ for various electron identification cuts.
      The cut conditions are summarized in the accompanying table (right).
      A larger ID number corresponds to a stricter condition.
      The values in the table represent the thresholds applied to the eID counters.
      The triangles correspond to the C target, while the circles correspond to the Cu targets.
    }
  \end{center}
\end{figure}

As depicted in Fig.~\ref{fig:sb_plot}, $S/B$ saturated when the cut condition was made strict enough.
The amount of hadron contamination became negligibly small under such a strict-cut condition.
Furthermore, given that $S/B_{ee}$ does not depend on the cut conditions, we have the following relation,
\begin{eqnarray}
   \frac{S'}{B'} = \frac{S'}{B'_{ee}} = \frac{S}{B_{ee}}
\end{eqnarray}
where $S'$ and $B'$ denote the amount of the signal and background, respectively, when $S/B$ is saturated.
The ratio of the contamination in the background can be obtained as follows
\begin{eqnarray}
   r_\mathrm{cont} \equiv \frac{B_\mathrm{cont}}{B_{ee}+B_\mathrm{cont}} = 1-\frac{S/B}{S'/B'}
\end{eqnarray}
Through calculations, the ratios of contamination in the background were obtained as $35 \pm 3$\% for the C target and $34 \pm 3$\% for the Cu target.

\subsection{Event selection\label{sec:selection}}
In addition to the $\chi^2$ cut, vertex cut, electron cut, and momentum cut ($0.4 < p < 2.7$~GeV/$c$),
certain event selections were also applied to the final samples.
If more than one electron or positron candidate was identified in either arm during the same event,
this event was excluded from the final samples.
Overall, such events accounted for approximately about 9\% of the total.

\begin{figure}[htbp]
 \begin{center}
  \subfigure[Mass]{
  \includegraphics[width=7cm]{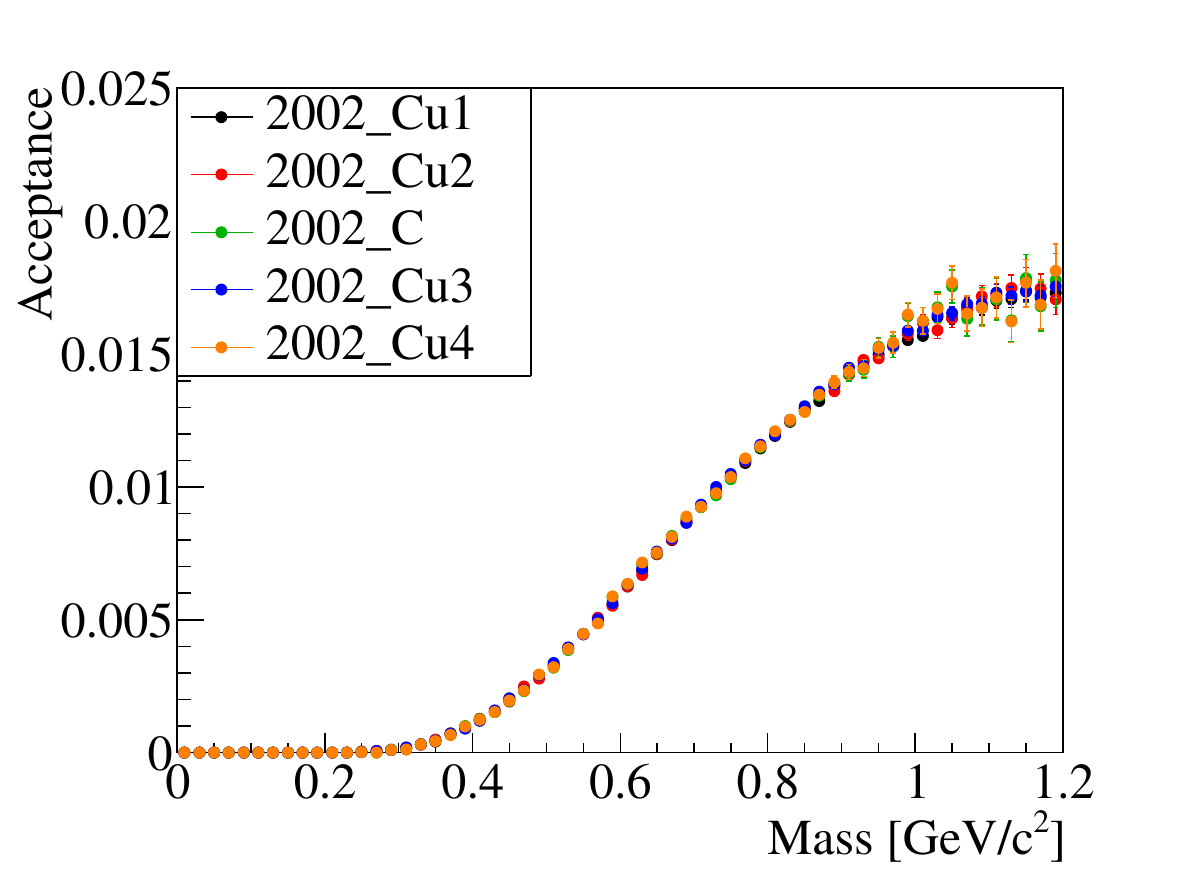}
  \label{fig:sim_acc_mass}}
  \subfigure[Transverse momentum]{
  \includegraphics[width=7cm]{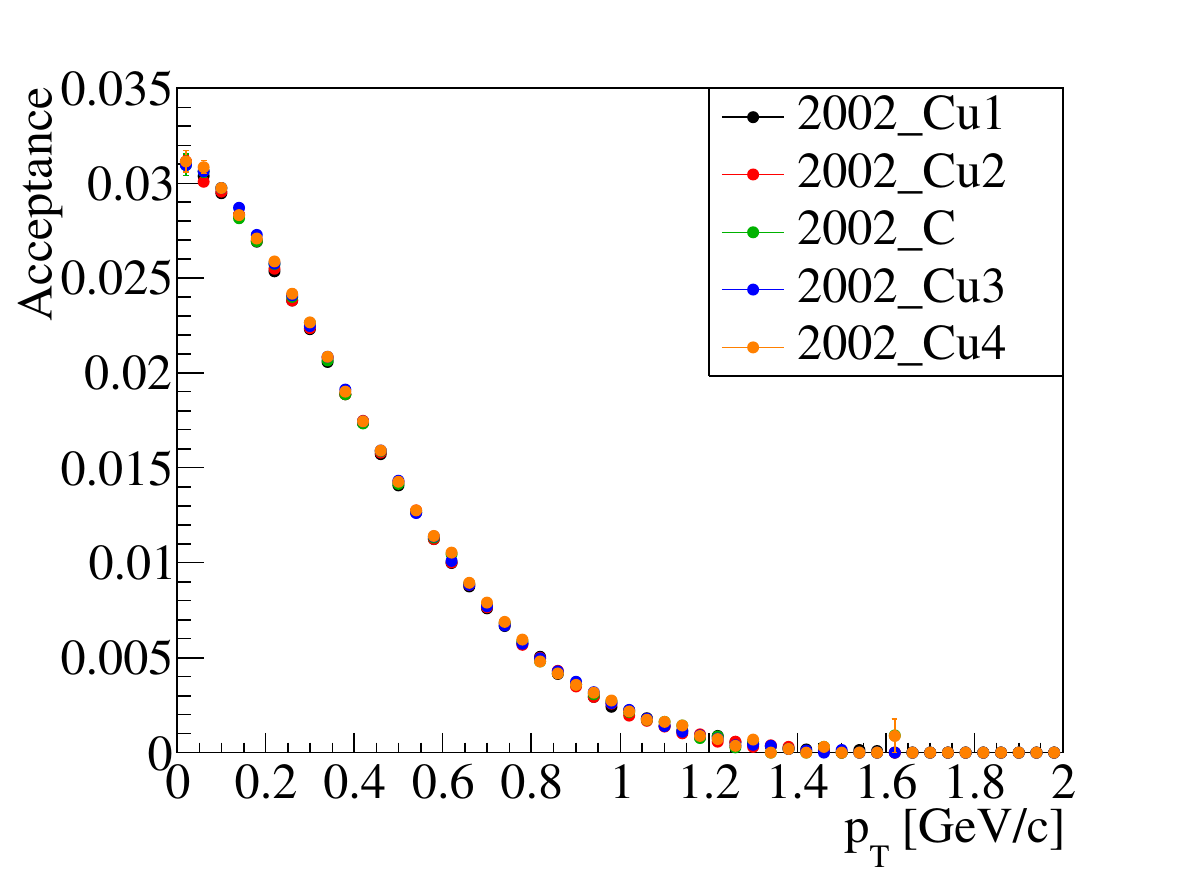}
  \label{fig:sim_acc_pt}}
  \subfigure[Rapidity]{
  \includegraphics[width=7cm]{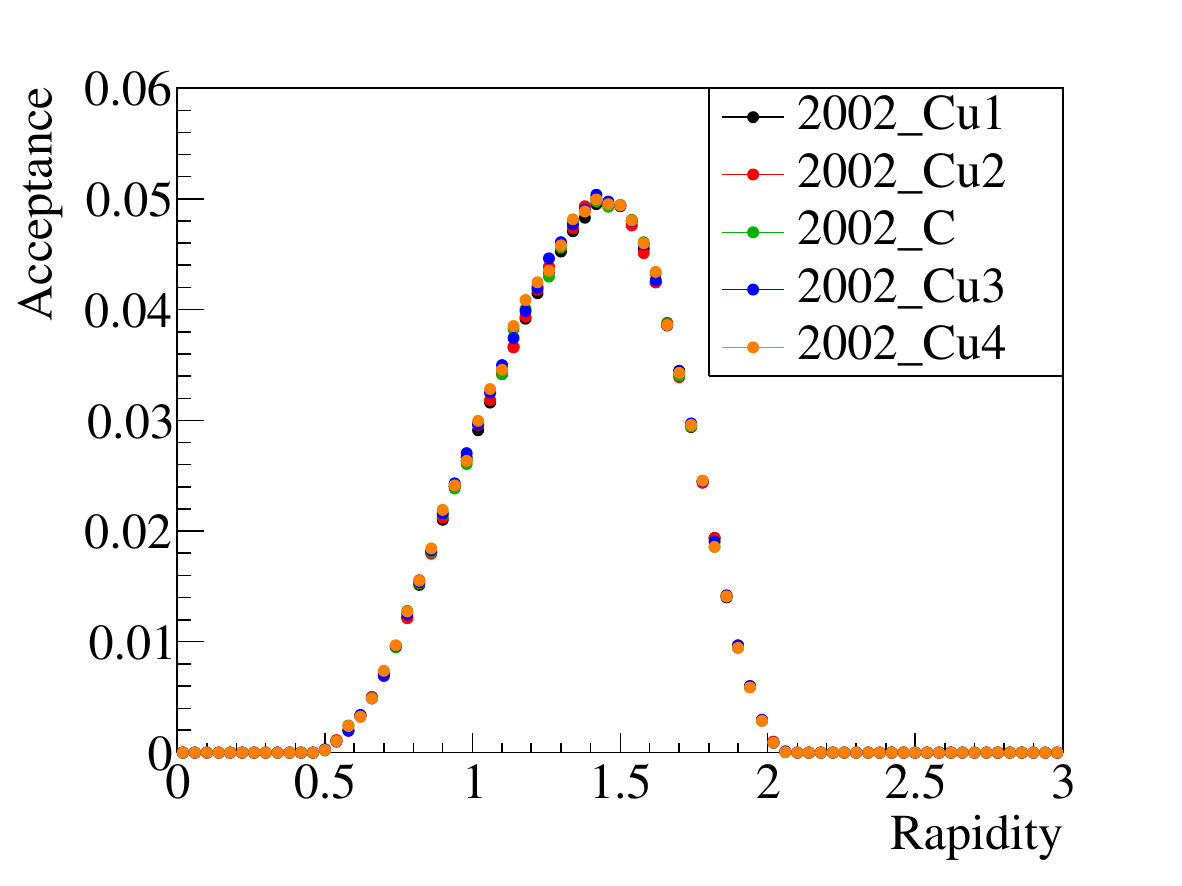}
  \label{fig:sim_acc_rapidity}}
  \subfigure[Mass and Momentum]{
  \includegraphics[width=7cm]{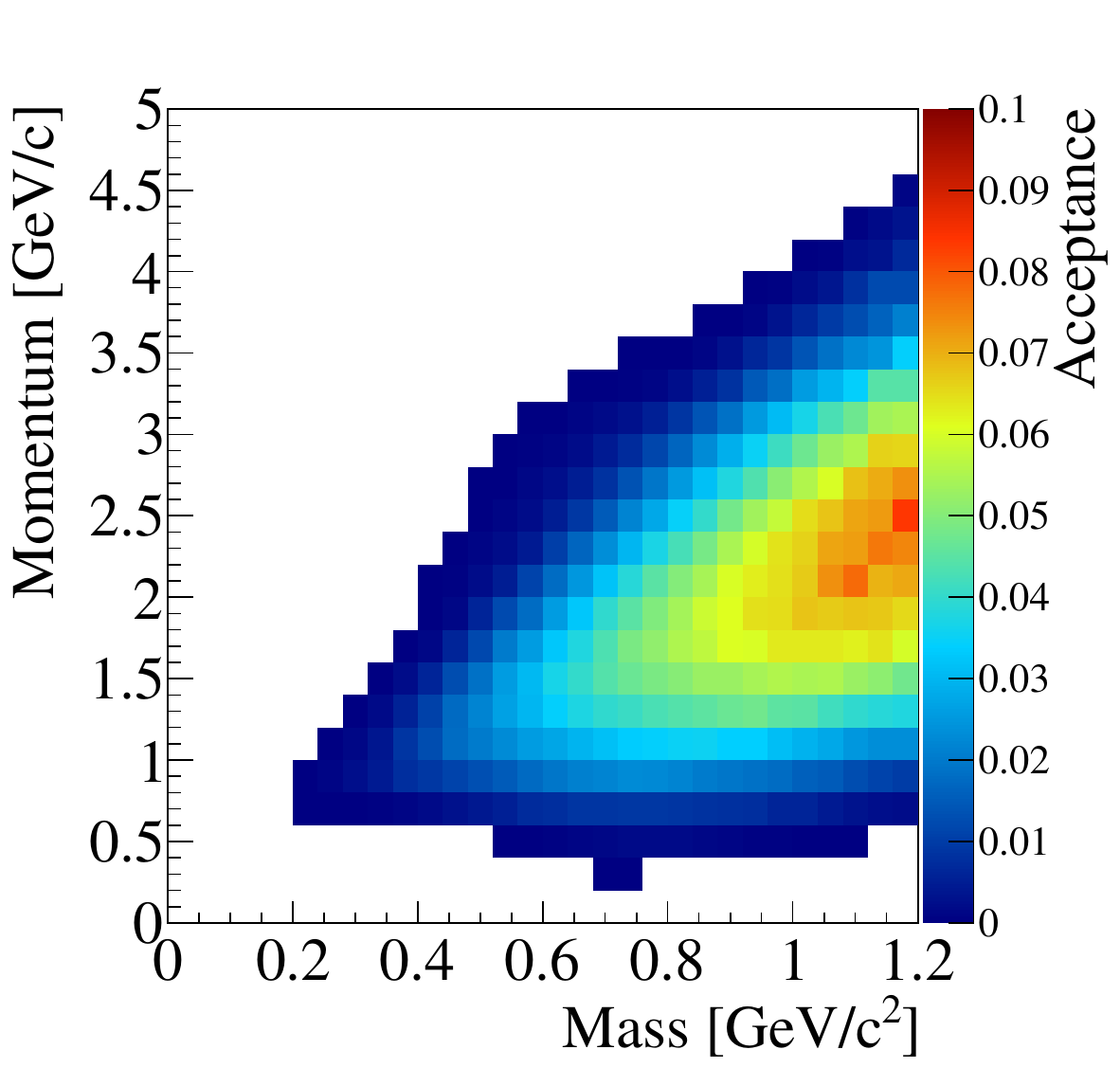}
  \label{fig:sim_acc2d_mass_mom}}
  \caption{Acceptance curves as functions of the mass (a), transverse momentum (b), and rapidity (c) after the angle cut.
  Different points represent different target positions.
  (d) shows the acceptance as a function of meson mass and momentum.
  }
  \label{fig:simulated_acceptance}
 \end{center}
\end{figure}

We simultaneously used three targets in the 2001 run and five targets in 2002 run.
Given the dependence of the view angles of the spectrometer on the target positions,
the acceptance value as a function of the invariant mass of each target differed.
To eliminate this effect, we applied acceptance cuts to the final samples based on the vertex momentum.
Notably, a downstream target exhibited a small acceptance value in the forward region and vice versa.
Hence, we restricted the forward and backward regions to equalize the acceptance value of all.
Figure \ref{fig:simulated_acceptance} depicts the acceptance distributions as functions of mass, transverse momentum, and rapidity for all targets. 
The cuts were applied to angles of momentum vectors at the vertex position.
Table~\ref{tab:acc_cut} lists the parameters for the angle cut.
The values were determined through the detector simulation described in Sec.~\ref{sec:simulation}.

\begin{table}[hbpt]
      \caption{Cut angles for electron and positron tracks.
      The cuts were applied to angles of momentum vectors at the vertex position.
      The listed values were determined based on the simulation described in Sec.~\ref{sec:simulation} such that the acceptance values for all targets became equal.
      }
      \label{tab:acc_cut}
   \begin{center}
      \begin{tabular}{cc}
         \hline
         Horizontal & $\pm 14.4$\degree\ -- $\pm 62.4$\degree\\
         \hline
         Vertical & $\pm 18.9$\degree\\
         \hline
      \end{tabular}
   \end{center}
\end{table}


\subsection{Mass Spectra\label{sec:mass_spectra}}

\begin{figure}[htbp]
 \begin{center}
   \includegraphics[width=15.5cm]{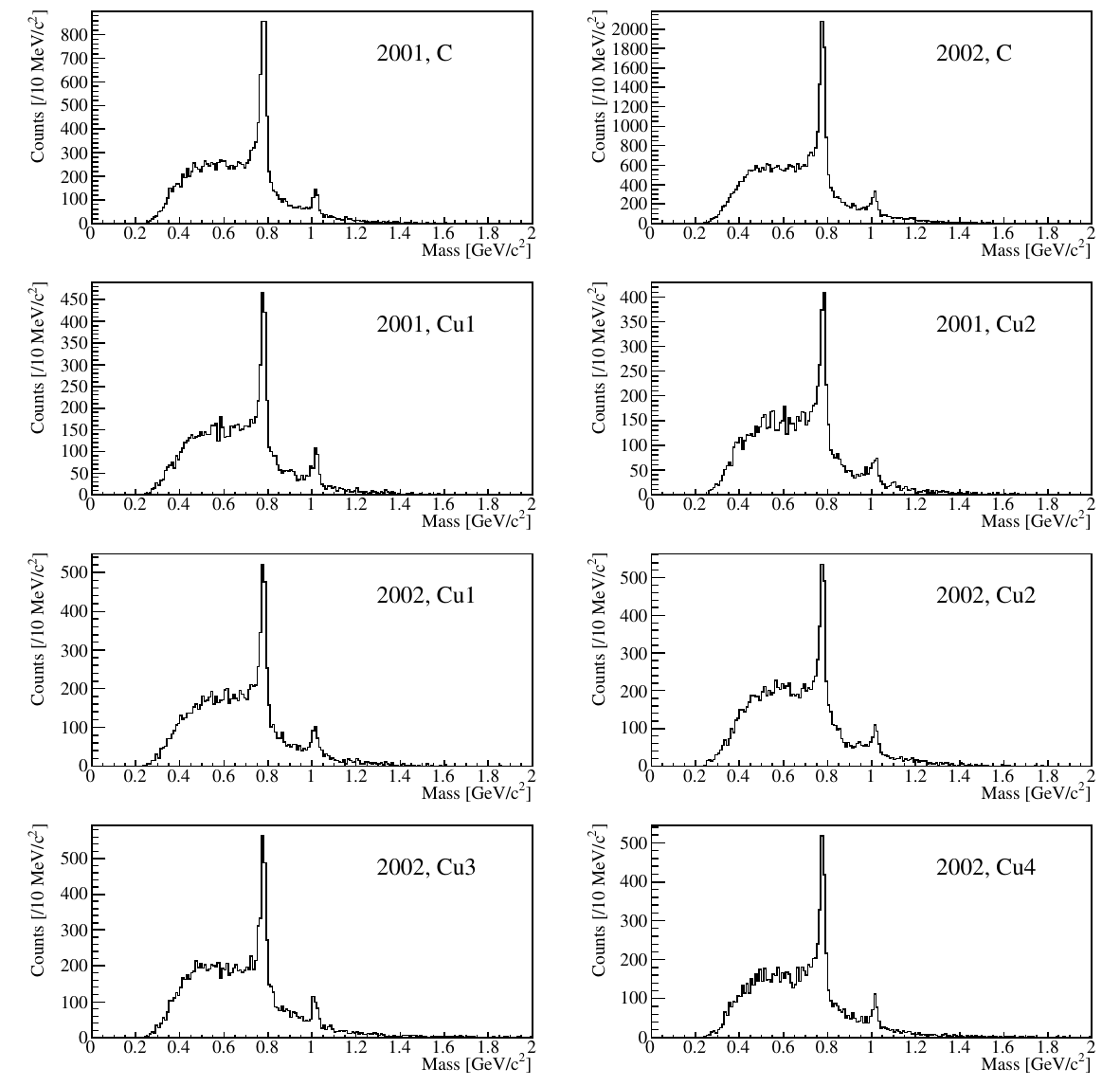}
   \caption{Obtained invariant mass spectra for each target.
    In total, both experimental runs used C targets (2001:1, 2002:1) and six Cu targets (2001:2, 2002:4).
    The fiducial cut aligns the mass acceptance values; hence, the shapes of the spectra are almost independent of the targets.
    }
   \label{fig:mass_each_target}
 \end{center}
\end{figure}

\begin{figure}[htbp]
   \begin{center}
      \subfigure[]{
      \includegraphics[width=7cm,keepaspectratio]{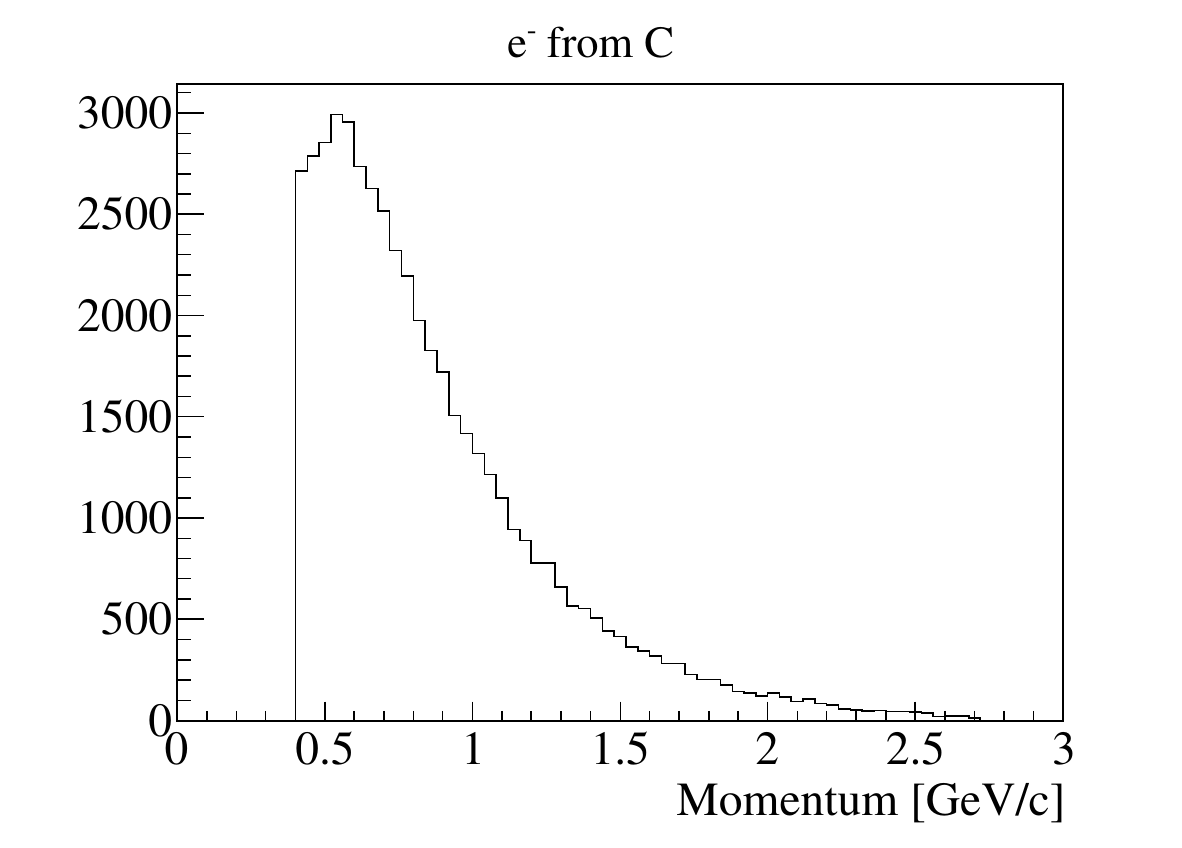}}
      \subfigure[]{
      \includegraphics[width=7cm,keepaspectratio]{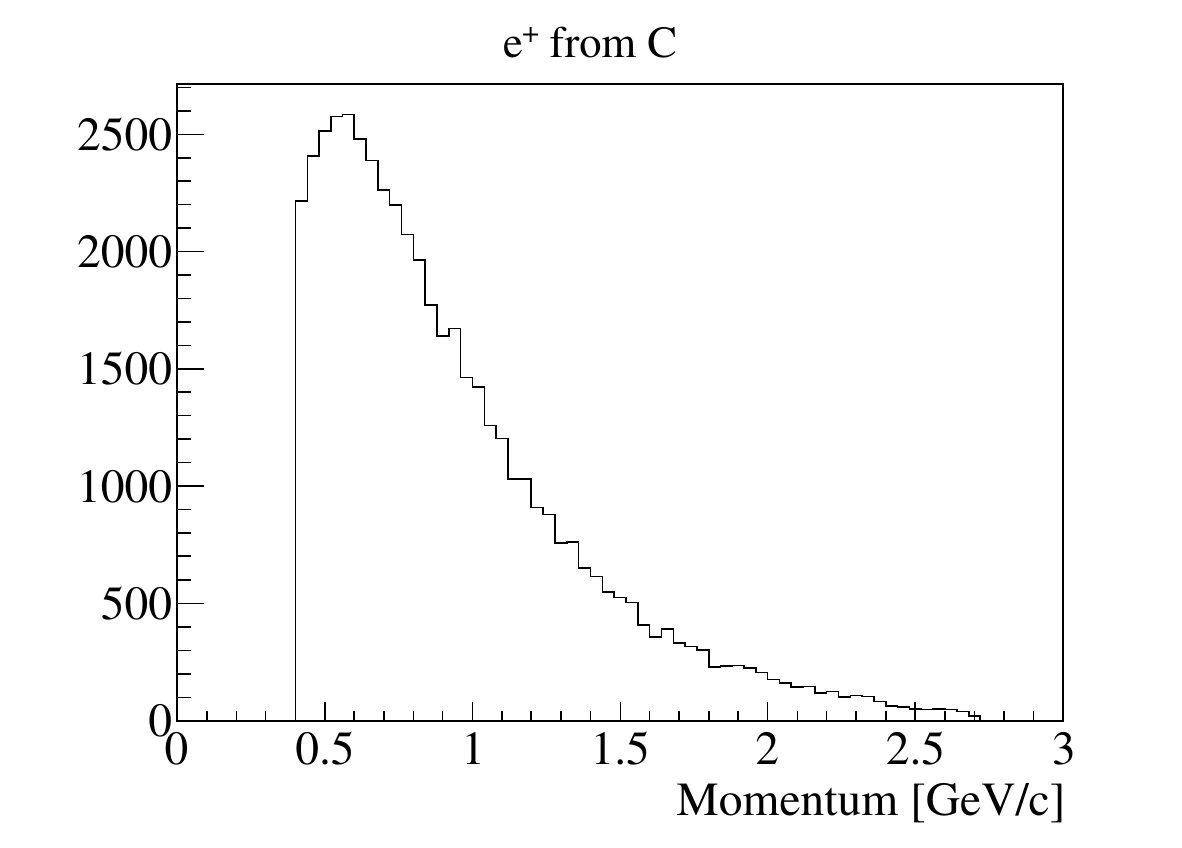}}
      \subfigure[]{
      \includegraphics[width=7cm,keepaspectratio]{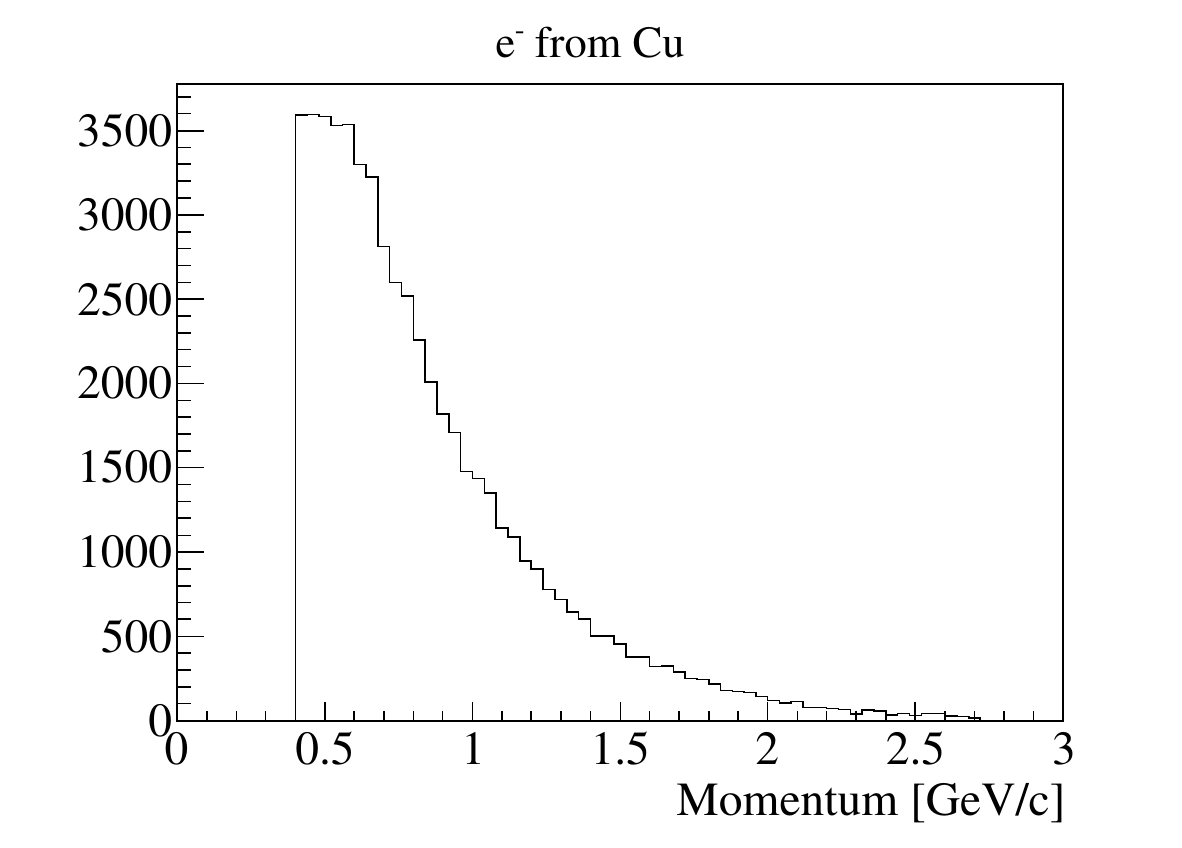}}
      \subfigure[]{
      \includegraphics[width=7cm,keepaspectratio]{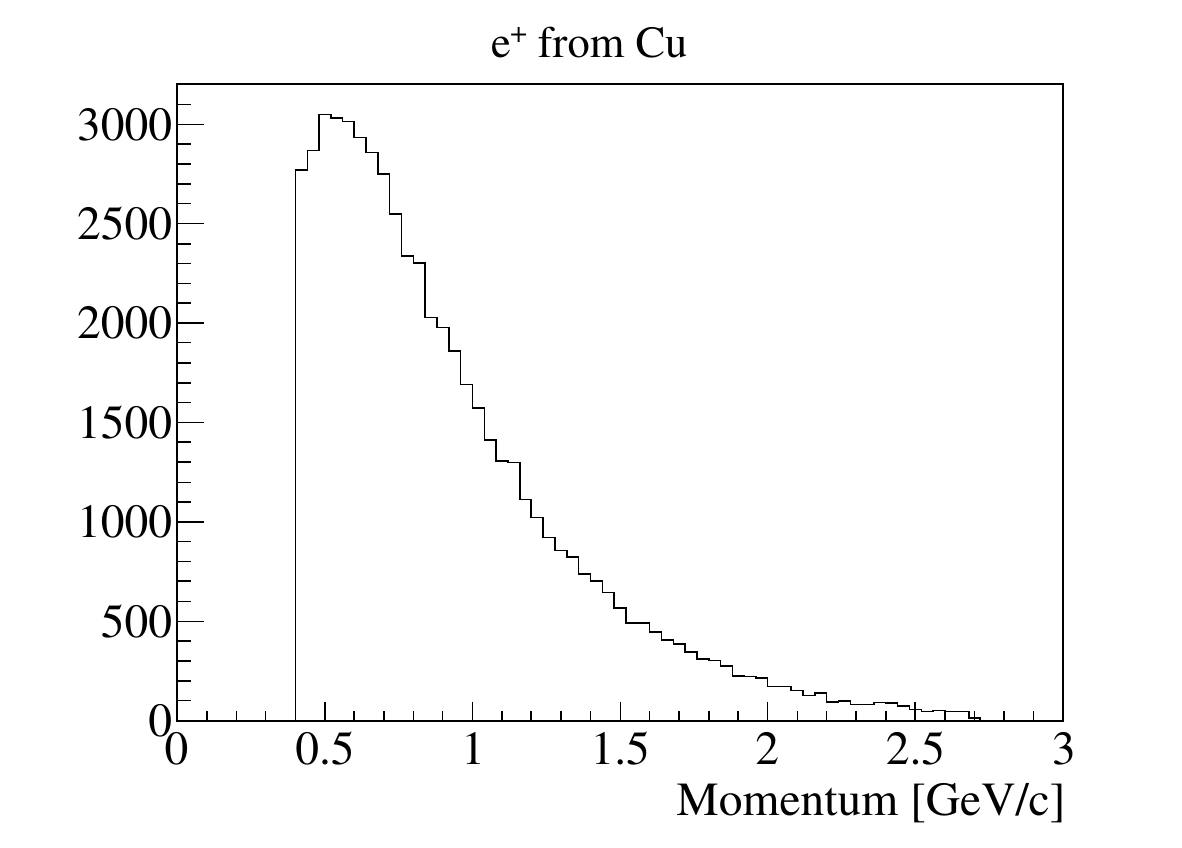}}
   \caption{Momentum distributions of $e^-$ (left) and $e^+$ (right) from the C targets (upper) and the Cu targets (lower).
      The event sets are identical to the obtained mass spectra depicted in Fig.~\ref{fig:mass_each_target}.
      }
      \label{fig:mom_dist_ee}
   \end{center}
\end{figure}

Figure~\ref{fig:mass_each_target} depicts the final invariant mass spectra for each target (two carbon and six copper targets) after implementing the above mentioned fiducial cuts.
The momentum distributions of the final electron and positron samples are depicted in Fig.~\ref{fig:mom_dist_ee}.
``LR events'' were defined as those in which a positron entered into the left arm and an electron entered into the right arm; meanwhile, those with the opposite pattern as ``RL events''.
We exclusively used LR events, as defined in Sec.~\ref{sec:selection}, as the number of RL events was significantly lower than the number of LR events.
Furthermore, the decrease in the acceptance value in the mass range below 0.3 also originates from our choice of LR events. In essence, we discarded pairs with relatively small opening angles.
Distinct peaks in the $\omega$ and $\phi$ mass regions can be observed in all spectra depicted in Fig.~\ref{fig:mass_each_target}. 

\subsection{Detector simulation\label{sec:simulation}}
 A detector simulation code was developed
 using the Geant4 toolkit~\cite{bib:geant4} to evaluate
 the following experimental effects on the mass spectra:
 the Bethe--Bloch type energy loss, known to shift the 
 peak position toward the low-mass side;
 the multiple Coulomb scattering, known to broaden the peak width;
 and bremsstrahlung, known to produces an elongated tail on the low-mass
 side of the resonance peak in the \ee\ channel.

We also created an event generator 
for mesons as $\rho$, $\omega$, $\phi$ and $\eta$
mesons, further using it as input for the detector simulation.
Additional details regarding this are presented in Sec.~\ref{sec:deconvolution}.

\subsubsection{Outline of the simulation\label{section-sim}}
All detector components within the acceptance range are listed in Table~\ref{table-material}
and were incorporated in the simulation.
 Detector efficiencies and tracking efficiencies 
 were also included in the simulation.
 For the eID counters, detailed detector
 responses were not simulated; however, efficiency curves were applied. 
 These efficiency curves as functions of the track momentum were evaluated,
 as described in Sec.~\ref{ssec:eid_efficiency}.
 Meanwhile, for the tracking chambers,
 hit positions were smeared with the obtained residual curves 
 approximated with the double-Gaussian shape.
 The shape parameters for each chamber layer,
 listed in Table~\ref{table-residual},
 were determined as described in Sec.~\ref{section-residual}.
 The detection efficiency of each layer, listed in 
 Table~\ref{table-chambereff}, was also applied.

\subsubsection{Mass scales and resolutions in the simulation\label{ssec:mass_scale}}
Mass scales and resolutions in the simulation were compared with 
the results of data analysis, with additional corrections applied as required.
Some effects, such as tracking chambers misalignments, were not entirely considered 
in the simulation.


To avoid possible meson mass modifications during comparisons, $e^+e^-$ pairs with
$\beta\gamma$ values greater than 1.5 were selected from the final samples;
these fast-moving mesons are anticipated to demonstrate negligible mass modification for $\phi$ mesons.
For $\omega$ mesons, the low-mass sides of the peaks (``excess region'' as described in Sec.~\ref{sec:fit_result}) were excluded from the fitting.
Events from three targets in the 2001 run (five targets in the 2002 run) were summed and
divided into ``LR-events'' and ``RL-events,'' as defined in Sec.~\ref{sec:selection},
as the acceptance values of both event types differed.
Each mass spectrum in the $\phi$ meson mass region was fitted 
against a quadratic background and a Gaussian.
The fit range was set to $2\sigma$, and the fitting process was repeated until convergence.

\begin{figure}[htbp]
   \begin{center}
      \subfigure[$\phi$ meson]{
         \includegraphics[width=7cm]{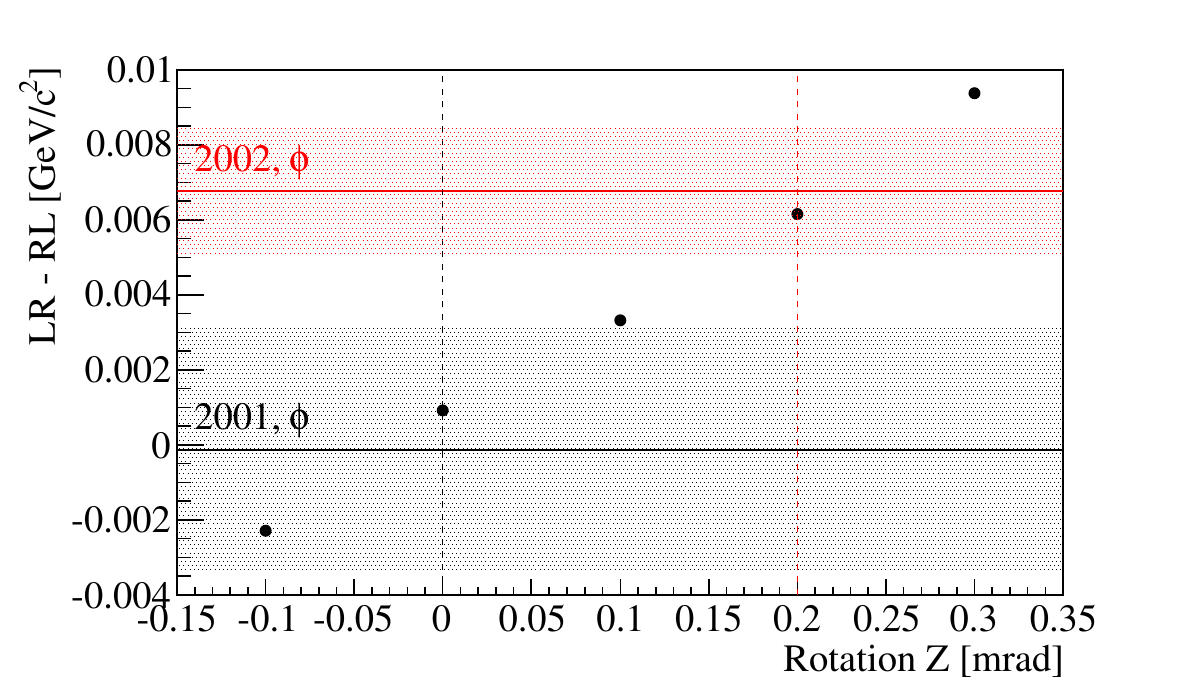}}
      \subfigure[$\omega$ meson]{
         \includegraphics[width=7cm]{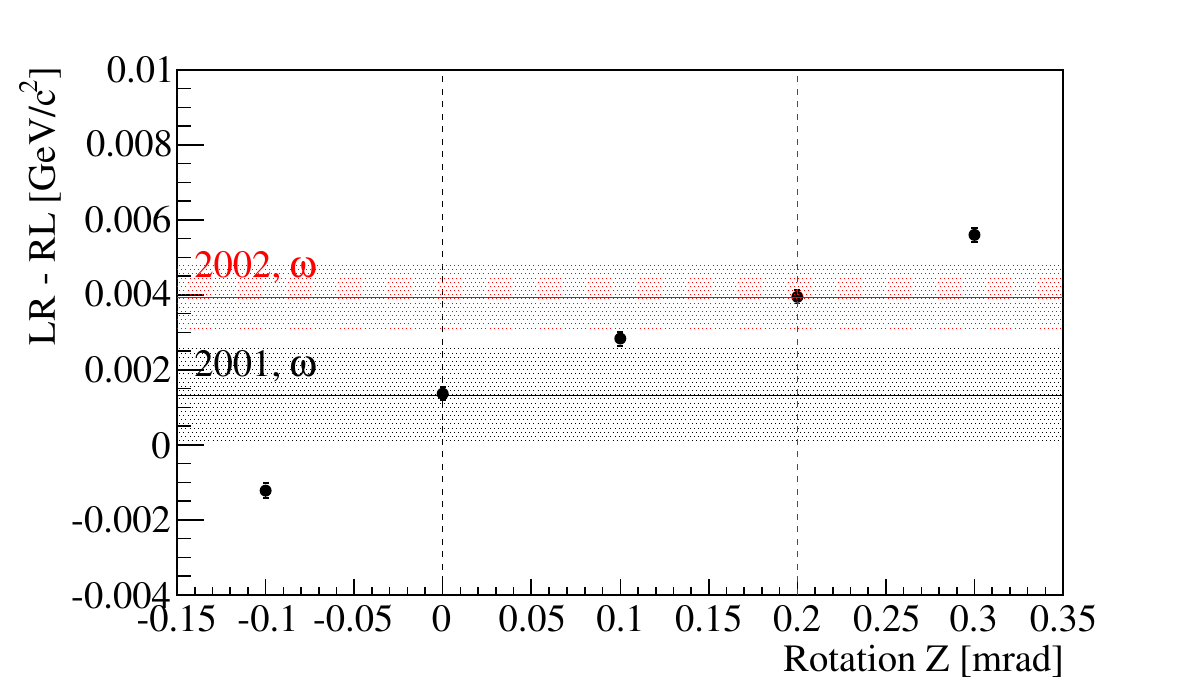}}
      \caption{
         Differences in peak positions between the LR and RL events for $\phi$ mesons (a) and $\omega$ mesons (b).
         Black points indicate the results of the simulation (difference as a function of the misaligned parameter).
         The black and red lines depict the results for the 2001 and 2002 runs, respectively.
         Statistical errors of the data are indicated by the hatched area.
      }
      \label{fig:lrdiff_vs_rotz}
   \end{center}
\end{figure}

First, the peak positions of mesons were checked,
and an additional correction for the rotation of the BDC around the $Z$-axis was applied.
Specifically, if the rotation angles of two BDCs around the $Z$-axis were correct, the peak positions of the LR and RL events expected to match. 
We evaluated the difference (and mean) in the peak position between the LR and RL events as a function of the rotation angle in the simulation, 
and corresponding results are depicted in Fig.~\ref{fig:lrdiff_vs_rotz}.
To reproduce the peak positions of the data, rotation angles of 0.0~mrad and 0.2~mrad were required in the 2001 and 2002 experimental runs, respectively.
Adopting these values in the simulation allowed us to simultaneously reproduce the peak positions of $\omega$ and $\phi$.
Because the peak positions were consistent with the PDG values,  
no additional mass scale factor was applied.
These rotation values were consistent within the uncertainty of the BDC alignment (0.2 - 0.4~mrad as described in Sec.~\ref{ssec:detector_alignment}) and the magnetic field (0.2\% as described in Sec.~\ref{section-magnet}).

\begin{figure}[htbp]
 \begin{center}
  \includegraphics[width=15cm]{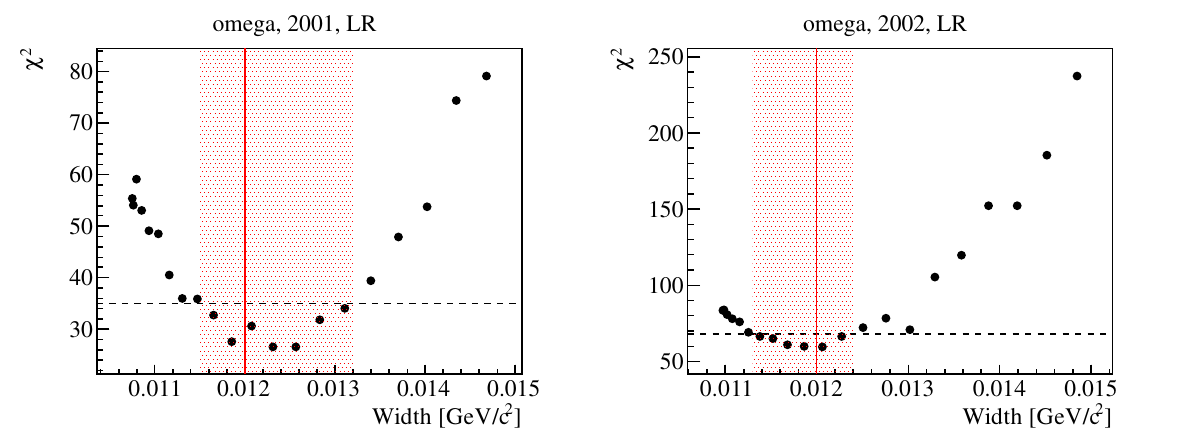}
  \caption{
     $\chi^2$ values obtained by fitting the measured mass spectrum of $\omega$ with the simulated spectra for which the width of $\omega$ is varied.
     Hatched regions correspond to $\varDelta\chi^2 < 9$, while red lines indicate 12.0~MeV/$c^2$, which is adopted for the analysis.
     This observed increase in width is attributed to the $X$ misalignment of the BDCs.}
  \label{fig:mass_width_fit}
 \end{center}
\end{figure}

Next, the width of the $\omega$ peak was evaluated.
To this end, the data were fitted with a peak signal and a background.
During this fitting, 
a convolution of the additional Gaussian component and the simulated spectrum was used for the peak, 
while a quadratic function was used for the background. 
The simulated spectrum already incorporated the effects of position detection resolutions and misalignment 
effects, except for the misalignment of BDCs along the X direction. 
The additional Gaussian component 
accounted for the effects of misalignment of BDCs along the X direction.
These fittings were performed by changing the width of the additional Gaussian component. 
The $\chi^2$ values of the fitting and the total width of the peak, including the 
natural width of the $\omega$ meson and mass resolution, were evaluated. 
Figure~\ref{fig:mass_width_fit} depicts the $\chi^2$ values as functions of the total widths.
However, the width corresponding to the minimum $\chi^2$ value could not be determined based on the $\phi$ meson peak owing to limited statistics.
Hence, we evaluated the width using only the $\omega$ meson data.
The range from the minimum $\chi^2$ to $+9$ is depicted as the hatched region in Fig.~\ref{fig:mass_width_fit}, corresponding to 11.5 to 13.2~MeV/$c^2$ for the 2001 run and 11.3 to 12.4~MeV/$c^2$ for the 2002 run.

\begin{figure}[htbp]
   \begin{center}
      \includegraphics[width=15cm]{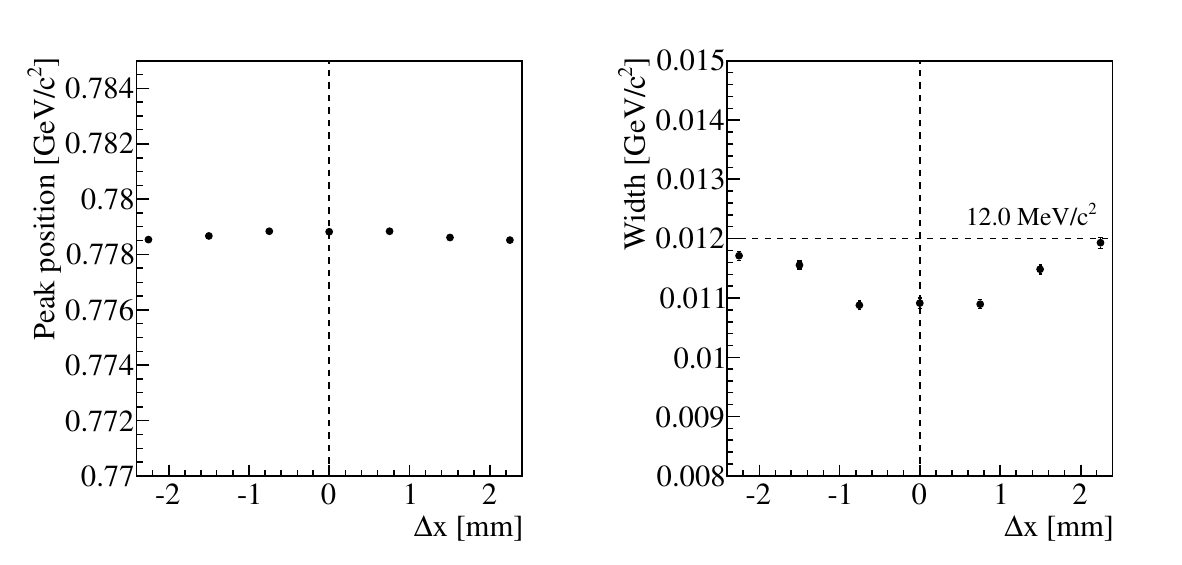}
      \caption{
      Peak position and width as functions of the magnitude of the BDC misalignment along the $X$-axis direction.
      Notably, the two BDCs were moved in opposite directions in the simulation.
      The peak width changed, whereas the peak position remained almost constant.
      }
      \label{fig:width_vs_deltax}
   \end{center}
\end{figure}

The effects of BDC misalignment in the $X$-axis direction were also evaluated.
For this, the positions of the BDCs in the simulation were shifted along the $X$-axis direction.
The peak position and width of the simulated data as functions of the shift ($\varDelta x$) are depicted in Fig.~\ref{fig:width_vs_deltax}.
These widths were estimated by fitting the simulation results against a Gaussian function;
hence, the values of widths included the natural widths of $\omega$ mesons and a mass resolution component.
Here, the shifts of two BDCs were applied in the opposite direction to maintain 
the peak position to reproduce the PDG value. 
As depicted in Fig.~\ref{fig:width_vs_deltax}, 
the considered misalignment in the $X$ direction increases the peak width without altering the peak position.
Specifically, the width increases from $11.0$~MeV/$c^2$ to $12.0$~MeV/$c^2$ if the magnitude of the misalignment is $3\sigma$ of the accuracy of the BDC alignment using straight tracks ($1\sigma \sim 0.75$~mm as described in Sec.~\ref{ssec:detector_alignment}).

In summary, the results of width evaluations obtained from the data and simulations for BDC misalignments were considered. 
The total width of the $\omega$ peak in the simulation was assumed to be 
12.0~MeV/$c^2$.
Furthermore, additional mass smearing was applied to reproduce this mass width in the simulation.
When evaluating the systematic error as described in Sec.~\ref{sec:systematic_errors}, we utilized the results obtained when values at the extremities of this range were adopted.
The obtained mass resolutions in the $\omega$ peak region 
for each $\beta\gamma$ bin are summarized in Table~\ref{tab:mass_resolution}.

\begin{table}[hbpt]
      \caption{Simulated mass resolutions at $\omega$ peaks
      for each $\beta\gamma$ bin and under not $\beta\gamma$ cuts (all $\beta\gamma$).}
      \label{tab:mass_resolution}
   \begin{center}
      \begin{tabular}{cc}
         \hline
              & $\omega$\\ \hline\hline
            $\beta\gamma < 2.1$       &  $6.4 \pm 0.1$~MeV/$c^2$ \\
            $2.1 < \beta\gamma < 2.7$ &  $8.0 \pm 0.1$~MeV/$c^2$  \\
            $2.7 < \beta\gamma$       &  $9.9 \pm 0.1$~MeV/$c^2$  \\
            all $\beta\gamma$         &  $8.2 \pm 0.1$~MeV/$c^2$  \\ \hline
      \end{tabular}
   \end{center}
\end{table}

\begin{figure}[hptb]
   \begin{center}
\includegraphics[width=12cm]{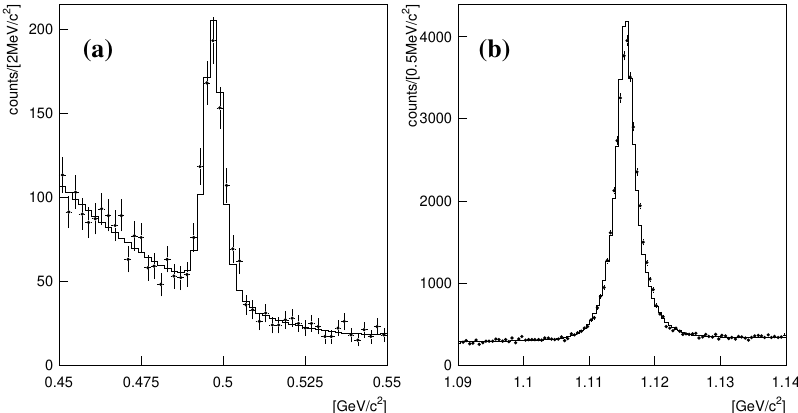}
 \caption{ 
 (a) Invariant masses of $\pi^+$ and $\pi^-$. 
Closed circles with error bars represent the data.  
 The histogram represents the best-fit
 result obtained using the simulated shape of $K_s^0$ and a quadratic background.
 (b) Invariant masses of $p$ and $\pi^-$. 
Closed circles with error bars depict the data.  
 The histogram denotes the fit result obtained using
 the simulated shape of $\Lambda$ and a linear background.
 The depicted data were recorded in the 2001 run and are summed for all targets.
 \FIGciteMuto  
 }
 \label{fig-kslambda}
    \end{center}
\end{figure}

Additionally, the mass scale and mass resolution were verified using $\Lambda \rightarrow p\pi^- $ and $K_s^0 \rightarrow \pi^+\pi^-$
resonances. The observed spectra were found to be consistent with
the simulated ones, as depicted in Fig.~\ref{fig-kslambda}. 
The mass centroid and Gaussian $\sigma$ of the $K_s^0$ data were obtained
as 496.8 $\pm$ 0.2 \MEV\ and 3.9 $\pm$ 0.4 \MEV\ using the 
Gaussian fit and as 496.9 \MEV\ and 3.5 \MEV, through the simulation, respectively. 
These values agree within the acceptable error range.
For the $\Lambda$ data, a Gaussian fit cannot well reproduce
the data shape owing to the asymmetric tail, as depicted in 
Fig.~\ref{fig-kslambda} (b). However, the simulated shape
well reproduces the data. 
When exclusively fitting around the peak using a Gaussian, 
the mass centroid and Gaussian $\sigma$ of the data
were obtained as 1115.62 $\pm$ 0.02 \MEV\ and 1.78 $\pm$ 0.04 \MEV\, respectively,
while the corresponding simulation results were 1115.52 \MEV\ and 1.63 \MEV, respectively.
All targets were summed for the fitting and for the simulation.

\begin{figure}[htbp]
 \begin{center}
  \includegraphics[width=8cm]{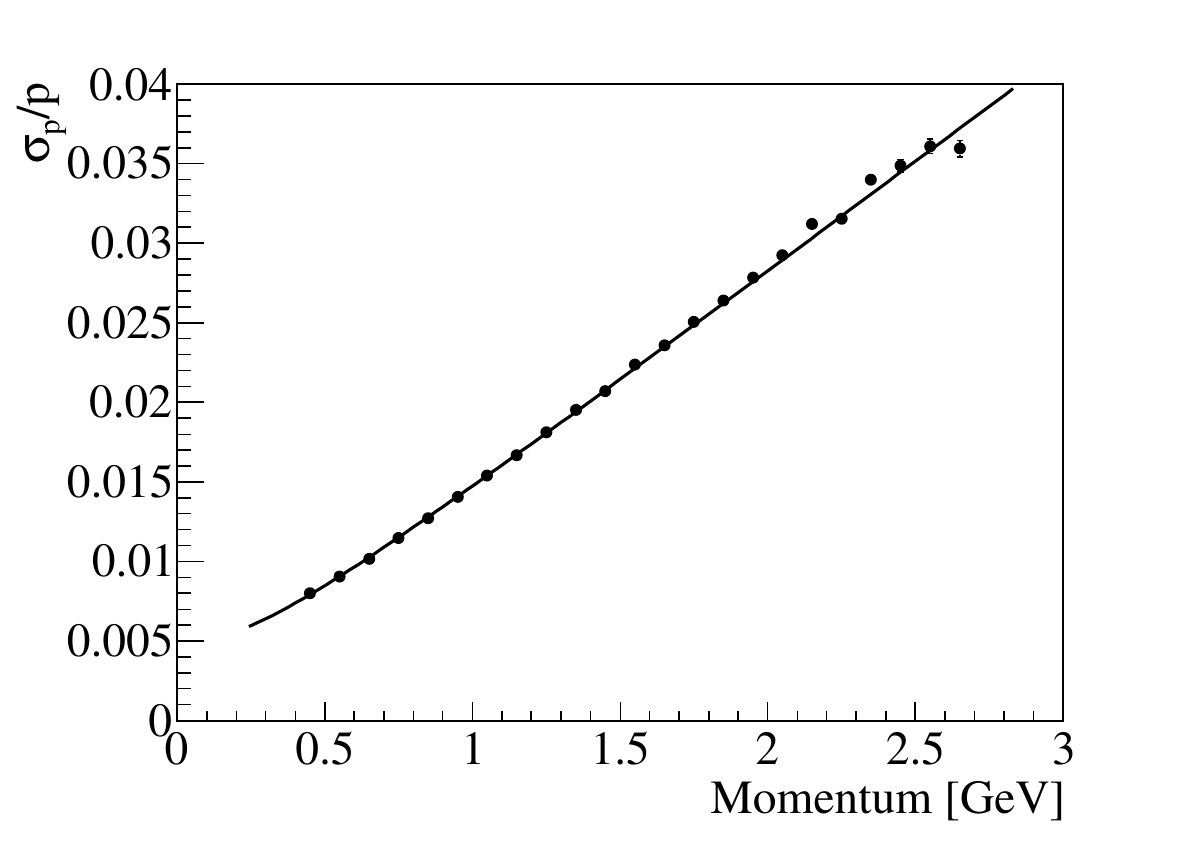}
  \caption{Momentum resolutions obtained in the simulation as a function of the momentum.
    Black line denotes the result of the fitting using Eq.~\ref{eqn:momentum_resolution}.
    }
  \label{fig:g4_mom_resolution}
 \end{center}
\end{figure}

The final momentum resolution in this simulation was evaluated
based on the following relation, and the obtained result is depicted in Fig.~\ref{fig:g4_mom_resolution}:
\begin{eqnarray}
   \sigma_p/p = \sqrt{(1.39\%\cdot p~[\mathrm{GeV}/c])^2 + (0.49\%)^2}
   \label{eqn:momentum_resolution}
\end{eqnarray}

\subsection{Backgound evaluation\label{sec:deconvolution}}
To extract the yields and mass shapes of vector mesons,
we fitted the data against combinatorial backgrounds and known hadronic sources,
including $\rho\rightarrow e^+e^-$,
$\omega\rightarrow e^+e^-$,
$\omega\rightarrow e^+e^-\pi^0$,
$\eta\rightarrow e^+e^-\gamma$,
and $\phi\rightarrow e^+e^-$.
The amplitudes of other hadronic sources, such as $\phi\rightarrow e^+e^-\eta$ and $\eta'\rightarrow e^+e^-\gamma$, were found to be zero during the fitting
and were thus excluded from the final analysis.
The combinatorial background represents combinations of uncorrelated $e^+e^-$ 
pairs emitted by different parent particles,
including misidentified charged pions.
The shape of this combinatorial background was evaluated using an event mixing method.
The spectral shapes of hadronic sources were evaluated based on the simulation described in Sec.~\ref{sec:simulation}.
The kinematic distributions of parent hadrons were obtained using the cascade code JAM~\cite{JAM}.
The amplitude of each component was determined through data fitting.
Notably, the ratio of $\omega\rightarrow e^+e^-$ and $\omega\rightarrow e^+e^-\pi^0$ was fixed by the branching ratio\footnote{
   $\mathrm{BR} = 7.7\times10^{-4}$ for $\omega\rightarrow e^+e^-\pi^0$ and $7.36\times10^{-5}$ for $\omega\rightarrow e^+e^-$ were used~\cite{bib:pdg2020}.
}.
Thus, we had five fitting parameters: the amount of $\rho$, $\omega$, $\eta$, and $\phi$ mesons, and the combinatorial background.

\subsubsection{Resonance shape and decay of hadronic resonance}

For the resonance shapes of $\rho$, $\omega$, $\phi$, $\eta$ mesons, we used the Breit--Wigner formula,
and relevant details are described in Sec.~\ref{sec:model_calculation}.
Two-body decays of mesons were assumed to be isotropic at the meson rest frame.

\begin{figure}[htbp]
   \begin{center}
      \subfigure[$\omega\rightarrow e^+e^-\pi^0$, Invariant mass]{
         \includegraphics[width=7cm,keepaspectratio]{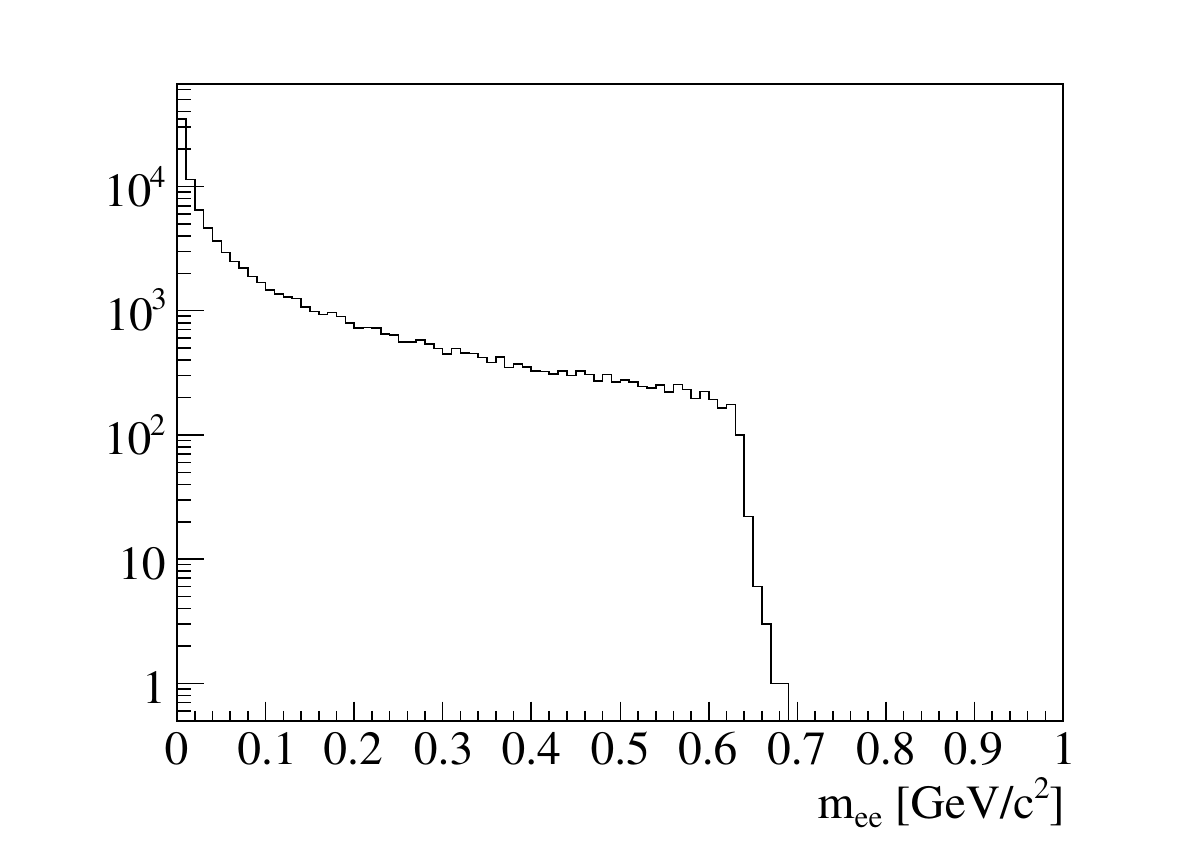}}
      \subfigure[$\eta\rightarrow e^+e^-\gamma$, Invariant mass]{
         \includegraphics[width=7cm,keepaspectratio]{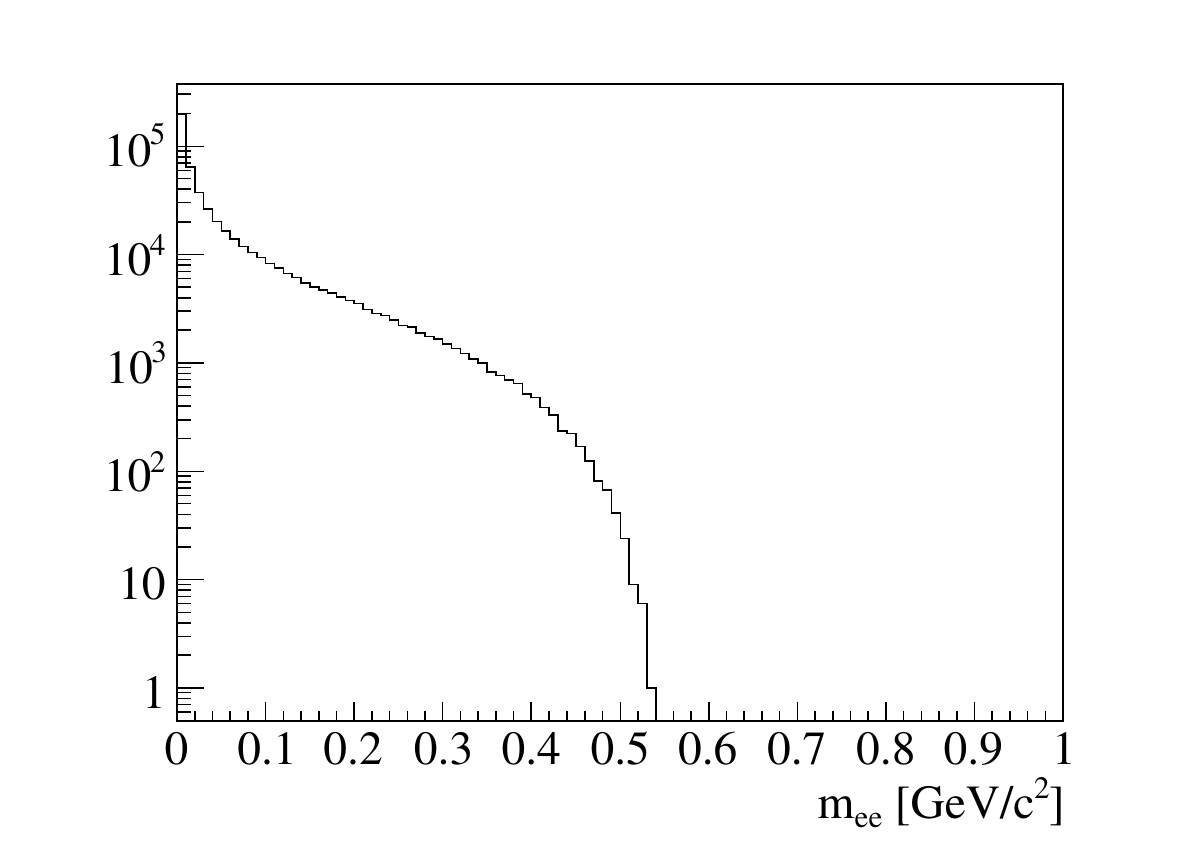}}
      \subfigure[$\omega\rightarrow e^+e^-\pi^0$, Form factor]{
         \includegraphics[width=7cm,keepaspectratio]{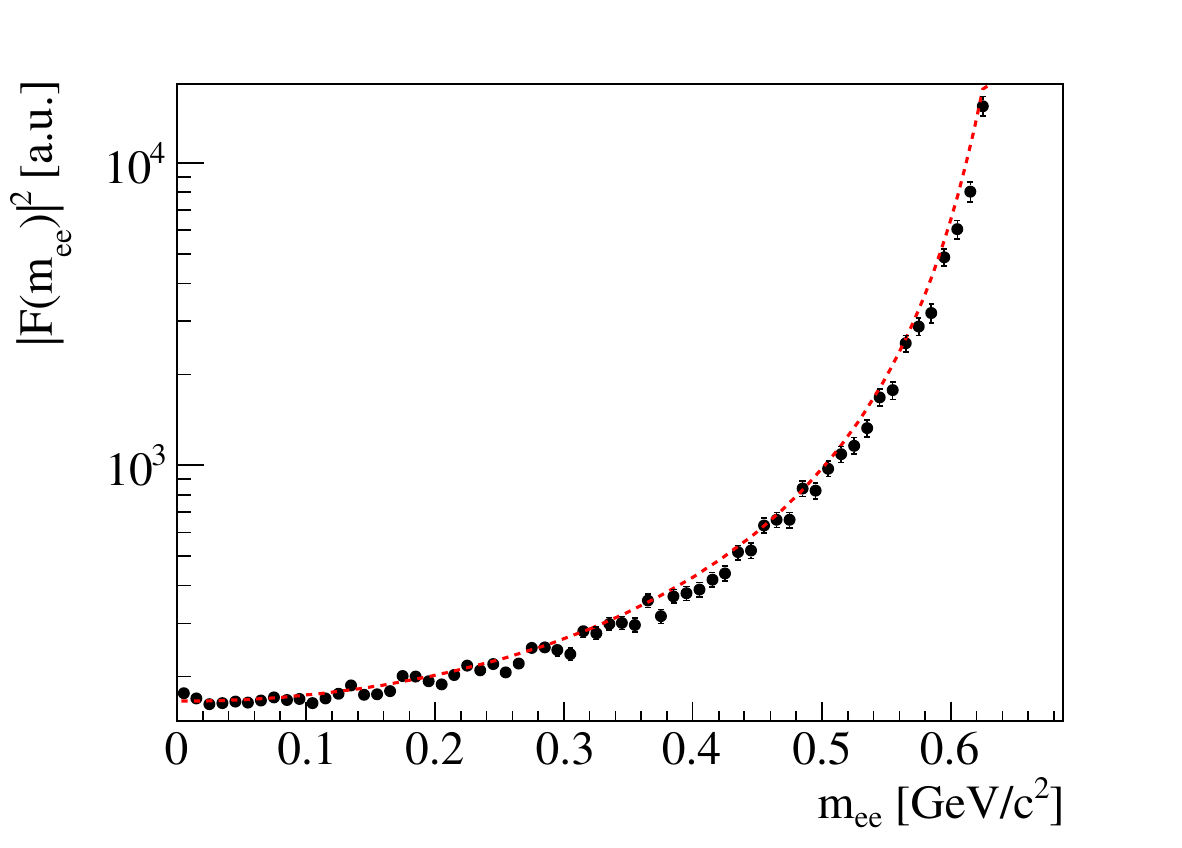}}
      \subfigure[$\eta\rightarrow e^+e^-\gamma$, Form factor]{
         \includegraphics[width=7cm,keepaspectratio]{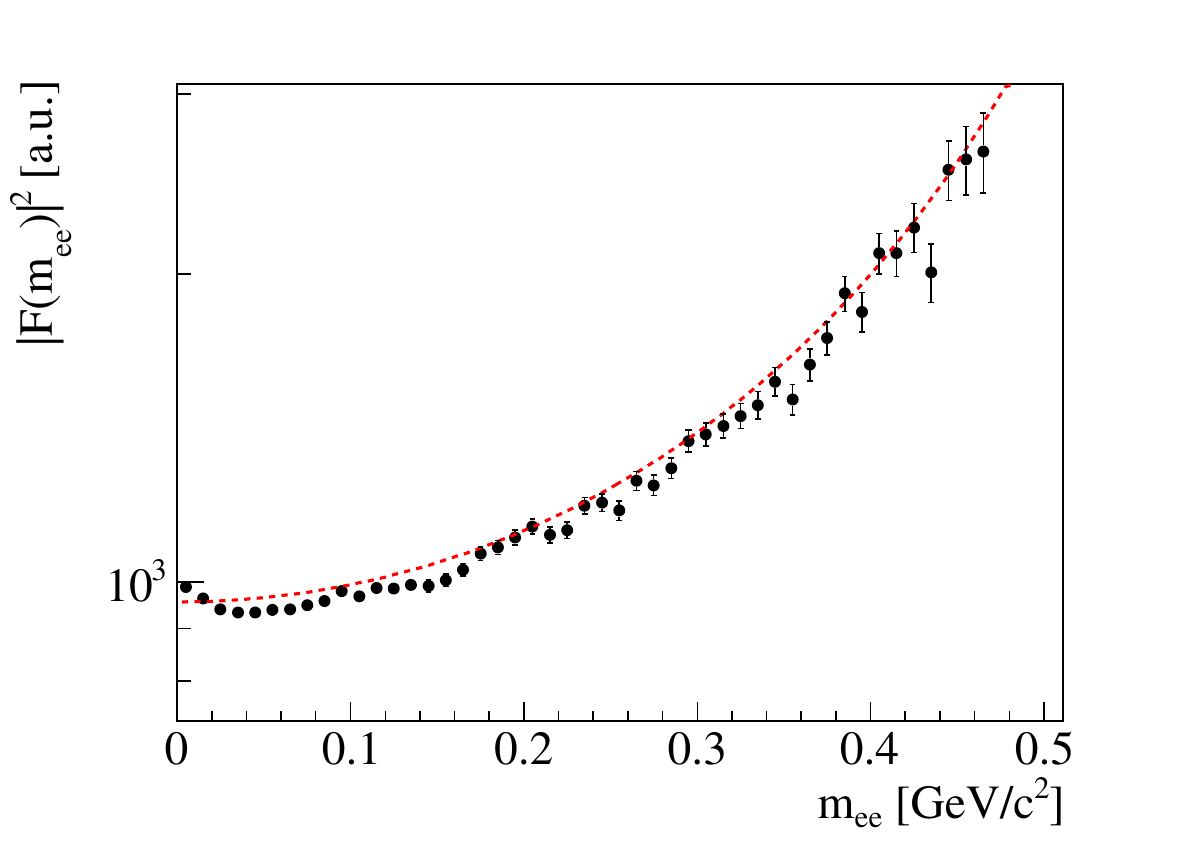}}
      \caption{
         Invariant mass distributions of electron pairs and form factors of the Dalitz decay of $\omega\rightarrow e^+e^-\pi^0$ and $\eta\rightarrow e^+e^-\gamma$.
         Black lines (a, b) and black circles (c, d) denote the results of the simulation, while red dashed lines (c, d) indicate those obtained using Eq.~\ref{eqn:dalitz_form_factor}.
      }\label{fig:dalitz_form_factor}
   \end{center}
\end{figure}

Meanwhile, three-body decays were generated using the event generator code Pluto~\cite{bib:pluto}, updated for the current analysis. 
The invariant mass distribution of the dielectron was generated to follow the given equation:
\begin{eqnarray}
   \frac{ \mathrm{d}\Gamma(m) }{ \mathrm{d}m }
   = &&\Gamma^{A\rightarrow B\gamma}
     \frac{2\alpha}{3\pi m}\sqrt{1-\frac{4m_e^2}{m^2}}
     \left(1+\frac{2m_e^2}{m^2}\right) \nonumber \\
     &&\times
     \left[
        \left(1+\frac{m^2}{m_A^2-m_B^2}\right)^2
        - \left(\frac{2m_Am}{m_A^2-m_B^2}\right)^2
     \right]^{\frac{3}{2}}
     \left|F_A(m)\right|^2
\end{eqnarray}
where $m$ denotes the invariant mass of the dilepton, $m_e$ represents the electron mass, 
$A$ and $B$ indicate the parent meson ($\omega$ or $\eta$) and decay products other than electron pairs ($\pi^0$ or $\gamma$),
and $F_A(m)$ represents the form factor of the Dalitz decay.
For the form factors, the following equation was used:
\begin{eqnarray}
   \left|F(m)\right|^2 = \frac{\Lambda^2(\Lambda^2+\gamma^2)}{(\Lambda^2-m^2)+\Lambda^2\gamma^2}
   \label{eqn:dalitz_form_factor}
\end{eqnarray}
Here, the following values were assumed: $\Lambda_{\omega} = 0.65$~GeV/$c^2$, $\gamma_{\omega} = 0.04$~GeV/$c^2$, $\Lambda_{\eta} = 0.72$~GeV/$c^2$, and $\gamma_{\eta} = 0.00$~GeV/$c^2$~\cite{bib:Landsberg}.
The invariant mass and the form factors of the Dalitz decay used are depicted in Fig.~\ref{fig:dalitz_form_factor}.
Moreover, the helicity angle distribution was generated to follow $1+\cos^2\theta$.

\subsubsection{Internal radiative correction of hadronic resonance}
\begin{figure}[htbp]
  \begin{center}
   \includegraphics[width=8cm]{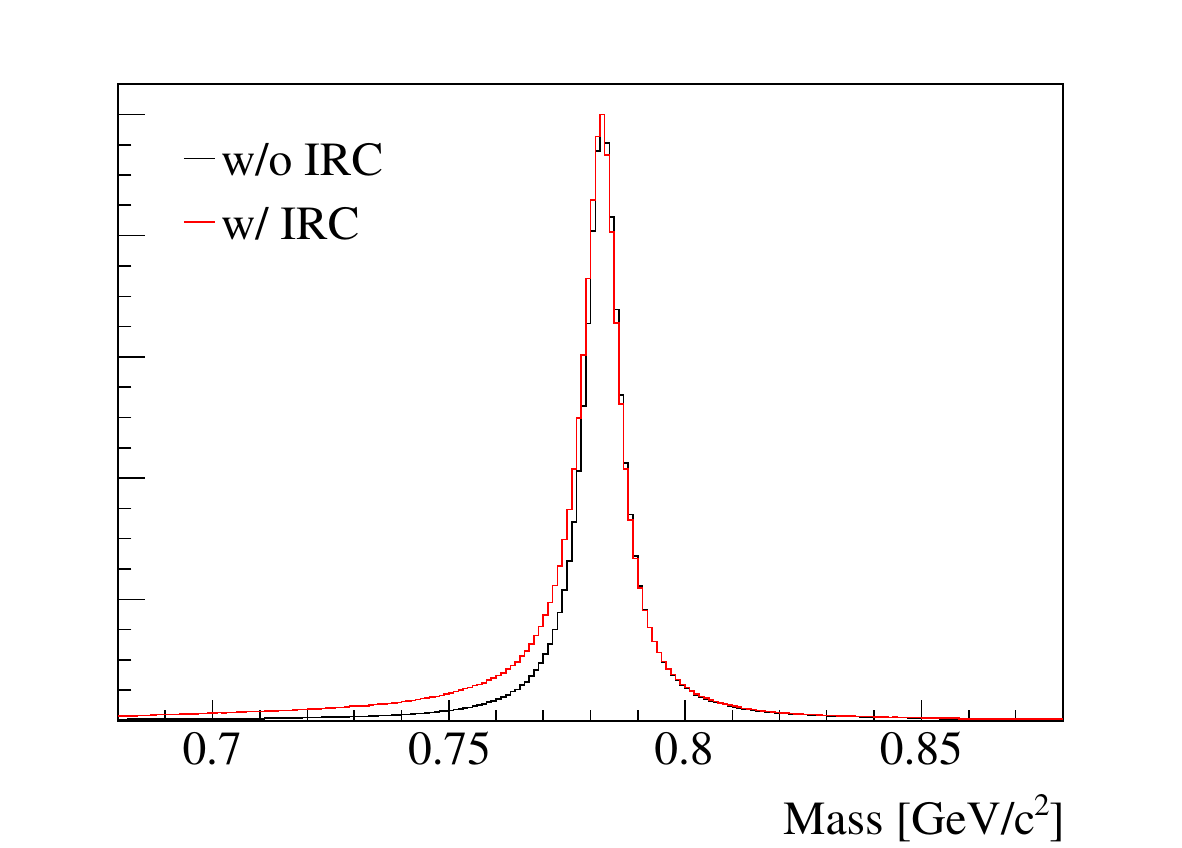}
   \caption{
   \label{fig:irc_omega_ee}
     Invariant mass spectra of $\omega\rightarrow e^+e^-$ before (black) and after (red)  internal radiative correction.
     Both spectra are scaled to the same peak height
     and do not incorporate experimental effects.
   }
  \end{center}
\end{figure}

The radiative corrections of hadronic resonances were updated in the current analysis.
Specifically, two types of radiative effects, internal and external, capable of distorting mass spectral shapes, were considered.
Among these, the external radiative effect, originating from interactions between electrons(positrons) and detector materials, including an interaction target, 
were evaluated using the Geant4 simulation described in Sec.~\ref{sec:simulation}.
The internal radiative effect, that is, quantum electrodynamics corrections at leptonic decays, was evaluated using the PHOTOS code~\cite{bib:photos}.
Here, additional effects such as vertex correction, vacuum polarization, and internal bremsstrahlung were considered.
The typical magnitude of this effect is depicted in Fig.~\ref{fig:irc_omega_ee}, which 
displays the simulated invariant mass spectra of $\omega\rightarrow e^+e^-$ decays before and after the correction.
This correction was also applied to other decays (
$\rho\rightarrow e^+e^-$,
$\omega\rightarrow e^+e^-\pi^0$,
$\eta\rightarrow e^+e^-\gamma$,
$\phi\rightarrow e^+e^-$
).
The magnitude of this effect is independent of materials.

%

\subsubsection{Combinatorial background}
The combinatorial background, representing the combinations of uncorrelated pairs of $e^+e^-$ and misidentified charged pions, constituted the primary background of the mass spectra.
To evaluate the shape of this background, we adopted an event mixing method, which is generally used in heavy-ion experiments.
In this method, one $e^+$ track from one event, and $e^-$ track from a different event.
 The usual event mixing method relies on the principle that such two tracks from different events exhibit no correlation.
 
However, since the particle multiplicity is lower and the signal to background
ratio is higher than that in usual heavy-ion experiments,
the different techniques were adopted as follows. 

In this experiment, the track multiplicity for electrons and positrons
in a triggered event is approximately two and
almost all events have only one electron and one positron.
Therefore, for the normalization of the obtained \ee\ shape,
we cannot use the formula $ N_{+-} = 2 \sqrt{N_{++}N_{--}}$,
which is based on the assumption that
the number of tracks with each charge follows
the binomial distribution~\cite{PHENIX-eventmix}.
Hence, we determined the normalization by the fit.

We did not classify events by multiplicity because only a few tracks existed in an event, unlike in heavy-ion experiments. 
Alternatively, we classified events based on the position of the interaction target and 
the LR/RL track pair configuration.
Mixing was restricted to the same classification.

Furthermore, the amount of correlated pairs was comparable to that of uncorrelated pairs.
The fundamental principle of the usual event mixing method was violated in this situation.
Tracks from resonances demonstrate strong correlations, and such tracks, even when originating 
from different events, can exhibit correlations 
and distort the background shape generated by the event mixing method.
Therefore, we used an ``iterative weighting method'' as follows:
\begin{enumerate}
   \item We fitted the data against an ordinary mixed background shape and hadronic sources.
   \item The background/data ratio $r(m)$ is determined in a bin-by-bin manner.
   \item Using this ratio, a weighted mixed shape was obtained:
      a mixed pair comprising $e^+$ from the $i$-th event and $e^-$ from $j$-th event is weighted by $\displaystyle{\sqrt{r(m_i)\times r(m_j)}}$, where $m_i$ denotes the mass of parent of $e^+e^-$ in the $i$-th event.
   \item We fitted the data against the weighted shape, iterating the process until the weight $r(m)$ converged.
\end{enumerate}

To evaluate the systematic errors of this method, we also used two additional types of mass shapes.
One was the usual mixed shape, while the other was the mixed shape obtained by excluding the $\omega$ meson resonance region, where the weight $r(m) = 0~(0.62 < m < 0.86~\mathrm{GeV}/c^2)$.
The corresponding result is described in Sec.~\ref{sec:systematic_errors}.

    \section{Results}
\subsection{Mass spectra with known hadronic sources\label{sec:fit_result}}
\begin{figure}[htbp]
 \begin{center}
  \includegraphics[width=15.5cm]{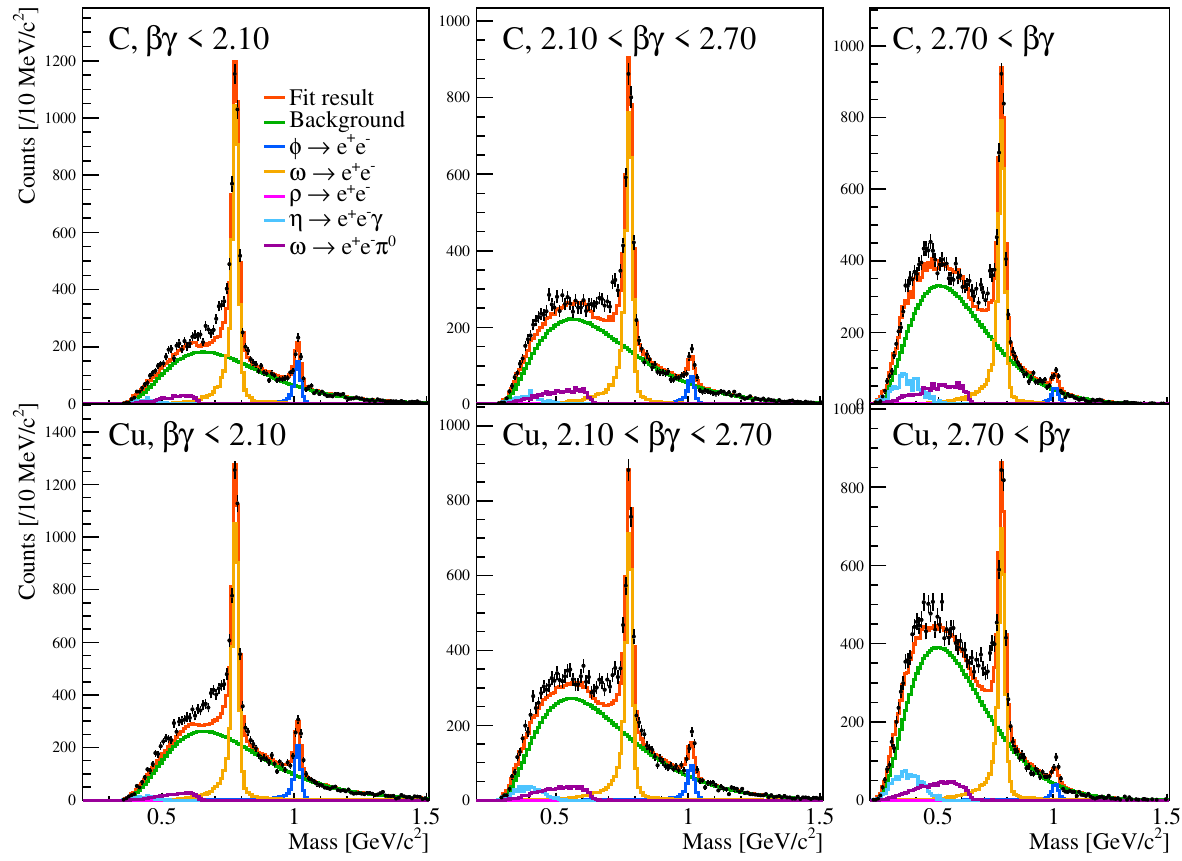}
  \caption{Fitting result obtained when considering all mass regions.
    Red lines depict the best fitting results, obtained as the summation of the backgrounds (green) and known sources,
    such as $\rho\rightarrow e^+e^-$ (magenta),
    $\omega\rightarrow e^+e^-$ (orange),
    $\omega\rightarrow e^+e^-\pi^0$ (purple),
    $\eta\rightarrow e^+e^-\gamma$ (cyan),
    and $\phi\rightarrow e^+e^-$ (blue).
    }
  \label{fig:fit_result_w_eta}
 \end{center}
\end{figure}

The obtained mass spectra were fitted by known hadronic sources and combinatorial backgrounds, as described in the previous section.
Such fittings were performed for each data set, target position, and $\beta\gamma$ region, following which 
the fitting results and errors were summed appropriately.
Figure~\ref{fig:fit_result_w_eta} illustrates
these summed spectra and fitting results for the C and Cu targets
across three $\beta\gamma$ regions.
Clear excesses from the fit results can be seen to the left of the $\omega$ peak in all targets and regions.

In the following discussions, the mass region from 0.0 to 0.55~GeV/$c^2$, 
corresponding to the $\eta$ region, was excluded from the fitting, because we focused on the mass modifications of $\rho$ and $\omega$ mesons. Therefore, the contribution of $\eta$ mesons was not included in the fit. 
The systematic uncertainty associated with the amount of excess caused by the $\eta$ region was evaluated later as described in Sec.~\ref{sec:systematic_errors}.

\subsection{Evaluation of excess\label{sec:eva_excess}}
\begin{figure}[htbp]
 \begin{center}
  \includegraphics[width=15.5cm]{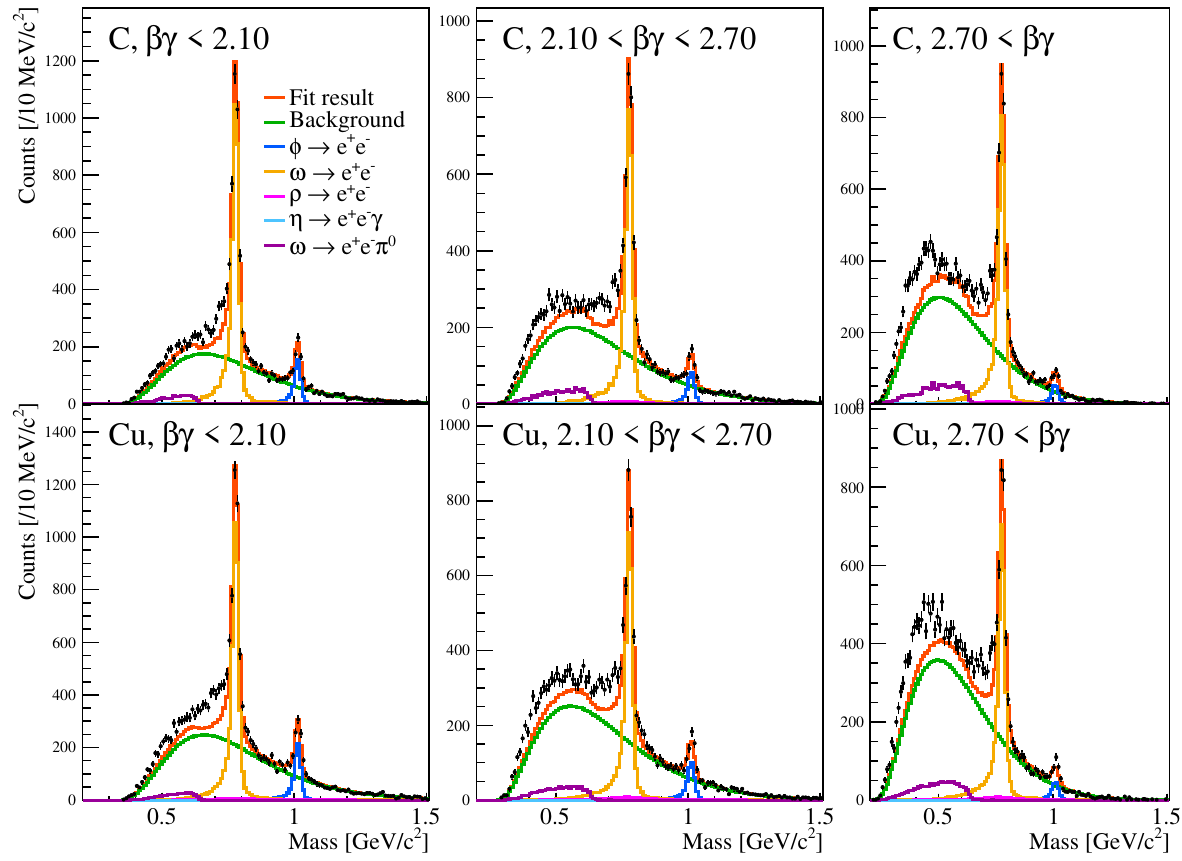}
  \caption{Fitting result of C and Cu target data.
    The excess region is excluded from the fitting.
    Red lines represent the fit results, obtained as the summation of the backgrounds (green) and  known sources,
    such as $\rho\rightarrow e^+e^-$ (magenta),
    $\omega\rightarrow e^+e^-$ (orange),
    $\omega\rightarrow e^+e^-\pi^0$ (purple),
    and $\phi\rightarrow e^+e^-$ (blue).
    }
  \label{fig:fit_result_bgall}
 \end{center}
\end{figure}

\begin{figure}[htbp]
 \begin{center}
  \includegraphics[width=13cm]{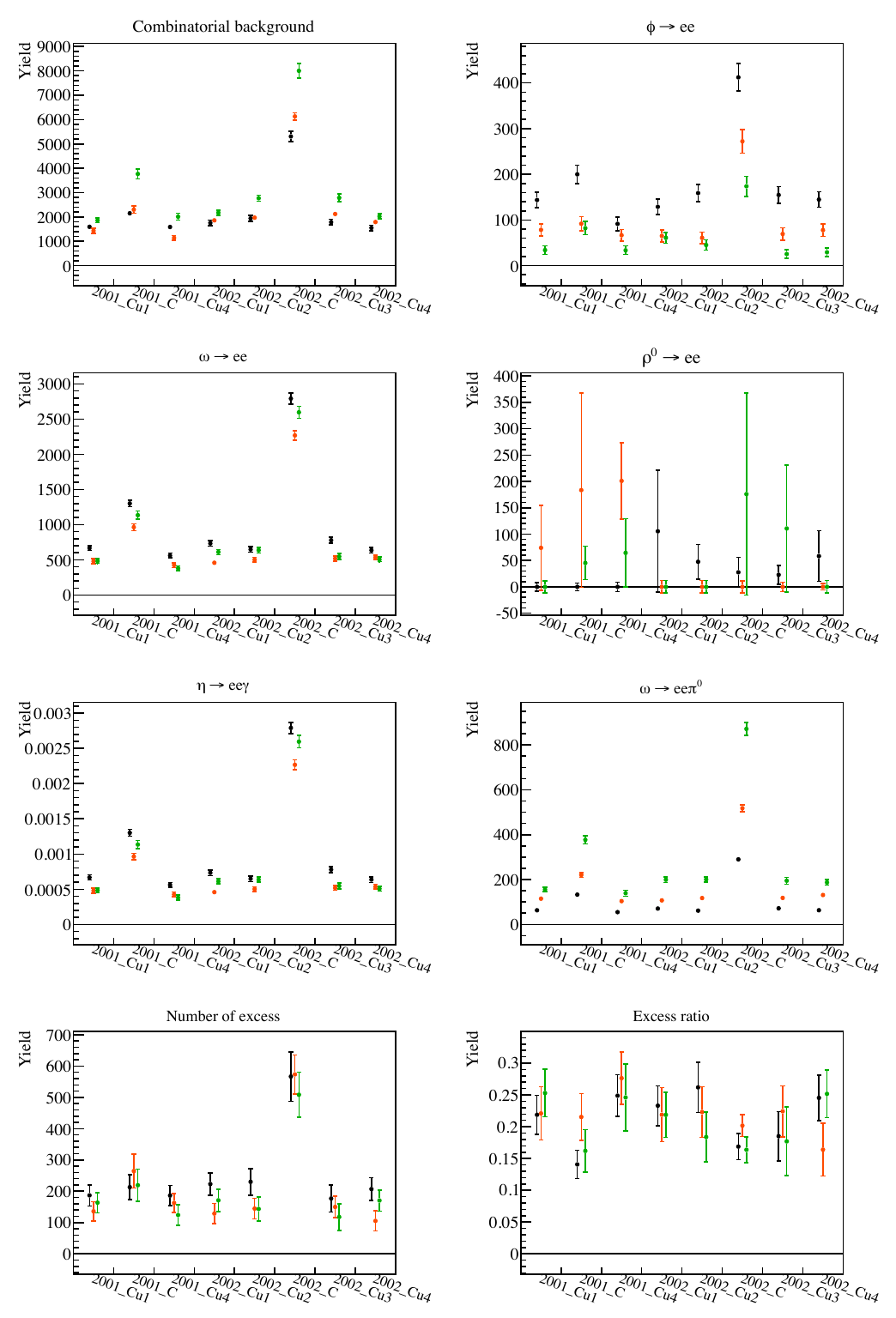}
  \caption{Yields for each component, target, and $\beta\gamma$ region.
    Target names are indicated along the horizontal axis.
    Black, red, and green points indicate the results corresponding to  $\beta\gamma<2.1$, $2.1<\beta\gamma<2.7$, and $2.7<\beta\gamma$, respectively. The excess region is excluded from the fitting.
    }
  \label{fig:yield_summary}
 \end{center}
\end{figure}

The significance of the excess amount was evaluated by comparing of the fitting results with and without the excess region. 
The mass region from 0.62 to 0.76~GeV/$c^2$ was defined as the excess region. 
Fittings both including and excluding the excess region were performed, and obtained $\chi ^2$ results are summarized in Table~\ref{tab:fit_chi2}.
Upon including  the mass excess region in the fit, assumed known sources were unable to reproduce the data across all  $\beta\gamma$ regions and targets at the 99\% confidence level,  owing to a significant excess 
in the low-mass side of the $\omega$ meson peak within each panel.
Subsequently, we excluded the excess region from the fit. Then, assumed known sources were found to reproduce the data well, as depicted
in Fig.~\ref{fig:fit_result_bgall}. The corresponding  $\chi^2$ values are summarized in Table~\ref{tab:fit_chi2}.
All the spectra were adequately reproduced in the fittings.
The yields of hadronic sources determined from the fit are summarized in Fig.~\ref{fig:yield_summary} and Table~\ref{tab:omega_yield}.

\begin{table}[hbpt]
      \caption{$\chi^2$/ndf and its probability values from the fitting process, (a) including and  (b) excluding the excess region.}
      \label{tab:fit_chi2}
      \begin{center}
      \begin{tabular}{cccccc}
         \hline
            & & $\chi^2/\mathrm{ndf}$~(a) & Probability~(a) & $\chi^2/\mathrm{ndf}$~(b) & Probability~(b) \\ \hline\hline
            & $\beta\gamma < 2.1$       &  $334.3/221$ & 1.3e-06 & $186.8/193$ & 0.61   \\
         C  & $2.1 < \beta\gamma < 2.7$ &  $322/206  $ & 4.1e-07 & $184.9/178$ & 0.35   \\
            & $2.7 < \beta\gamma$       &  $247.2/160$ & 1.2e-05 & $132.9/132$ & 0.46   \\ \hline
            & $\beta\gamma < 2.1$       &  $892.9/628$ & 1.5e-11 & $613.6/544$ & 0.020  \\
         Cu & $2.1 < \beta\gamma < 2.7$ &  $707.7/537$ & 9.6e-07 & $519.9/453$ & 0.016   \\
            & $2.7 < \beta\gamma$       &  $507.8/418$ & 0.0017   & $333.7/334$ & 0.49    \\ \hline
      \end{tabular}
   \end{center}
\end{table}

\begin{table}[hbpt]
      \caption{Yield of $\omega$ mesons, the number of excess instances, significance of excess, and excess ratio.}
      \label{tab:omega_yield}
   \begin{center}
      \begin{tabular}{cccccc}
         \hline
            & & $N_{\omega}$ & $N_\mathrm{excess}$ & significance & excess ratio \\ \hline\hline
            & $\beta\gamma < 2.1$       & $4278 \pm 98$ & $ 836 \pm 98$ & $8.5 \sigma$ & $0.16 \pm 0.019$ \\
         C  & $2.1 < \beta\gamma < 2.7$ & $3394 \pm 84$ & $ 802 \pm 82$ & $9.8 \sigma$ & $0.19 \pm 0.019$ \\
            & $2.7 < \beta\gamma$       & $3848 \pm106$ & $ 771 \pm 90$ & $8.6 \sigma$ & $0.17 \pm 0.020$ \\ \hline
            & $\beta\gamma < 2.1$       & $4271 \pm 96$ & $1310 \pm 98$ & $13.4 \sigma$ & $0.23 \pm 0.017$ \\
         Cu & $2.1 < \beta\gamma < 2.7$ & $3127 \pm 82$ & $ 890 \pm 84$ & $10.6 \sigma$ & $0.22 \pm 0.021$ \\
            & $2.7 < \beta\gamma$       & $3348 \pm 90$ & $ 909 \pm 97$ & $9.4 \sigma$ & $0.21 \pm 0.022$ \\ \hline
      \end{tabular}
   \end{center}
\end{table}

\begin{figure}[htbp]
 \begin{center}
  \includegraphics[width=12cm]{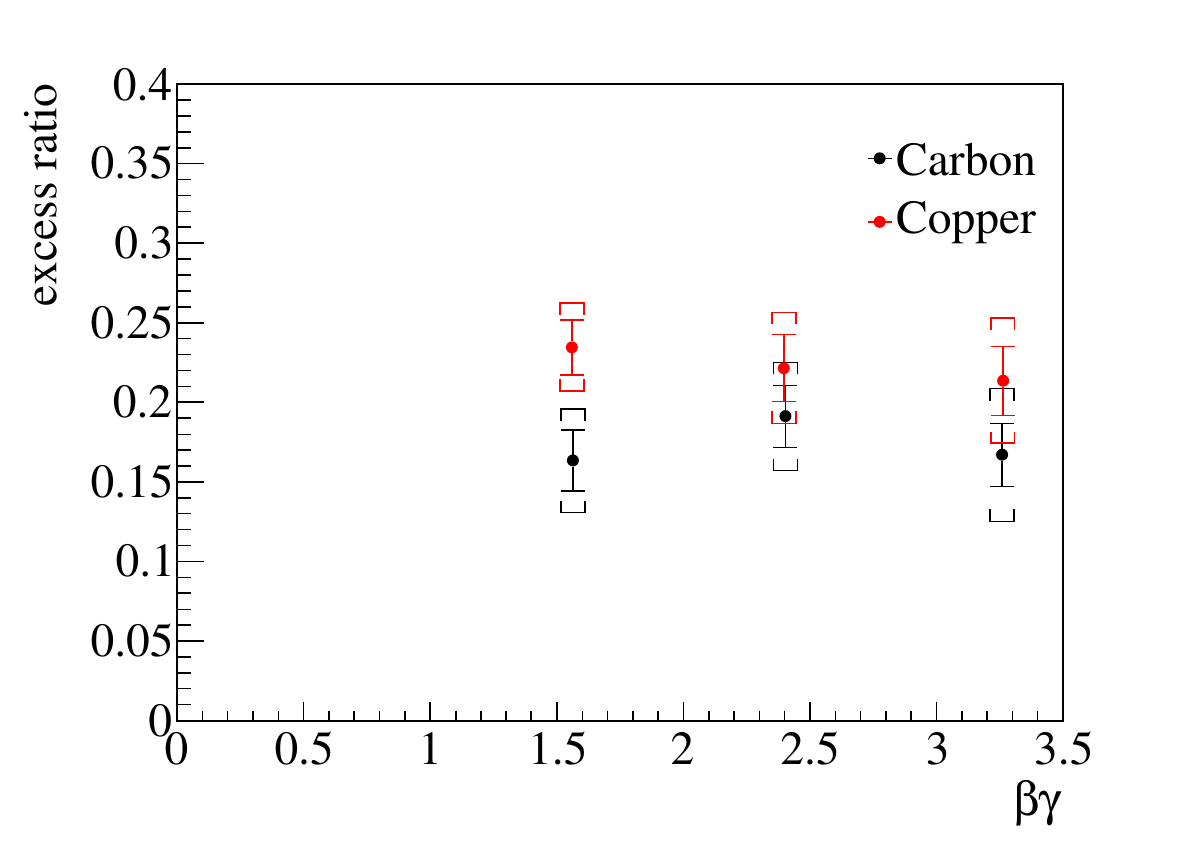}
  \caption{Excess ratio vs $\beta\gamma$.
    Black and red points depict the results of the C and Cu targets, respectively.
    Statistical errors are indicated by error bars, while systematic errors are indicated by square brackets.
    }
  \label{fig:excess_ratio_vs_bg_wbar}
 \end{center}
\end{figure}

Furthermore, the excess amount was evaluated by subtracting the fitted curve from the data in the excluded region.
The excess ratios, $N_\mathrm{excess}/(N_\omega + N_\mathrm{excess})$, are listed in Table~\ref{tab:omega_yield},
and those as functions of the $\beta\gamma$ are depicted in Fig.~\ref{fig:excess_ratio_vs_bg_wbar}.


\begin{figure}[htbp]
   \begin{center}
      \includegraphics[width=10cm]{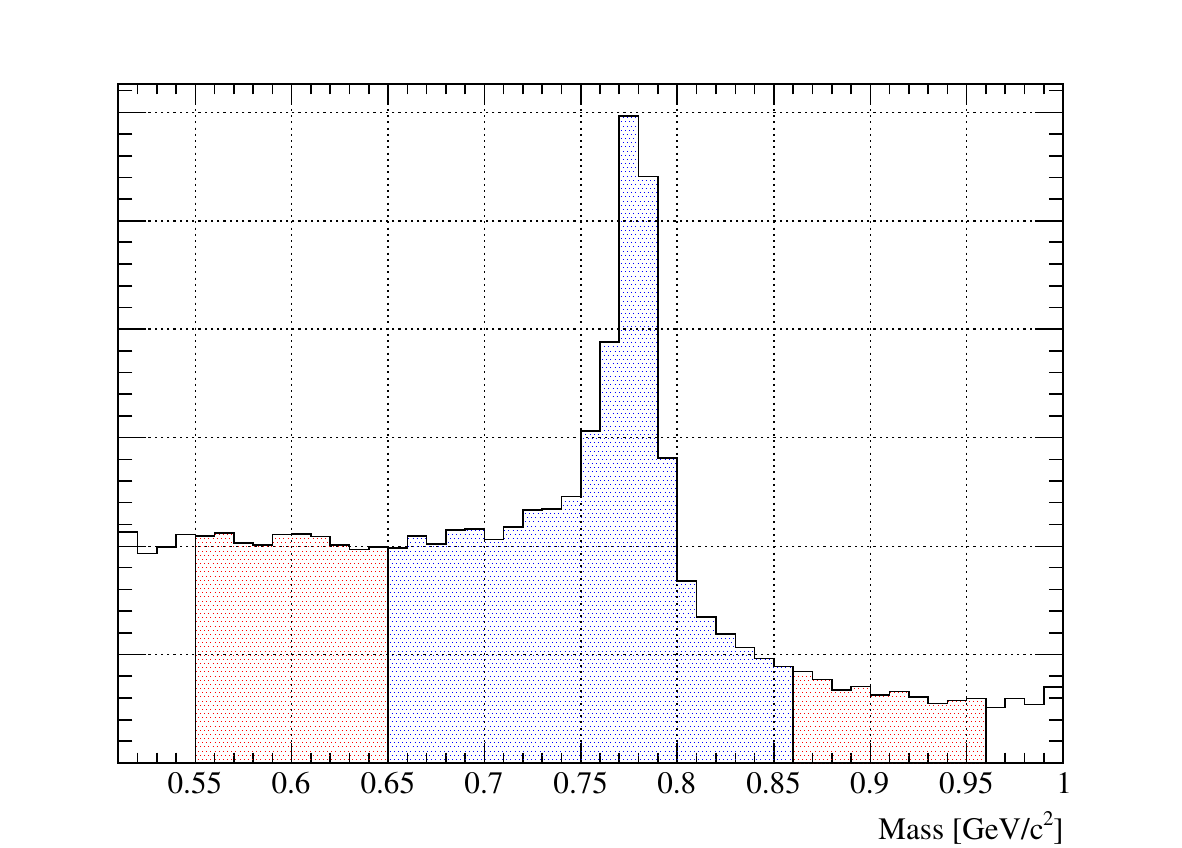}
      \caption{Definitions of the signal and side regions.}
      \label{fig:side_band_definition}
   \end{center}
\end{figure}

\begin{figure}[htbp]
   \begin{center}
      \subfigure[Carbon, $\beta\gamma$]{
      \includegraphics[width=7cm,keepaspectratio]{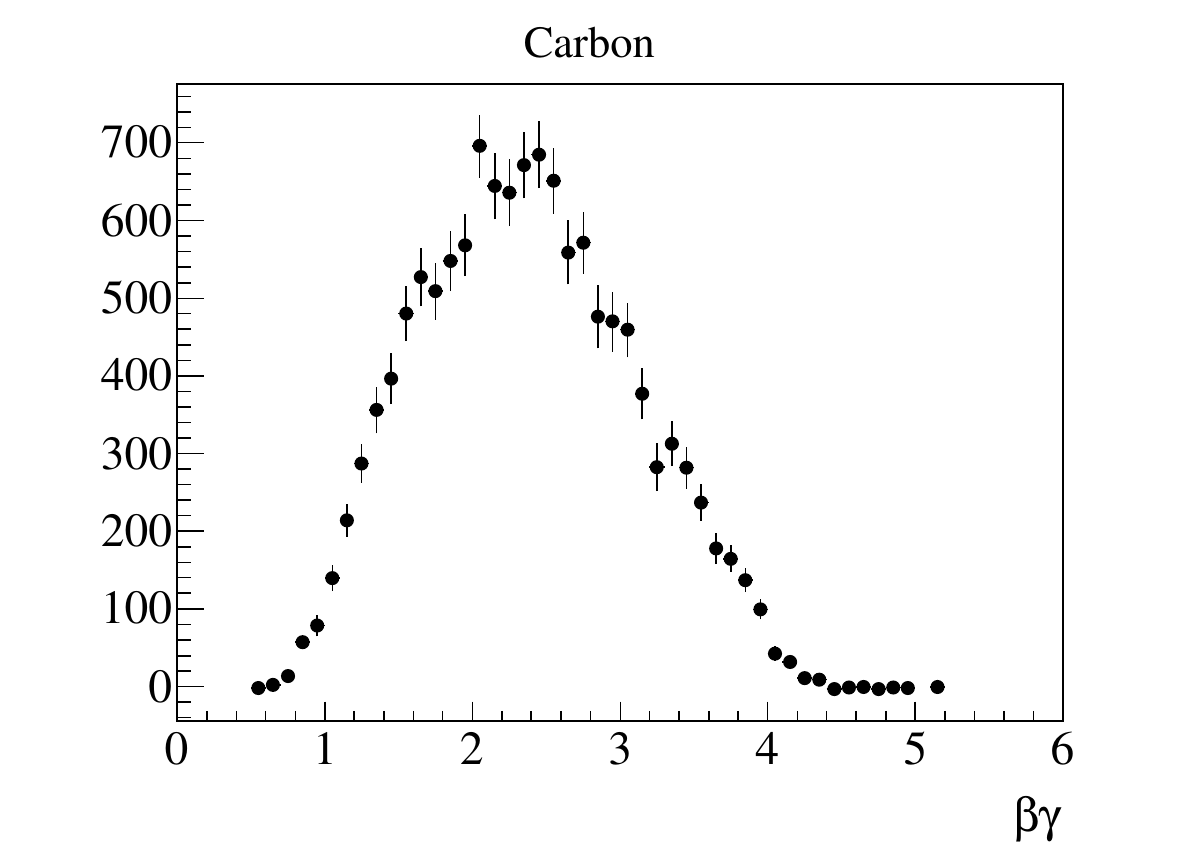}}
      \subfigure[Copper, $\beta\gamma$]{
      \includegraphics[width=7cm,keepaspectratio]{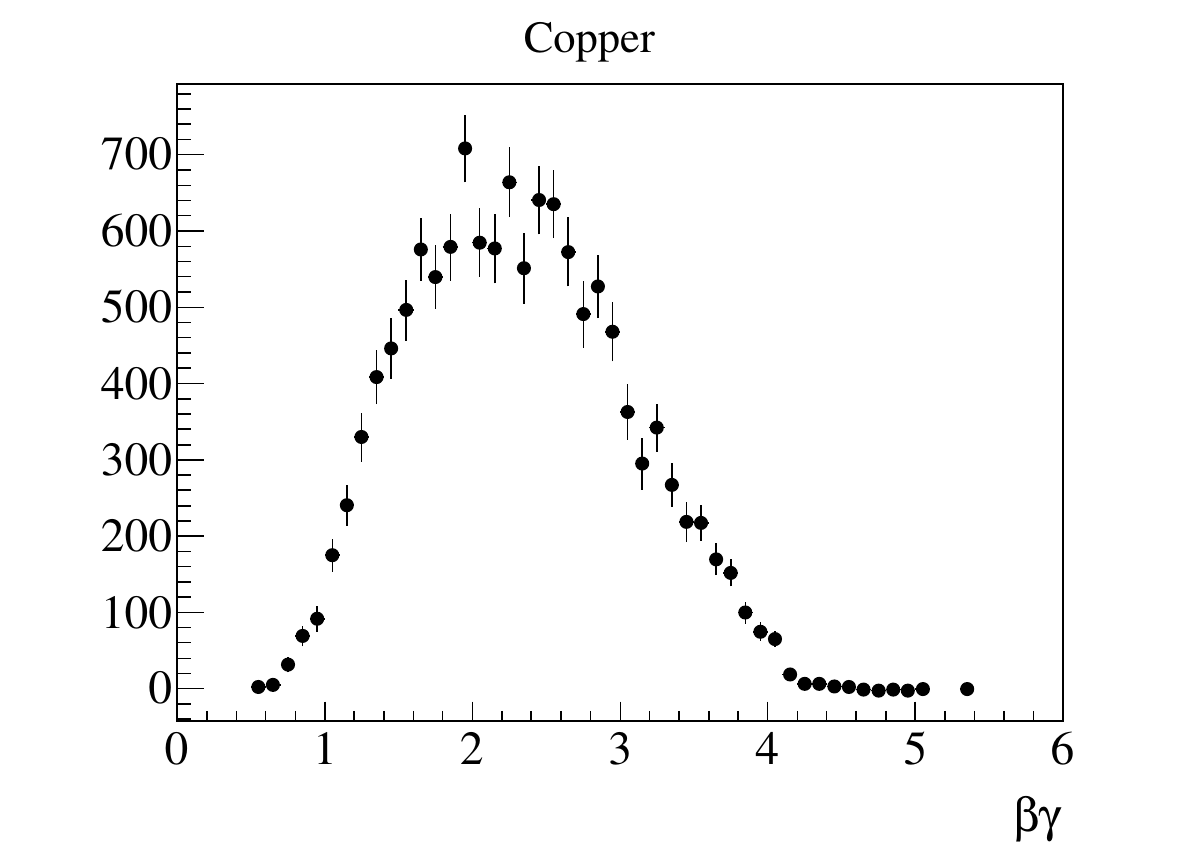}}
      \subfigure[Carbon, $p_T$]{
      \includegraphics[width=7cm,keepaspectratio]{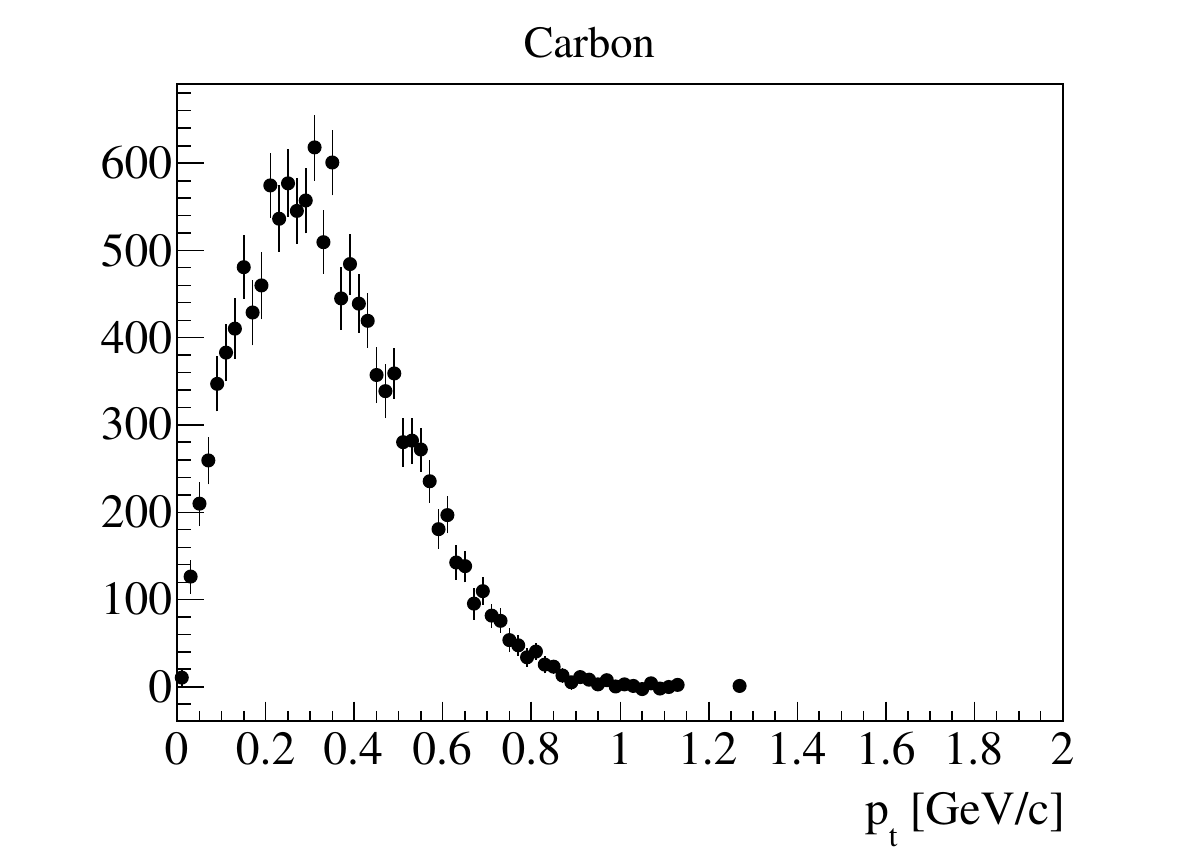}}
      \subfigure[Copper, $p_T$]{
      \includegraphics[width=7cm,keepaspectratio]{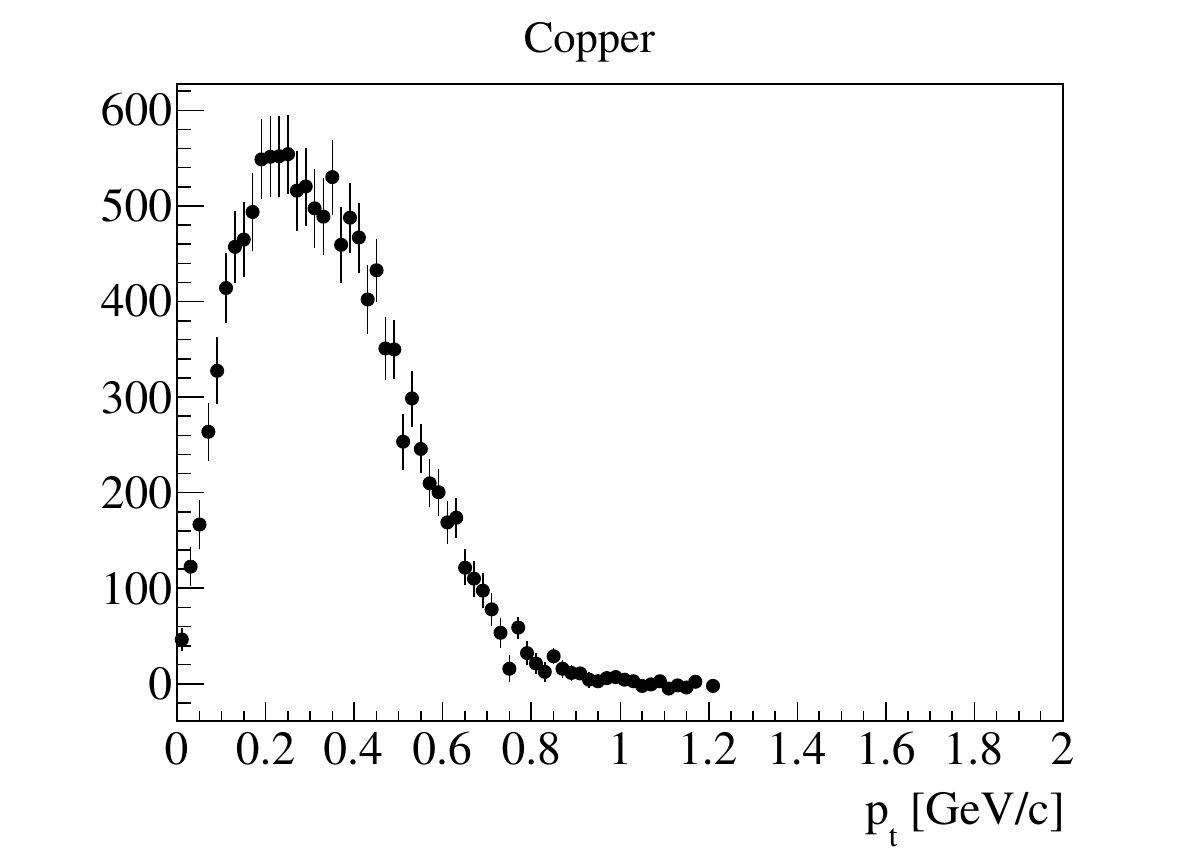}}
      \subfigure[Carbon, rapidity]{
      \includegraphics[width=7cm,keepaspectratio]{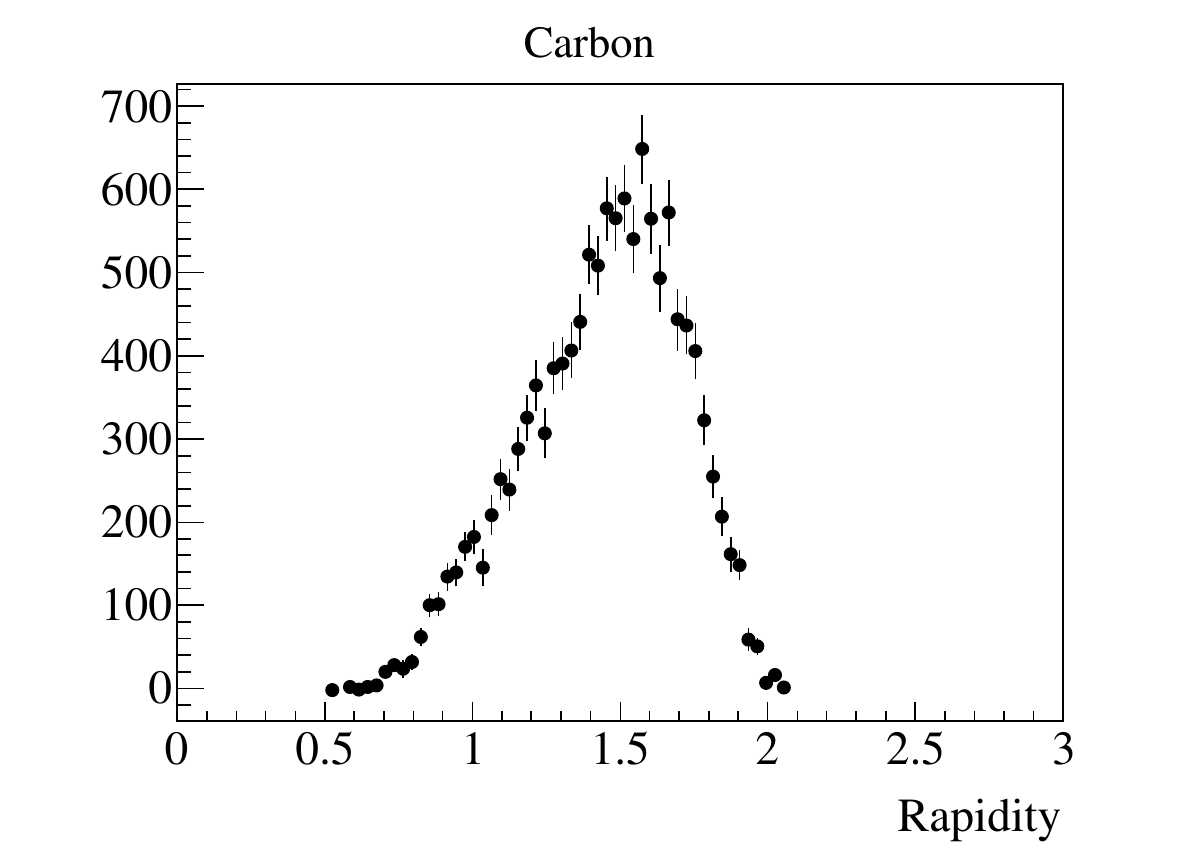}}
      \subfigure[Copper, rapidity]{
      \includegraphics[width=7cm,keepaspectratio]{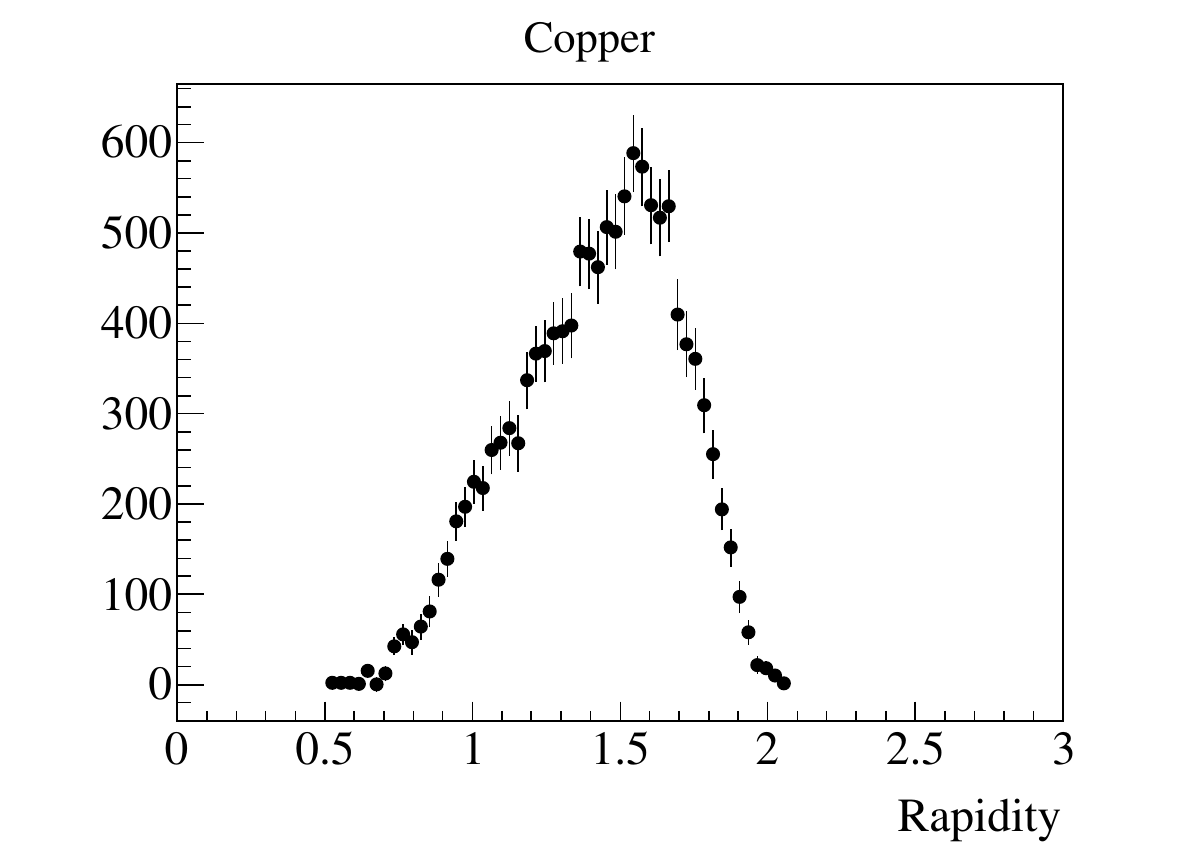}}
   \caption{Kinematic distributions of dilepton pairs in the signal region. (a) $\beta\gamma$ distribution of the C target, (b) $\beta\gamma$ of the Cu target,
      (c) $p_T$ of the C target, (d) $p_T$ of the Cu target, (e) rapidity of the C target, and (f) rapidity of the Cu target.
      Backgrounds are subtracted employing the side-band method as described in the text. }
      \label{fig:side_band_subtraction}
   \end{center}
\end{figure}

The $\beta\gamma$ distributions of $\omega$ mesons were evaluated based on the data employing a side-band subtraction method.
The signal ($\omega$ meson) region, spanned from 
0.65 to 0.86~GeV/$c^2$, the left-side region spanned from 0.55 to 0.65~GeV/$c^2$,
and the right-side region, spanned from 0.86 to 0.96~GeV/$c^2$, 
as depicted in Fig.~\ref{fig:side_band_definition}.
Subsequently, the kinematic distributions of $\omega$ mesons were evaluated
by subtracting the means of the left- and right-side distributions from those of the signal region, and the obtained results are depicted
in Fig.~\ref{fig:side_band_subtraction}.


\subsection{Systematic errors in the excess ratio\label{sec:systematic_errors}}
Systematic errors in the excess ratio were estimated by re-fitting with the following conditions:
\begin{enumerate}
   \renewcommand{\labelenumi}{\Alph{enumi}.}
    \item The fitting region was modified from 0.55 - 2.0~GeV/$c^2$ to 0.0 - 2.0~GeV/$c^2$ (A1), and then to 0.55 - 1.2~GeV/$c^2$ (A2). Notably, A1 includes $\eta\rightarrow e^+e^-\gamma$.
   \item The bin width of the mass spectra was modified from 10.0~MeV/$c^2$ to 6.67~MeV/$c^2$ (B1) or 15.0~MeV/$c^2$ (B2).
   \item The mass scale was changed from 1.000 to 0.998 (C1) or 1.002 (C2).
   \item The mass resolution considered in the simulation was changed to its minimum and maximum values within the uncertainty range.
      The nominal value of Gaussian width was 12.0~MeV/$c^2$ for both data sets,
      with minimum and maximum values of  11.5 (D1) and 13.2~MeV/$c^2$ (D2) for 2001 run, and 11.3 (D1) and 12.4~MeV/$c^2$ (D2) for 2002 run, respectively.
   \item The conditions of event mixing method were altered. The weights of all mass regions were set to one (E1). Furthermore, the weights of the omega region were set to 0 (E2), 
   and the weights of the omega region and the exclude region were also set to 0 (E3).
   \item The evaluation method for the electron identification efficiency was changed. ADC cut efficiency was evaluated using pure electron samples, while the position and time difference cut efficiency was evaluated using final samples (F1).
\end{enumerate}

Maximum deviations were derived for each category,  and the root of the sum of their squares was calculated to evaluate systematic errors. Table~\ref{tab:systematic_error} summarizes the relative deviations and
relative systematic errors for each condition.

\begin{table}[hbpt]
      \caption{
         Relative errors of the excess ratios under different fitting conditions.
         The squares of the maximum errors of each category (e.g., comparison between A1 and A2) were summed to obtain systematic errors (total).
      }
      \label{tab:systematic_error}

   \begin{center}
      \begin{tabular}{cccc}
         \hline
            & \multicolumn{3}{c}{C} \\
            Condition & $\beta\gamma < 2.1$ & $2.1 < \beta\gamma < 2.7$ & $2.7 < \beta\gamma$ \\ \hline\hline
A1      &       1.1 \%  &       6.9 \%  &       7.9 \%  \\
A2      &       7.2 \%  &       9.3 \%  &       12.8 \% \\
B1      &       13.8 \% &       10.3 \% &       16.0 \% \\
B2      &       3.2 \%  &       1.4 \%  &       10.5 \% \\
C1      &       7.1 \%  &       7.3 \%  &       12.2 \% \\
C2      &       9.3 \%  &       8.5 \%  &       13.3 \% \\
D1      &       3.4 \%  &       0.5 \%  &       0.8 \%  \\
D2      &       5.0 \%  &       6.2 \%  &       3.7 \%  \\
E1      &       2.6 \%  &       1.7 \%  &       2.6 \%  \\
E2      &       7.9 \%  &       5.9 \%  &       5.0 \%  \\
E3      &       1.5 \%  &       0.6 \%  &       2.5 \%  \\
F1      &       1.3 \%  &       0.9 \%  &       1.7 \%  \\ \hline\hline
Total   &       20.4 \% &       18.4 \% &       25.2 \% \\ \hline
            & & & \\ \hline
            & \multicolumn{3}{c}{Cu} \\
            Condition & $\beta\gamma < 2.1$ & $2.1 < \beta\gamma < 2.7$ & $2.7 < \beta\gamma$ \\ \hline\hline
A1      &       2.1 \%  &       9.0 \%  &       10.3 \% \\
A2      &       4.7 \%  &       7.4 \%  &       8.8 \%  \\
B1      &       7.0 \%  &       9.0 \%  &       13.1 \% \\
B2      &       2.0 \%  &       5.4 \%  &       4.1 \%  \\
C1      &       4.3 \%  &       6.3 \%  &       7.1 \%  \\
C2      &       5.2 \%  &       5.9 \%  &       7.0 \%  \\
D1      &       2.1 \%  &       0.2 \%  &       1.2 \%  \\
D2      &       3.0 \%  &       4.0 \%  &       3.5 \%  \\
E1      &       2.0 \%  &       1.8 \%  &       1.1 \%  \\
E2      &       5.9 \%  &       6.0 \%  &       2.8 \%  \\
E3      &       2.4 \%  &       0.1 \%  &       2.0 \%  \\
F1      &       0.7 \%  &       0.9 \%  &       0.8 \%  \\ \hline\hline
Total   &       11.9 \% &       16.0 \% &       18.7 \% \\ \hline
      \end{tabular}
   \end{center}
\end{table}

    \section{Evaluations of spectral modifications}
Significant excesses were observed across all targets and all $\beta\gamma$ regions, indicating
considerable mass modifications of $\rho$ and $\omega$ mesons within nuclei.
The $\beta\gamma$ dependence of these $\rho$ and $\omega$ mass modifications was derived, revealing statistically significant excesses across all $\beta\gamma$ regions.
For an interpretation of the $\beta\gamma$ dependence of meson modifications based on the contents of Fig.~\ref{fig:excess_ratio_vs_bg_wbar},
it should be noted that the magnitudes of the excess are influenced not only by meson mass modifications,
but also by other experimental effects such as meson production kinematics and detector acceptance.
Thus, model calculations were performed to understand the characteristics of the observed excesses.


To interpret the obtained results, 
we parameterized the modifications of the meson mass spectra as described in Sec.~\ref{sec:model_calculation} and conducted a Monte-Carlo type model calculation.
The model incorporated the nuclear density distributions of the target nuclei, the momentum distributions of the generated mesons, and experimental effects within the detectors.
All contributions other than $\rho\rightarrow e^+e^-$ and $\omega\rightarrow e^+e^-$ (including $\omega\rightarrow \pi^0e^+e^-$) were subtracted from the obtained mass spectra,
and the spectra after subtraction were fitted against the simulated shapes of $\omega\rightarrow e^+e^-$ and $\rho\rightarrow e^+e^-$ for several modification parameters.
The optimal parameters for reproducing the mass spectra
were obtained as detailed in Sec.~\ref{sec:modification_parameter},
where mass modification parameters  
for $\rho$ and $\omega$ mesons were determined for each $\beta\gamma$ bin.
The $\beta\gamma$ dependence of mass modifications for $\rho$ and $\omega$ mesons was obtained for the first time in our analysis.
In addition, new model analyses were conducted for comparisons with the CLAS results, incorporating assumptions about the asymmetric resonance shape.

\subsection{Model calculations\label{sec:model_calculation}}
\begin{figure}[htbp]
 \begin{center}
 \subfigure[$\beta\gamma$, C]{
\includegraphics[width=4.5cm,keepaspectratio]{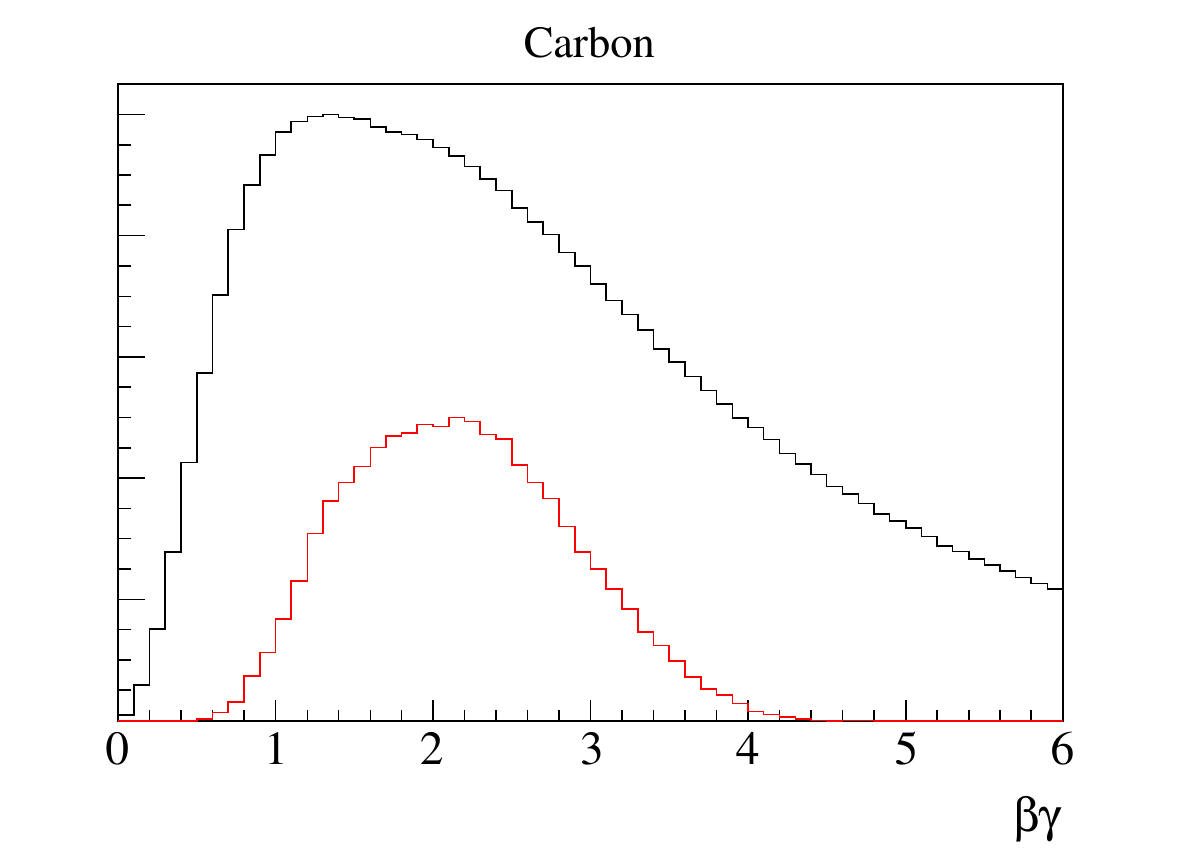}}
  \subfigure[$p_T$, C]{
  \includegraphics[width=4.5cm,keepaspectratio]{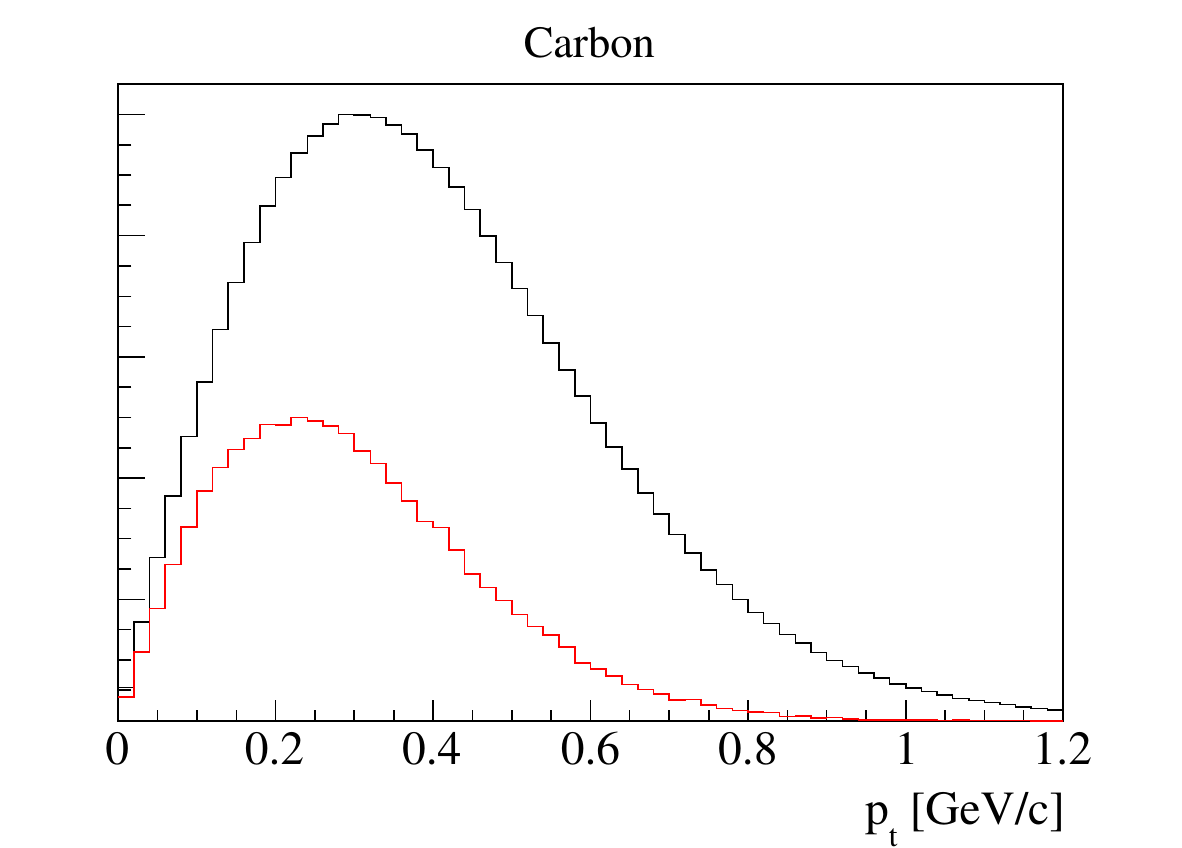}}
  \subfigure[Rapidity, C]{
  \includegraphics[width=4.5cm,keepaspectratio]{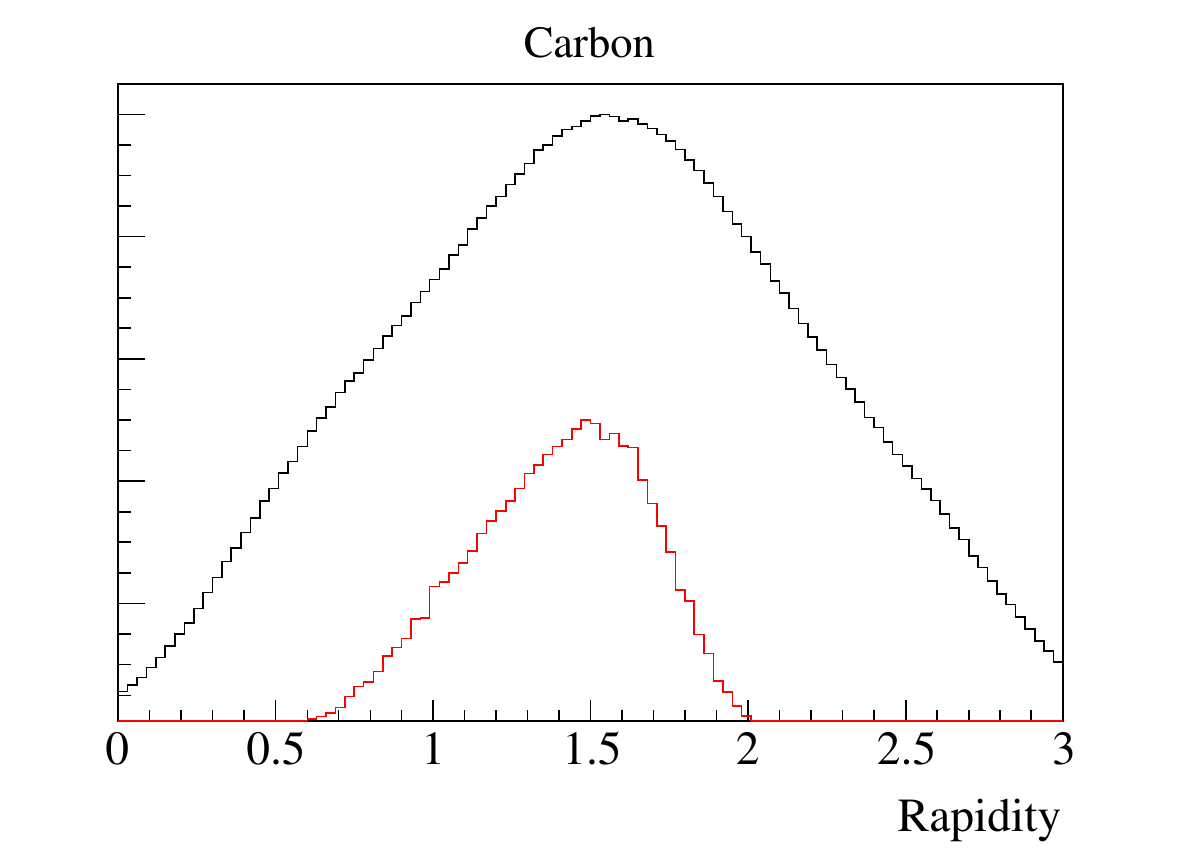}}
\subfigure[$\beta\gamma$, Cu]{
  \includegraphics[width=4.5cm,keepaspectratio]{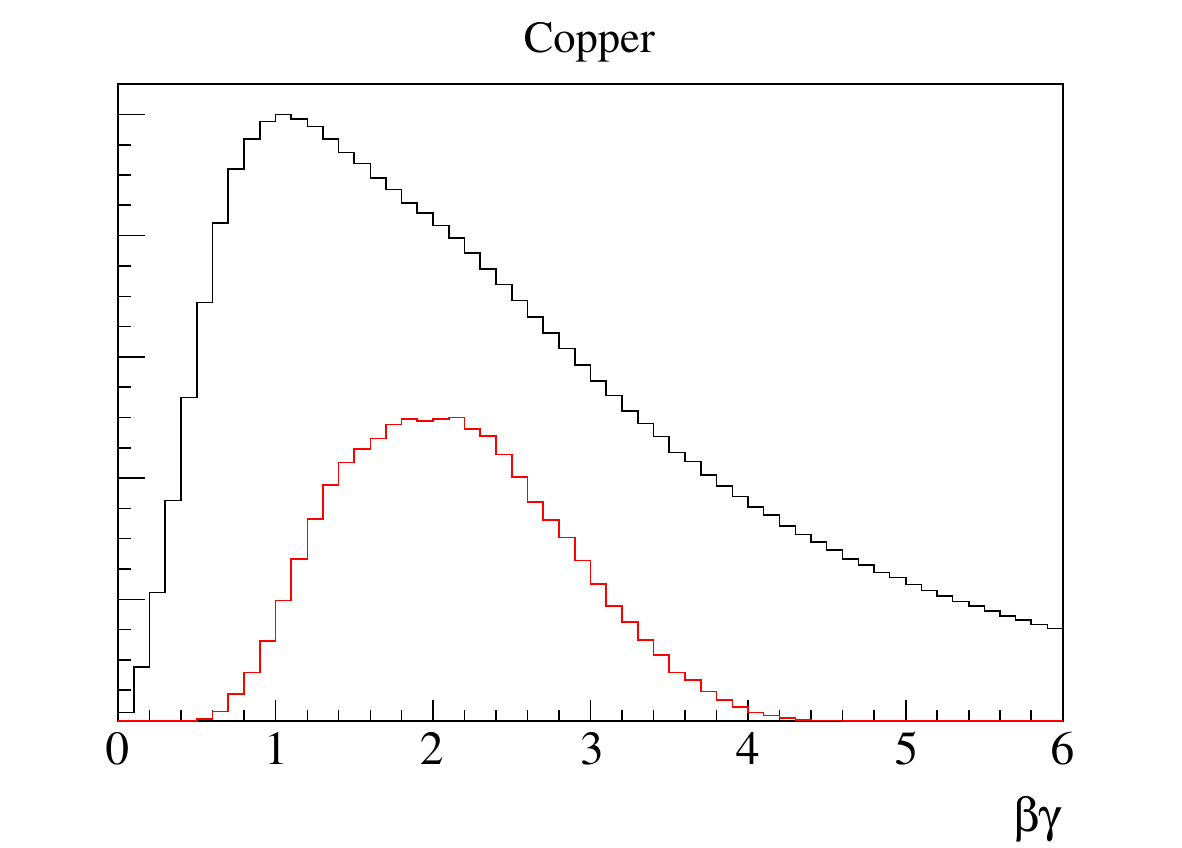}}
  \subfigure[$p_T$, Cu]{
  \includegraphics[width=4.5cm,keepaspectratio]{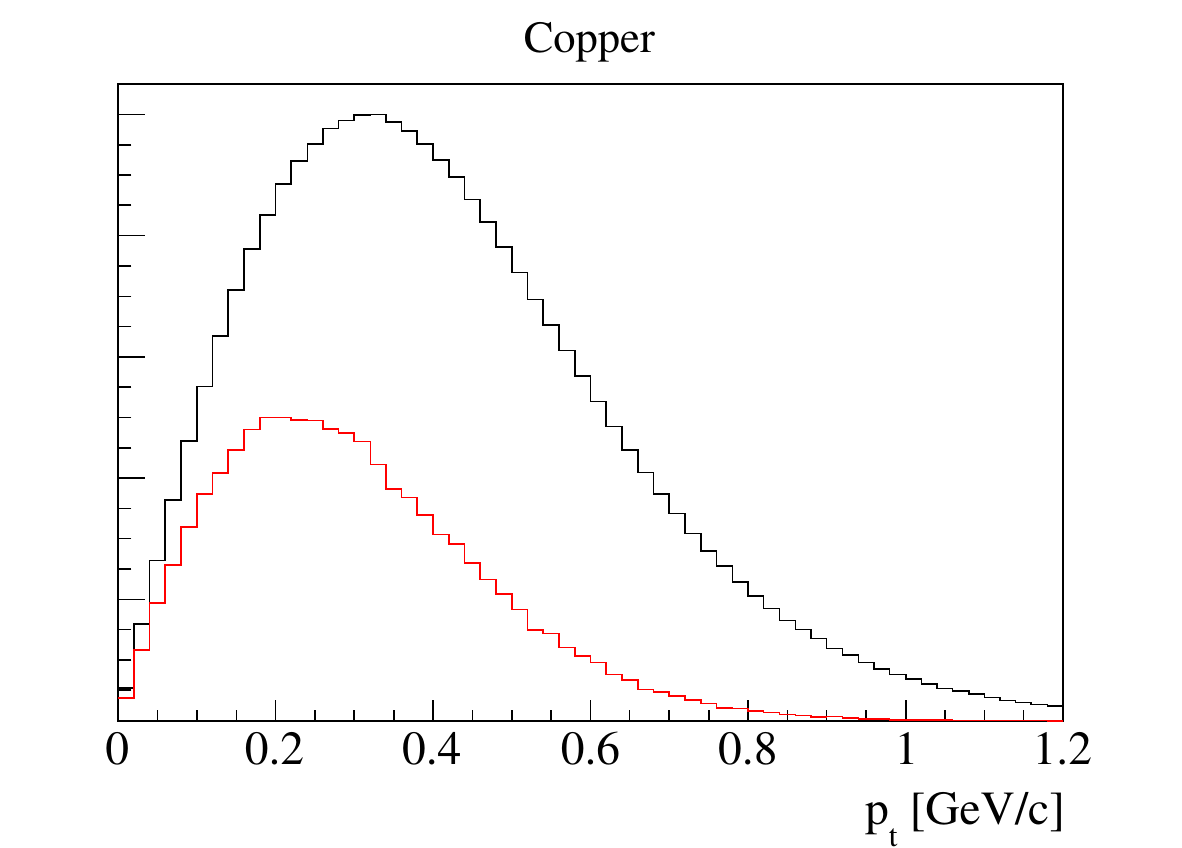}}
  \subfigure[Rapidity, Cu]{
  \includegraphics[width=4.5cm,keepaspectratio]{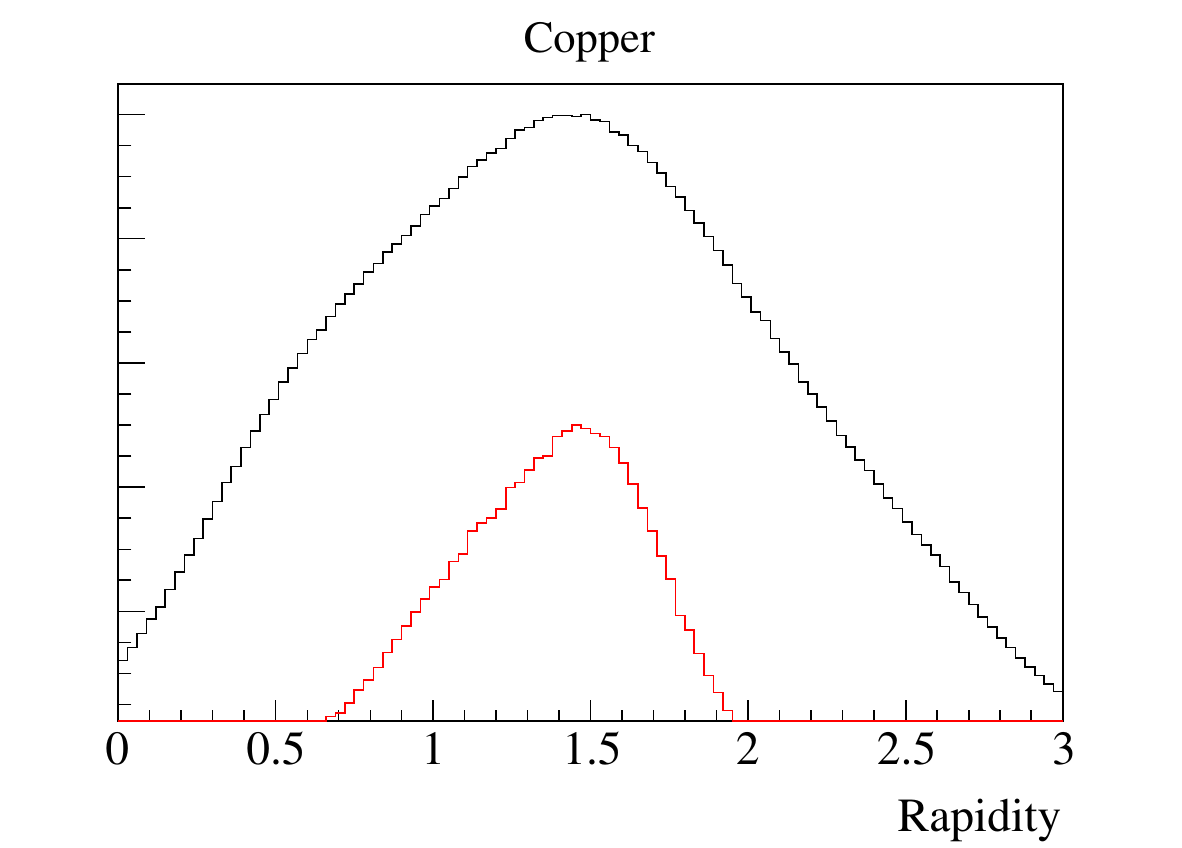}}

  \caption{
     Assumed kinematic distributions of $\omega$ mesons.
     Black lines indicate the distributions at generation points,  while red lines depict those after acceptance cuts.
     The red lines are scaled arbitrarily for visibility.
  }
  \label{fig:beta_gamma_model}
 \end{center}
\end{figure}

In the model calculations, 
the flight paths of the mesons were traced stepwise, while conserving
their initial three momentum. Each step size was 0.1 fm.
The decay probability of a meson in a given step depends on its
momentum and its density-dependent width and mass
at the middle point of the step. In other words,
the decay probability of the meson depends on the  $\beta\gamma c \tau$ and 
the flight length of the meson.
When a meson decays during a step, the decayed mass
was determined on the basis of a mass shape, where the pole
and width correspond to the middle-point mass
and width used above, respectively.
The Breit--Wigner-type probability distribution was used for the mass shape, and details are described later.
Mesons that survive a flight of 1000 steps, that is, 100 fm, decay regardless.
The momentum distributions of the mesons are generated using the code JAM~\cite{JAM}, which
approximately reproduces the observed momentum distribution of $\omega$
within the detector acceptances~\cite{bib:tabaru}. The assumed momentum
distribution is depicted in Fig.~\ref{fig:beta_gamma_model}.

As described in \cite{bib:tabaru}, JAM overestimated the production cross sections and the
$\alpha$ parameters for $\omega$ and $\phi$ compared to our measurements. 
However, the
significant difference between $\omega$ and $\phi$ was reproduced qualitatively.
It suggests that the measured difference of $\alpha$
is due to the different production mechanisms in JAM, 
namely, the secondary
processes are dominant for $\phi$, while the primary
processes are dominant for $\omega$.
On the production points of mesons, we did not use JAM because
they distributed even outside the target nuclear radius,
because the nucleons are also distributed in time evolution
due to the lack of the bounding force.
Thus, we did not use the distribution by JAM directly, 
but used a simpler model:
a static Woods-Saxon distribution for the density of the production and decay points
of mesons.
To reflect the possible different production mechanisms to explain
the different $\alpha$ parameter, it is assumed that
the production points of the $\rho$ and $\omega$ mesons are
uniformly distributed on the surface of the incident-side hemisphere
at the half-density of the target nucleus~\cite{bib:naruki_prl}, and for $\phi$, 
distributed in the entire volume of the target nucleus, proportionally to the density~\cite{bib:muto_prl}.

For the in-medium spectral shape, we adopted the following two formulas for the pole mass and width of the meson.
The pole mass was assumed to depend linearly on the density, $\rho$, at the decay point:
\begin{eqnarray}
   M(\rho) = \left(1-k_1\frac{\rho}{\rho_0}\right)M(0)
   \label{eqn:k1_definition}
\end{eqnarray}
The in-medium total decay width was assumed to follow the relation:
\begin{eqnarray}
   \Gamma_\mathrm{tot}(\rho) = \left( 1+k_2\frac{\rho}{\rho_0}\right) \Gamma_\mathrm{tot}(0),
   \label{eqn:k2_definition}
\end{eqnarray}
where $\rho$ denotes the nuclear density, and $\rho_0$ represents the normal nuclear matter density.
Here, we assumed two extreme cases for the branching ratio,
the branching ratio is constant or the partial decay width is constant in the medium,
i.e., $\Gamma_{ee}(\rho)/\Gamma_\mathrm{tot}(\rho) = \Gamma_{ee}(0)/\Gamma_\mathrm{tot}(0)$ or $\Gamma_{ee}(\rho) = \Gamma_{ee}(0)$,
owing to the lack of clear experimental observations or 
theoretical predictions regarding the partial width of $e^+ e^-$ decays in the medium.
Additionally, we assumed constant values for the modification parameters, $k_1$ and $k_2$, in
each $\beta\gamma$ region. 


The Woods--Saxon function was adopted to model the distribution of the nuclear density, $\rho$,
\begin{eqnarray}
   \rho(r) = \displaystyle{\frac{\rho_0N}{1+\exp{\left((r-R)/\tau\right)}}},
\end{eqnarray}
where $\rho_0 = 0.17$~fm$^{-3}$, $\tau = 0.57~(0.50)$~fm, $R = 2.29~(4.08)$~fm, and 
$N = 0.9~(1.2)$ were used for the C(Cu) target.
The values of $R$ and $\tau$ were obtained from \cite{bib:woods_saxon}.
The normalization factor, $N$, was determined to equalize the volume integral of the distribution and the mass number.

On the mass distributions of mesons, we adopted three cases:
\begin{itemize}

\item Case (i):
Non-relativistic Breit--Wigner formula (nBW) as
\begin{eqnarray}
   \mathrm{nBW}(m) = \frac{\Gtot}{2\pi}\frac{1}{(m-m_0)^2+\Gtot^2/4},                      \label{form-nBW}             
\end{eqnarray}
where $m$ denotes the invariant mass, $m_0$ represents the pole mass, $\Gtot$ indicates the
total decay width of the meson.

\item Case (ii):
The asymmetric mass shape nBW/$m^3$, which was used in the analysis of the CLAS experiment~\cite{bib:clas_g7}.
Left side of the peak is higher than the right side, which could
reduce the amount of possible excess in the left-side.

\item Case (iii):
Asymmetric mass shape nBW/$m^3$ and the constant partial decay width of the dielectron channel were assumed.

\end{itemize}

\begin{figure}[htbp]
 \begin{center}
  \includegraphics[width=15cm]{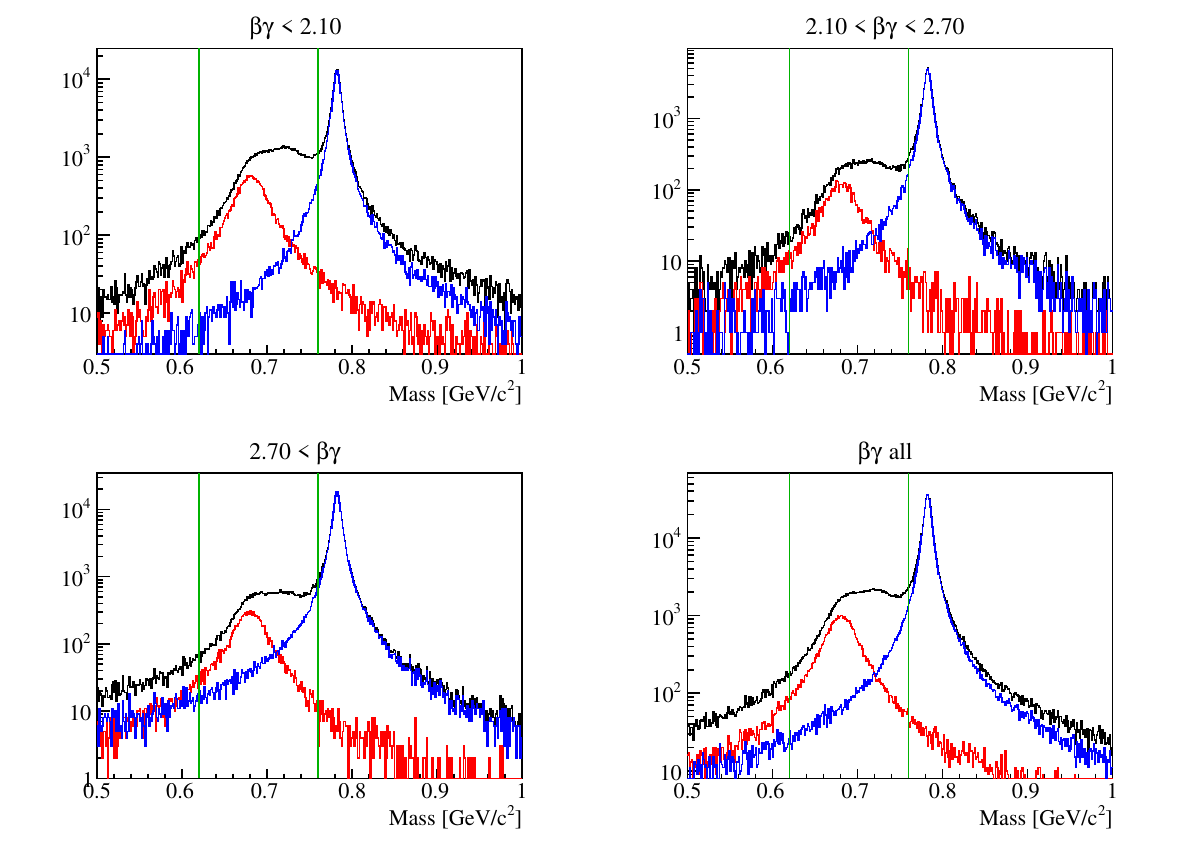}
  \caption{Example of a model mass distribution of $\omega$ 
  mesons with the non-relativistic Breit--Wigner shape for the Cu target (without the acceptance cut and mass smearing).
  The black lines depict the mass distributions corresponding to $k_1 = 0.12$ and $k_2 = 3.0$.
  The blue lines depict the mass distributions corresponding to the case exclusively selecting the decays at $\rho/\rho_0 < 0.1$, that is, outside the nucleus. The red lines depict the mass distributions considering decays at $\rho/\rho_0 > 1.0$.
  Vertical green lines indicate the region excluded from the fit.}
  \label{fig:model_k2_3_generated_mass}
 \end{center}
\end{figure}

In each case, the same shape was used for $\rho$ and $\omega$.
Detailed plots and discussion on the various possible shapes
are shown in Appendix~\ref{section-App-massform}.
An example of the generated mass distribution for the Case (i)
is depicted in Fig.~\ref{fig:model_k2_3_generated_mass}.

Table~\ref{tab:decay_fraction} shows the fractions of in-medium decays of $\rho$ and $\omega$ mesons for $k_2=0$ and $k_2=1$.
The results are before the correction for the detector acceptance.
The decay probability of $\rho$ mesons in the nucleus is larger than that of $\omega$ mesons.
A large mass modification effect for $\rho$ mesons is expected.
Here, the definition of ``in-medium'' is that the density at the decay point is larger than half that at the center of the nucleus.
\begin{table}[hbpt]
   \begin{center}
   \caption{Fractions of in-medium decays ($\rho > 0.5\rho_0N$) of $\rho$ and $\omega$ mesons for each target and each $\beta\gamma$ region.
   The branching ratio is assumed to be constant in case (i) and (ii), and the partial decay width is constant in case (iii).
   The detector acceptance is not taken into account.
   }
\begin{tabular}{ccccc}
   \hline
   & & $\beta\gamma < 2.1$ & $2.1 < \beta\gamma < 2.7$ & $2.7 < \beta\gamma$ \\ \hline\hline
   &$\rho$~(C)   & 67\% & 50\% & 33\% \\
   case (i), (ii), (iii) & $\rho$~(Cu)  & 76\% & 65\% & 48\% \\
   $k_2=0$ & $\omega$~(C) & 8.6\% & 4.3\% & 2.3\% \\
   &$\omega$~(Cu)& 14\% & 7.3\% & 4.1\% \\ \hline
   &$\rho$~(C)    & 76\% & 63\% & 45\% \\
   case (i), (ii) &$\rho$~(Cu) & 83\% & 79\% & 65\% \\
   $k_2=1$ &$\omega$~(C) & 13\% & 6.5\% & 3.7\% \\
   &$\omega$~(Cu) & 24\% & 13\% & 7.4\% \\ \hline
   &$\rho$~(C)   & 72\% & 56\% & 36\% \\
   case (iii) & $\rho$~(Cu)  & 79\% & 72\% & 53\% \\
   $k_2=1$ & $\omega$~(C) & 8.7\% & 4.2\% & 2.4\% \\
   &$\omega$~(Cu)& 15\% & 7.3\% & 4.2\% \\ \hline
\end{tabular}
  \label{tab:decay_fraction}
   \end{center}
\end{table}
\subsection{Determination of the modification parameters\label{sec:modification_parameter}}
As described in the previous section, the modified shapes 
of the mass spectra were determined using
the $k_1$ and $k_2$ parameters. 
To identify the optimal $k_1$ and $k_2$ parameters that best reproduces the data, we evaluated the $\chi^2$ obtained from fitting 
as a function of the $k_1$ and $k_2$ parameters.
The $k_1$ and $k_2$ parameters were changed manually from 0.04 to 0.18 with a step size of 0.02 and from 0.0 to 9.0 with a step size of 1.0, respectively, 
and the mass spectra were fitted against the modified mass spectra obtained from the simulation using each $(k_1, k_2)$ value.

The data were categorized into three $\beta\gamma$ regions and analyzed as described in Sec.~\ref{sec:fit_result}.
From the fit results in Sec.~\ref{sec:fit_result}, contributions other than $\rho\rightarrow e^+e^-$ and $\omega\rightarrow e^+e^-$ (including $\omega\rightarrow \pi^0e^+e^-$) were subtracted from the spectra, and the data after subtraction were fitted with simulated shapes.
For each $\beta\gamma$ region,  data for the same target nucleus but different data sets, ---corresponding to different target locations and data-collection periods--- were summed before fitting.
The C and Cu data were simultaneously fitted and $\chi^2$ value was computed as the sum of the two fittings considering common $(k_1, k_2)$ values.
Here, the fitting parameters included the quantities of $\rho$ and $\omega$ for each target nucleus.

The optimal $k_1$ and $k_2$ parameters as well as their error values were obtained 
using the $\chi^2$ contour plot as a function of $k_1$ and $k_2$.
The minimum $\chi^2$ point was obtained by fitting the nine points surrounding the minimum using a parabolic surface.
Since the local minimum is not necessarily a single point within the parameter range, we also determined the areas that could be potentially excluded at the 99\% confidence level based on the $\varDelta\chi^2$ values derived from the minimum points.
The parameters in the fitting against the mass distributions were $k_1$, $k_2$, 
the numbers of $\omega$ mesons from the C and Cu targets, and 
the numbers of $\rho$ mesons from the C and Cu targets.
Consequently, the number of parameters in the fitting was six. 
In the six parameters fitting,
$\varDelta\chi^2$ values of 16.81 
corresponded to the 99\% confidence level. 

\begin{figure}[htbp]
 \begin{center}
  \includegraphics[width=15.5cm]{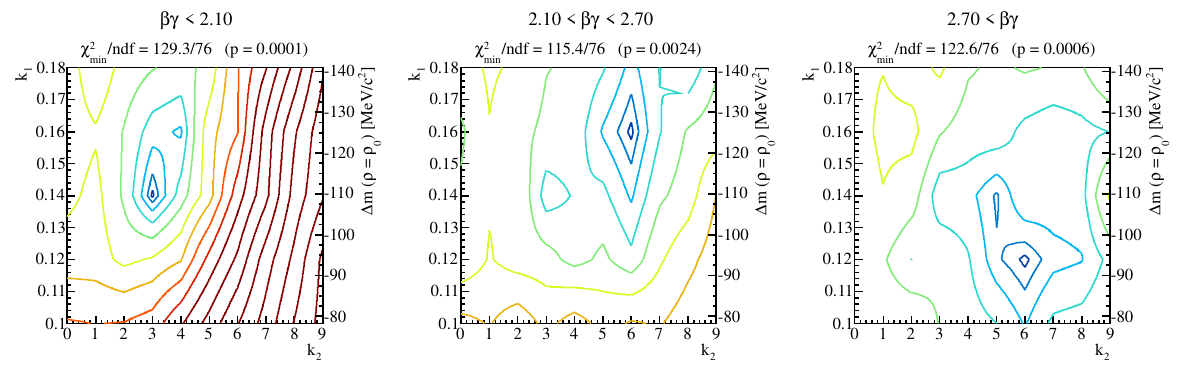}
    \caption{Contour lines of $\chi^2$ values as functions of the parameters $k_1$ and $k_2$ in case (i).
    Each line corresponds to $\varDelta\chi^2 (=\chi^2-\chi^2_\mathrm{min}) = n^2, n = 1, 2, 3, \cdots$.
    Local minimum points surround $(k_1, k_2)=(0.12\text{--}0.16, 3.0\text{--}6.0)$.
    }
  \label{fig:chi2_vs_k1k2_case1}
 \end{center}
\end{figure}

\begin{figure}[htbp]
 \begin{center}
  \includegraphics[width=15.5cm]{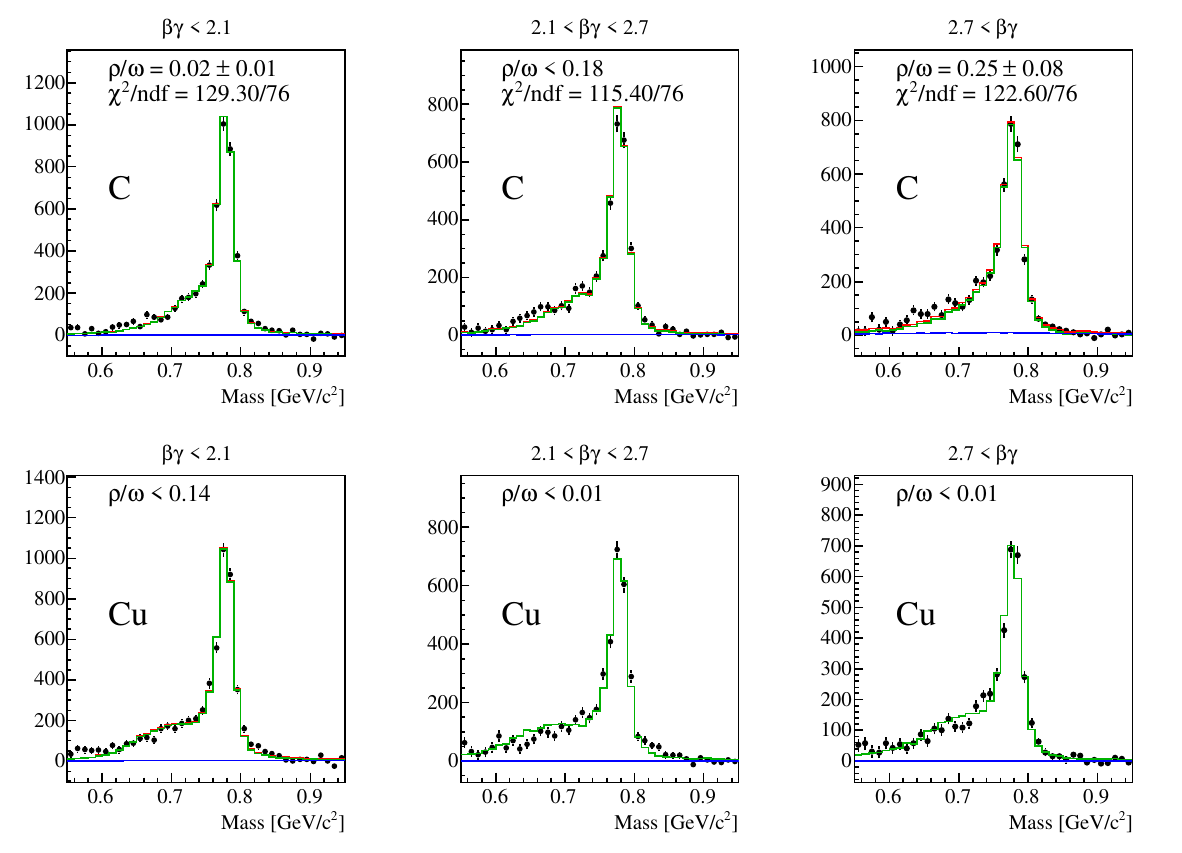}
    \caption{Fitting results near the $\chi^2$ minimum points in case (i). 
    The combinatorial background has been subtracted.
    The points correspond to $(k_1, k_2) = (0.14, 3.0)$ for $\beta\gamma<2.1$, $(0.14, 3.0)$ for $2.1<\beta\gamma<2.7$, and $(0.12, 6.0)$ for $2.7<\beta\gamma$.
    The values of the $(k_1, k_2)$ parameters are common for C (upper) and Cu (lower) targets.
    Red lines indicate the fitting results obtained as the sum of $\omega\rightarrow e^+e^-$ (green) and $\rho\rightarrow e^+e^-$ (blue).
    }
  \label{fig:model_fit_result_case1}
 \end{center}
\end{figure}

The results for case (i) are depicted in Fig.~\ref{fig:chi2_vs_k1k2_case1}.
The model assumed in the case (i) was the same as the one utilized in our previous study, except for the mass shape~\cite{bib:naruki_prl},
in which the $k_1$ parameter had a finite value  
even in the absence of mass broadening ($k_2 \sim 0$). 
Whereas the current analysis yields different fitting results, since both the data analysis and 
simulation of mass spectra have been updated. 
In the updated analysis, at minimum $\chi^2$ values, the value of $k_1$ parameter is greater compared to that in the previous study
with the finite value of $k_2$ parameter, indicating a larger mass shift and finite width broadening. 
Figure~\ref{fig:model_fit_result_case1} depicts the fitted mass spectra near the minimum values of the $k_1$ and $k_2$ 
parameters. 
 These spectra demonstrate that the excess in the mass spectra was reproduced solely by $\omega$ meson modifications. 
The ratio of the number of generated $\rho$ mesons to that of $\omega$ mesons is $\rho/\omega \sim 0$.
Importantly, the obtained results contradict the value 
of the $\rho/\omega$ ratio observed in $pp$ reactions;
previous observations in $pp$ reactions conducted at 
an incident momentum of 12~GeV/$c$ yielded $\rho/\omega = 1.0 \pm 0.2$, as evidenced in Ref.~\cite{bib:pp_12GeV}.
These inconsistencies suggest that the simulated model must be further refined to accurately reproduce the data. 

\begin{figure}[htbp]
 \begin{center}
  \includegraphics[width=15.5cm]{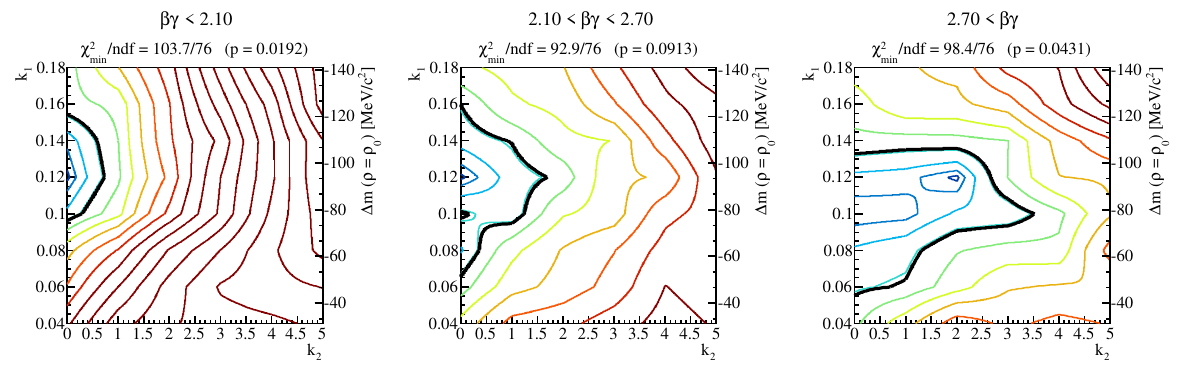}
    \caption{Contour lines of $\chi^2$ values as functions of $k_1$ and $k_2$ for case (ii).
    Each line corresponds to $\varDelta\chi^2 (=\chi^2-\chi^2_\mathrm{min}) = n^2, n = 1, 2, 3, \cdots$.
    Black lines depict the 99\% confidence interval.
    }
  \label{fig:chi2_vs_k1k2_case2}
 \end{center}
\end{figure}

\begin{figure}[htbp]
 \begin{center}
  \includegraphics[width=15.5cm]{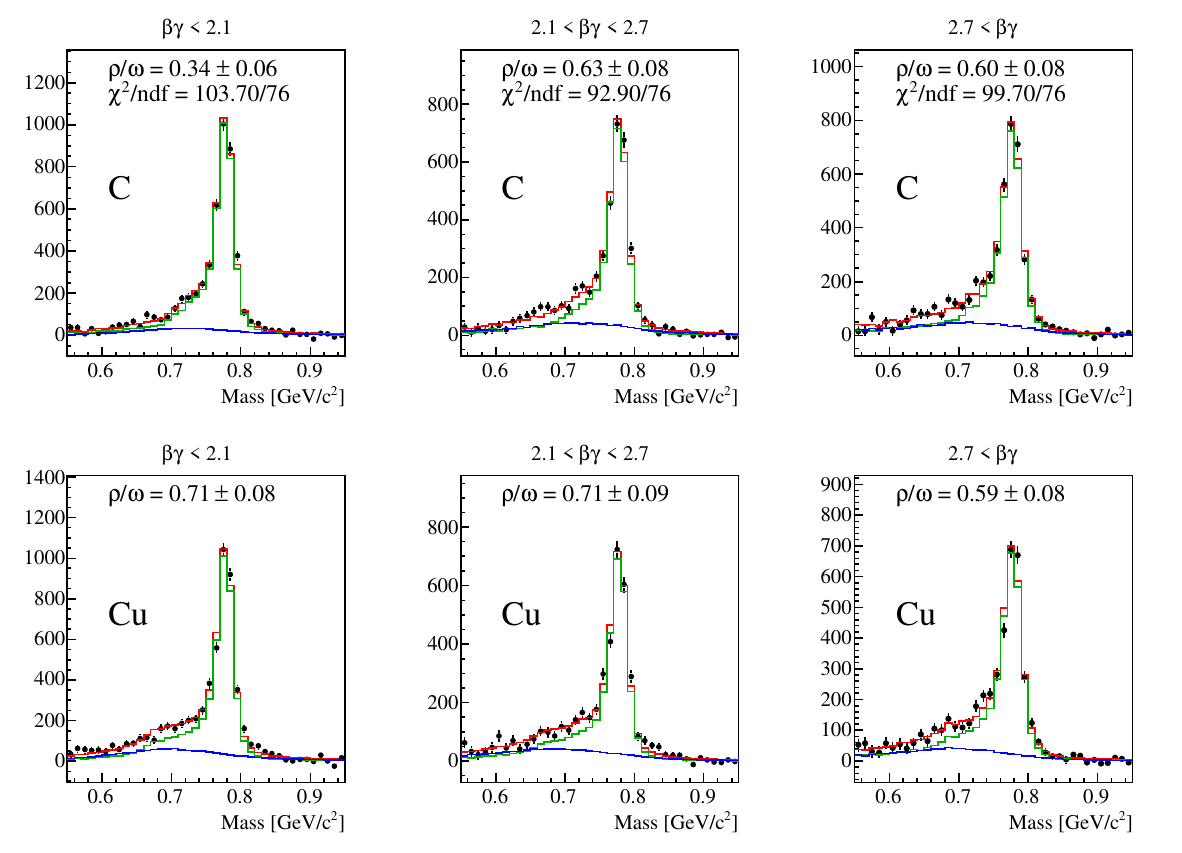}
    \caption{Fitting results near minimum $\chi^2$ values in case (ii). 
    The combinatorial background has been subtracted.
    The depicted points correspond to $(k_1, k_2) = (0.12, 0.0)$ for $\beta\gamma<2.1$, $(0.12, 0.0)$ for $2.1<\beta\gamma<2.7$, and $(0.10, 0.0)$ for $2.7<\beta\gamma$.
    The values of the $(k_1, k_2)$ parameters are common for C (upper) and Cu (lower) target.
    Red lines depict the fitting results obtained as the sum of $\omega\rightarrow e^+e^-$ (green) and $\rho\rightarrow e^+e^-$ (blue).
    }
  \label{fig:model_fit_result_case2}
 \end{center}
\end{figure}

\begin{figure}[htbp]
 \begin{center}
  \includegraphics[width=15.5cm]{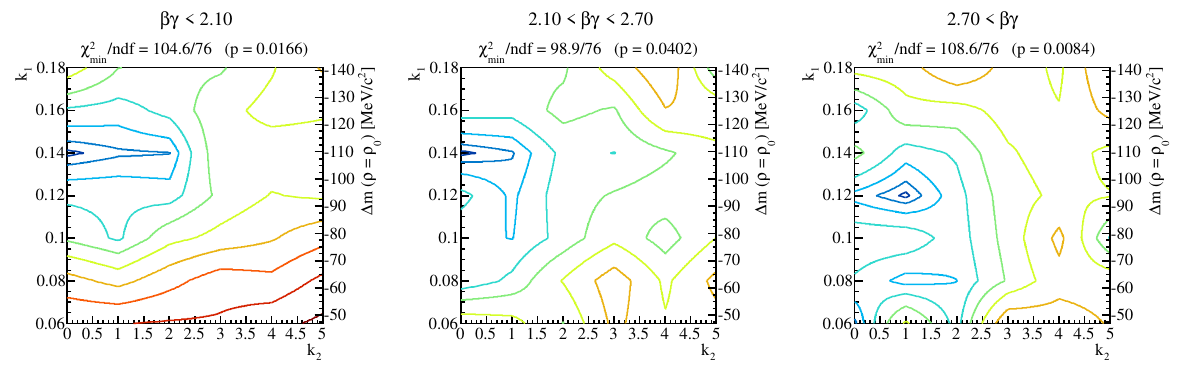}
    \caption{Contour lines of $\chi^2$ values as functions of $k_1$ and $k_2$ for case (iii).
    Each line corresponds to $\varDelta\chi^2 (=\chi^2-\chi^2_\mathrm{min}) = n^2, n = 1, 2, 3, \cdots$.
    }
  \label{fig:chi2_vs_k1k2_case6}
 \end{center}
\end{figure}

\begin{figure}[htbp]
 \begin{center}
  \includegraphics[width=15.5cm]{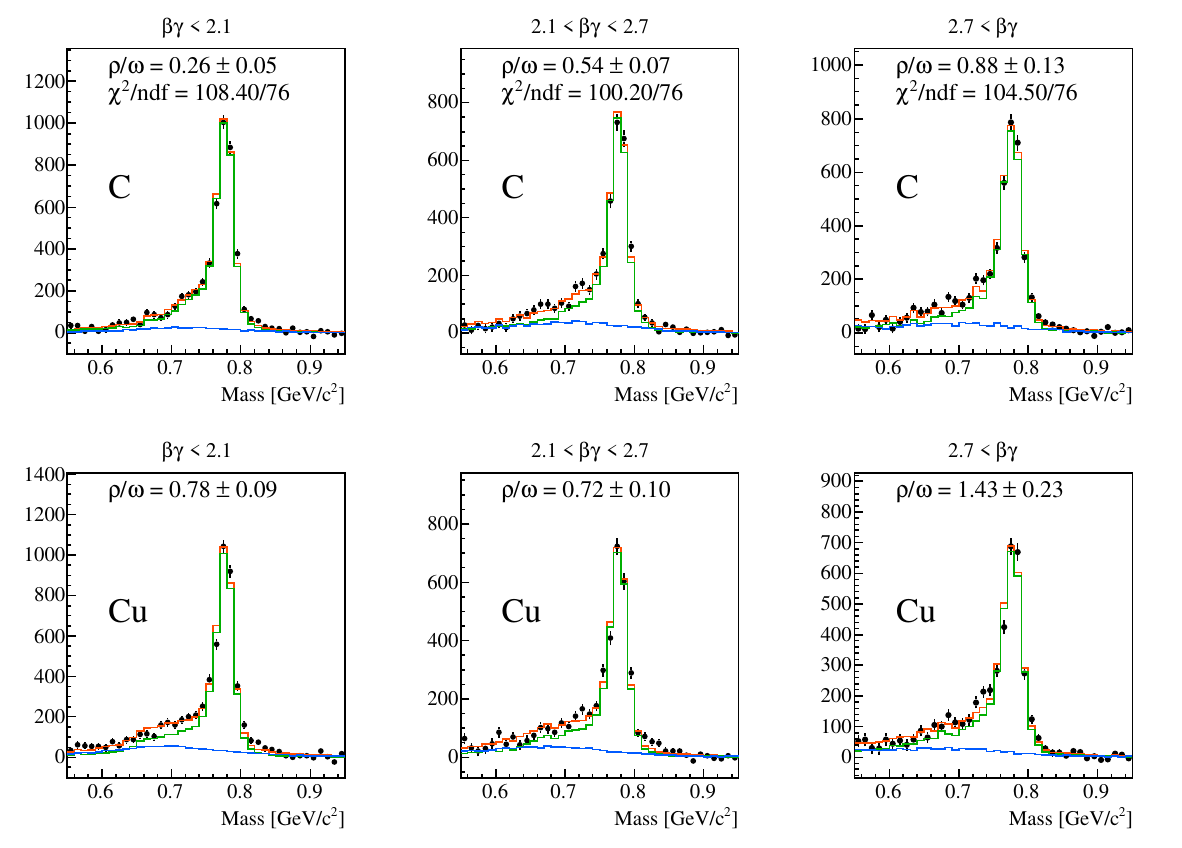}
    \caption{Fitting results near minimum $\chi^2$ values in case (iii). 
    The combinatorial background has been subtracted.
    The depicted points correspond to $(k_1, k_2) = (0.14, 0.0)$ for $\beta\gamma<2.1$, $(0.14, 0.0)$ for $2.1<\beta\gamma<2.7$, and $(0.12, 1.0)$ for $2.7<\beta\gamma$.
    The values of the $(k_1, k_2)$ parameters are common for C (upper) and Cu (lower) target.
    Red lines depict the fitting results obtained as the sum of $\omega\rightarrow e^+e^-$ (green) and $\rho\rightarrow e^+e^-$ (blue).
    }
  \label{fig:model_fit_result_case6}
 \end{center}
\end{figure}

The absence of the broadened $\rho$ component from the fitting results of case (i) 
indicates that the high-mass tail of the largely broadened $\rho$ meson no longer matches the data.
When the mass width of the $\rho$ meson is increased to account for the excess observed on the 
low-mass side of $\rho$/$\omega$ mesons, the symmetric mass distribution 
results in a large tail in the high-mass region as well, failing to reproduce the data. 
This highlights the necessity for incorporating an asymmetric mass shape.  
In case (ii), we assumed the mass shape distribution 
modeled by the function nBW/$m^3$, and the corresponding results are depicted in Fig.~\ref{fig:chi2_vs_k1k2_case2}.
The optimal values of $k_1$ in this case ranged from 0.10--0.12, while $k_2$ values were zero across all $\beta\gamma$ bins.
The fitted mass spectra near the minimum $k_1$ and $k_2$ values in case (ii) are depicted in Fig.~\ref{fig:model_fit_result_case2}. 
As depicted, the $\rho/\omega$ ratios exhibit finite nonzero values, which largely agree with previous experimental results ($\rho/\omega = 1.0 \pm 0.2$), albeit slightly deviating by $2\sigma$--$3\sigma$ at the maximum.
These results are also consistent with the previous findings~\cite{bib:naruki_prl}.
In case (iii), where the asymmetric mass shape and the constant partial decay width were assumed,
the results of the optimal $k_1$ and $k_2$ were almost the same as case (ii) as shown in Fig.~\ref{fig:chi2_vs_k1k2_case6} and Fig.~\ref{fig:model_fit_result_case6}.
With respect to the $\chi^2$ values, the results of case (ii) are slightly better than those of case (iii).


%

The obtained values of the $k_1$ and $k_2$ parameters near the $\chi^2$ minimum are listed in Table~\ref{tab:summary_model_fit1}, 
while the calculated $\rho/\omega$ ratios and $\chi^2$ values obtained near the minimum $k_1$ and $k_2$ values are
summarized in Table~\ref{tab:summary_model_fit2}. 
\begin{table}[hbpt]
   \caption{Obtained values of $k_1, k_2$ near the $\chi^2$ minimum.
   For the $k_2$ parameter in case (ii), the obtained values are consistently zero with the presented numbers indicating the upper limits of the 99\% confidence level. 
   The shown error values represent statistical errors.}
   \label{tab:summary_model_fit1}
   \begin{center}
      \begin{tabular}{cccc}
         \hline
         & & $k_1$ & $k_2$ \\ \hline\hline
                    & $\beta\gamma < 2.1$       & $0.15^{+0.02}_{-0.02}$ & $3.3^{+1.0}_{-1.0}$  \\
         case (i)   & $2.1 < \beta\gamma < 2.7$ & $0.16^{+0.02}_{-0.04}$ & $5.9^{+2.6}_{-3.1}$ \\
                    & $2.7 < \beta\gamma$       & $0.12^{+0.05}_{-0.02}$ & $5.5^{+3.5}_{-3.6}$ \\
         \hline
                    & $\beta\gamma < 2.1$       & $0.12^{+0.03}_{-0.03}$ & $<0.7$    \\
         case (ii)  & $2.1 < \beta\gamma < 2.7$ & $0.12^{+0.04}_{-0.06}$ & $<1.7$    \\
                    & $2.7 < \beta\gamma$       & $0.10^{+0.03}_{-0.05}$ & $<3.5$    \\
         \hline
                    & $\beta\gamma < 2.1$       & $0.14^{+0.02}_{-0.02}$ & $<2.3$    \\
         case (iii) & $2.1 < \beta\gamma < 2.7$ & $0.14^{+0.02}_{-0.06}$ & $<1.7$    \\
                    & $2.7 < \beta\gamma$       & $0.12^{+0.02}_{-0.05}$ & $<2.3$    \\
         \hline
         \hline
      \end{tabular}
   \end{center}
\end{table}
\begin{table}[hbpt]
   \caption{
   Obtained $\rho/\omega$ ratios and $\chi^2$ values.
   The values of the $(k_1, k_2)$ parameters utilized in the evaluations are listed as well. 
   The listed error values are statistical errors.}
   \label{tab:summary_model_fit2}
   \begin{center}
      \begin{tabular}{cccccc}
         \hline
         & &  $\rho/\omega$~(C) & $\rho/\omega$~(Cu) & $\chi^2_\mathrm{min}$/ndf 
         & used $(k_1, k_2)$\\ \hline\hline
                    & $\beta\gamma < 2.1$       &  $0.02 \pm 0.01$ & $<0.14$ & $129.3/76$ & $(0.14, 3.0)$\\
         case (i)   & $2.1 < \beta\gamma < 2.7$ &  $<0.18$         & $<0.01$ & $115.4/76$ &     $(0.14, 3.0)$\\
                    & $2.7 < \beta\gamma$       &  $0.25 \pm 0.08$ & $<0.01$ & $122.6/76$ & 
                    $(0.12, 6.0)$\\
        \hline
                    & $\beta\gamma < 2.1$       &   $0.34 \pm 0.06$  &  $0.71 \pm 0.08$  &  $103.7/76$ & $(0.12, 0.0)$\\
         case (ii)  & $2.1 < \beta\gamma < 2.7$ &  $0.63 \pm 0.08$  &  $0.71 \pm 0.09$  &  $ 92.9/76$ & $(0.12, 0.0)$\\
                    & $2.7 < \beta\gamma$       &   $0.60 \pm 0.08$  &  $0.59 \pm 0.08$  &  $ 99.7/76$ & $(0.10, 0.0)$\\
         \hline
                    & $\beta\gamma < 2.1$       &   $0.26 \pm 0.05$  &  $0.78 \pm 0.09$  &  $108.4/76$ & $(0.14, 0.0)$\\
         case (iii) & $2.1 < \beta\gamma < 2.7$ &  $0.54 \pm 0.07$  &  $0.72 \pm 0.10$  &  $100.2/76$ & $(0.14, 0.0)$\\
                    & $2.7 < \beta\gamma$       &   $0.88 \pm 0.13$  &  $1.43 \pm 0.23$  &  $ 104.5/76$ & $(0.12, 1.0)$\\
         \hline
         \hline
      \end{tabular}
   \end{center}
\end{table}

\subsubsection{Systematic errors of the modification parameters}
\label{Systematic errors of the modification parameters}
For case (ii), the systematic errors of the modification parameters $k_1$ and $k_2$ were evaluated following 
the methodology adopted for the excess ratio as described in Sec.~\ref{sec:systematic_errors}.
The fitting and evaluation of minimum $\chi^2$ values were 
repeated with variations in the following conditions (More details are described in Sec.~\ref{sec:systematic_errors}) :
\begin{enumerate}
   \renewcommand{\labelenumi}{\Alph{enumi}.}
   \item Fit region
   \item Bin width of mass spectra
   \item Mass scale
   \item Mass resolution in the simulation
   \item Event mixing method
   \item The evaluation method for the electron identification efficiency
\end{enumerate}

\begin{figure}[htbp]
 \begin{center}
  \includegraphics[width=10cm]{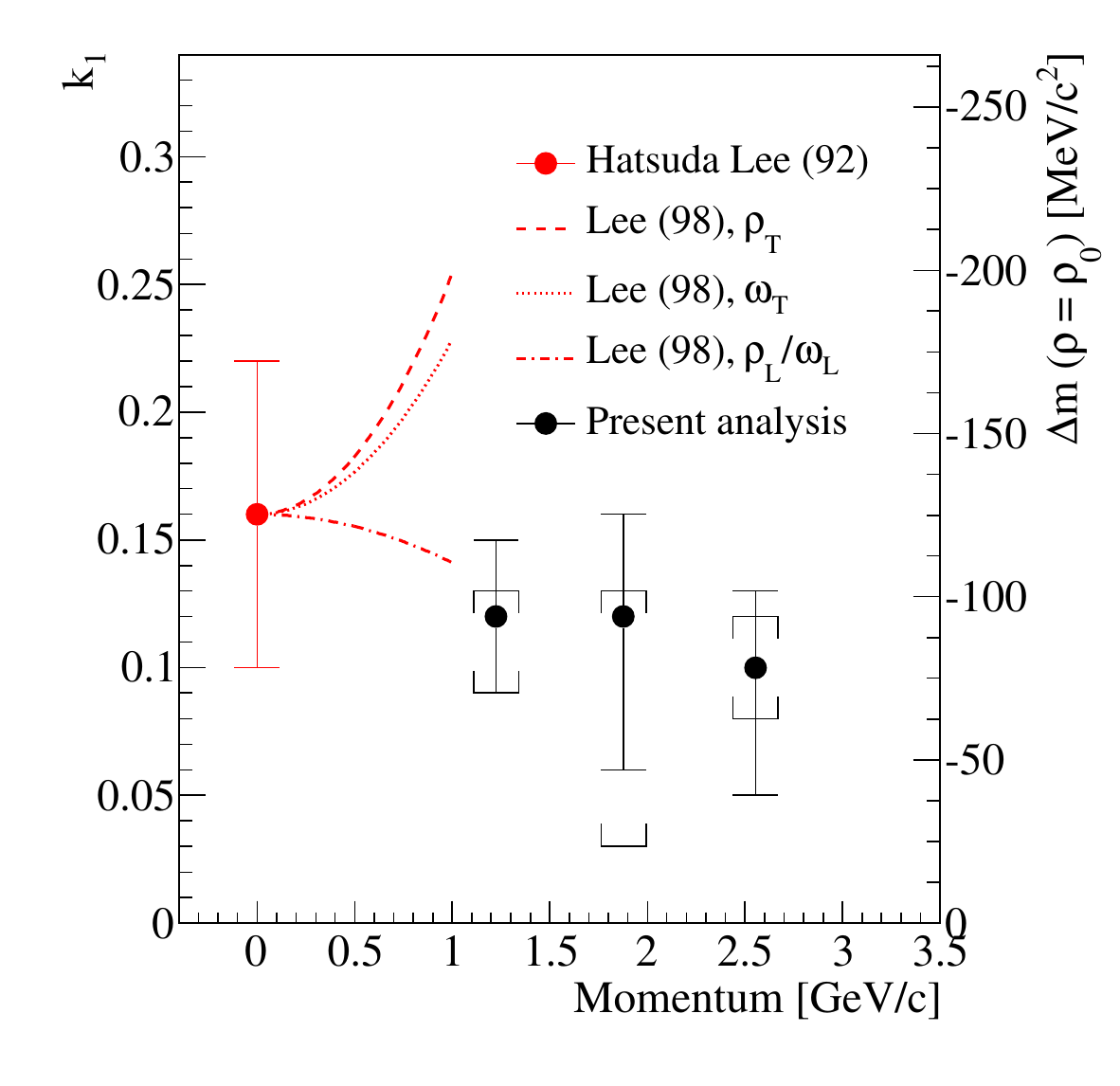}
  \caption{Optimal values of the $k_1$ parameter as functions of the meson momentum.
    Error bars indicate statistical errors, while square brackets denote systematic errors.
    The red point corresponds to the value obtained theoretically by Hatsuda and Lee~\cite{bib:HL92,bib:HL95}.
     Since the calculation assumes mesons to be at rest relative to the nucleus, the momentum at that point is zero.
    The red lines show the momentum dependence of the shift~\cite{bib:shlee98} in terms of $k_1$.
    }
  \label{fig:k1_vs_momentum}
 \end{center}
\end{figure}

The difference between each parameter value and its nominal value was subsequently evaluated, and the squared sum of the maximum and minimum values within each category was used to determine the systematic error.
The corresponding results are summarized in Table~\ref{tab:k1k2_systematic_errors_case2}.
For $\beta\gamma < 2.1$, $2.1 < \beta\gamma < 2.7$, and $2.7 < \beta\gamma$, we obtained 
$k_1 = 0.12 ^{+0.03}_{-0.03}\mathrm{(stat.)} ^{+0.01}_{-0.03}\mathrm{(sys.)}$,
   $k_1 = 0.12 ^{+0.04}_{-0.06}\mathrm{(stat.)} ^{+0.01}_{-0.09}\mathrm{(sys.)}$, and
   $k_1 = 0.10 ^{+0.03}_{-0.05}\mathrm{(stat.)} ^{+0.02}_{-0.02}\mathrm{(sys.)}$, respectively.
The obtained results are depicted in Fig.~\ref{fig:k1_vs_momentum}.
\begin{table}[hbpt]
   \caption{Errors in $k_1$ values under different fitting conditions.
   The squares of the maximum values in each category were 
   summed to obtain systematic errors.}
   
   \label{tab:k1k2_systematic_errors_case2}
   \begin{center}
      \begin{tabular}{c cc cc cc}
         \hline
         & \multicolumn{2}{c}{$\beta\gamma < 2.1$} & \multicolumn{2}{c}{$2.1 < \beta\gamma < 2.7$} & \multicolumn{2}{c}{$2.7 < \beta\gamma$} \\
         \hline\hline
         A1 & $+ 0     $ & $- 0.0013$    & $+ 0.0009$ & $- 0     $ & $+ 0.0086$ & $- 0     $   \\
         A2 & $+ 0.0004$ & $- 0     $    & $+ 0.0001$ & $- 0     $ & $+ 0.0002$ & $- 0     $   \\
         B1 & $+ 0.0041$ & $- 0     $    & $+ 0     $ & $- 0.0414$ & $+ 0     $ & $- 0.0078$   \\
         B2 & $+ 0.0049$ & $- 0     $    & $+ 0.0002$ & $- 0     $ & $+ 0     $ & $- 0.0002$   \\
         C1 & $+ 0     $ & $- 0.0017$    & $+ 0     $ & $- 0.0389$ & $+ 0.0040$ & $- 0     $   \\
         C2 & $+ 0     $ & $- 0.0157$    & $+ 0     $ & $- 0.0014$ & $+ 0     $ & $- 0.0048$   \\
         D1 & $+ 0.0008$ & $- 0     $    & $+ 0     $ & $- 0.0430$ & $+ 0     $ & $- 0.0179$   \\
         D2 & $+ 0.0004$ & $- 0     $    & $+ 0     $ & $- 0.0006$ & $+ 0.0119$ & $- 0     $   \\
         E1 & $+ 0.0053$ & $- 0     $    & $+ 0.0013$ & $- 0     $ & $+ 0.0013$ & $- 0     $   \\
         E2 & $+ 0     $ & $- 0.0177$    & $+ 0     $ & $- 0.0475$ & $+ 0     $ & $- 0.0077$   \\
         E3 & $+ 0     $ & $- 0.022 $    & $+ 0     $ & $- 0.0509$ & $+ 0     $ & $- 0.0091$   \\
         F1 & $+ 0.0078$ & $- 0     $    & $+ 0.0024$ & $- 0     $ & $+ 0.0010$ & $- 0     $   \\ \hline\hline
         Total &  $+ 0.01$ & $- 0.03$
               &  $+ 0.01$\footnote{The actual calculation result is 0.0016 but rounded up to the second decimal place to match the number of digits of statistical error.} & $- 0.09$
               &  $+ 0.02$ & $- 0.02$\\
         \hline
      \end{tabular}
   \end{center}
\end{table}

The systematic errors for the upper limit of $k_2$ were also evaluated.
For all fitting conditions, the range of $k_2$ satisfying $\varDelta\chi^2<16.81$ (corresponding to the 99\% confidence level) was determined, and the maximum value was adopted as the upper limit.
The following results were obtained: $k_2 < 0.9$, $k_2 < 2.4$, and $k_2 < 4.5$
for $\beta\gamma < 2.1$, $2.1 < \beta\gamma < 2.7$, and $2.7 < \beta\gamma$.

In this model fitting, the amounts of $\rho$ and $\omega$ mesons are free parameters in each $\beta\gamma$ bin.
Thus, the $\rho/\omega$ ratios are not necessarily the same for each $\beta\gamma$ bin.
Assuming that the momentum distributions of the mesons by JAM are correct, the $\rho/\omega$ ratio should be the same for all $\beta\gamma$ ranges.
Since this can be considered one of the sources of systematic error, we also performed a fitting to make the $\rho/\omega$ ratio common for all $\beta\gamma$ bins.
As a result, the minimum points of $k_1$ and $k_2$ did not change from the nominal case,
and the difference was negligibly small compared to other errors.
Therefore, it was not treated as a systematic error.

\subsection{Comparisons with previous experimental results and theoretical calculations}
\subsubsection{Comparison with the results of our previous E325 experiment analysis}

Several updates  have been implemented compared to our previous analysis~\cite{bib:naruki_prl}:
\begin{itemize}
   \item Optimization of kinematic parameters cut to enhance detector acceptance coverage and improved fiducial selections to maintain consistent acceptance levels across target positions.
   \item Inclusion of internal radiative correction in the mass distribution.
   \item Introduction of an asymmetric mass distribution function, specifically the one used for the analysis of the CLAS experiment with BW$/m^3$ dependence to improve the data modeling accuracy.
   \item Application of a more recent form factor for the Dalitz decay using the event generator PLUTO~\cite{bib:pluto}.
\end{itemize}

Owing to the above updates, the optimal value of $k_1$ changed from $k_1 \sim 0.09$ to $k_1 \sim 0.12$ aligning closely
 with the results of previous studies within acceptable error ranges.
Meanwhile, the values of $k_2$ parameter consistently remain at zero, in agreement with the findings of previous studies. 

\subsubsection{Comparison with the result of the CLAS experiment}
Similar to the CLAS experiment, our data indicate that the BW$/m^3$ function better reproduces the results compared to non-relativistic BW, whose shape is almost the same as that of a constant-width relativistic one.
However, a discrepancy is observed between the results of the two analyses regarding the mass shift. While our findings 
suggest a finite mass shift (approximately 10\% reduction at the normal nuclear density), the CLAS results exhibit non-significant shifts ($|\varDelta m| = 10\pm10$~MeV/c${}^2$)~\cite{bib:clas_g7}.
The origin of this difference remains unclear and must be evaluated in future experiments. 

The CLAS experiment reported an approximately 1.4-fold increase in the widths of $\rho$ mesons for the Fe-Ti target ($\Gamma\sim217.7$~MeV)~\cite{bib:clas_g7}. 
The results of the E325 experiment are also consistent within the current error range, exhibiting only slight differences, as the sensitivity to width broadening measurements remains limited in the E325 experiment.

\begin{figure}[htbp]
 \begin{center}
  \includegraphics[width=15.5cm]{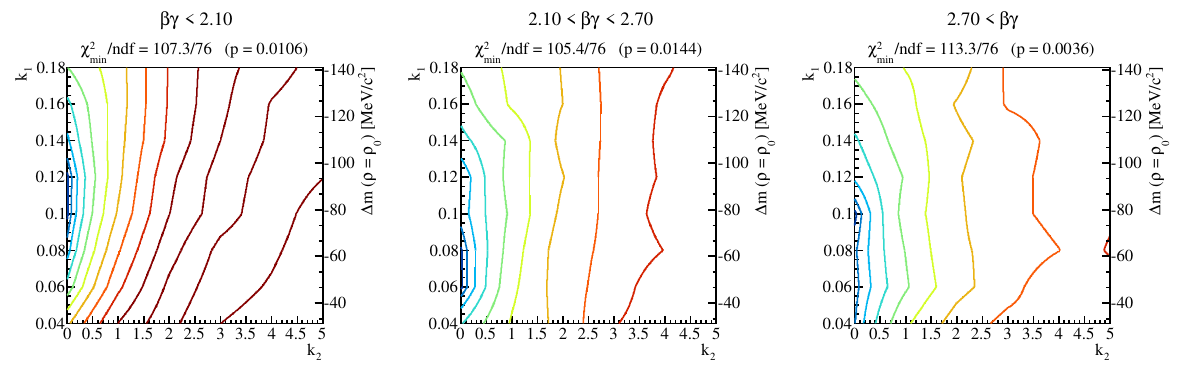}
    \caption{Contour lines of $\chi^2$ values as functions of $k_1$ and $k_2$ assuming no modification of $\omega$ as in the CLAS experiment.
    Each line corresponds to $\varDelta\chi^2 (=\chi^2-\chi^2_\mathrm{min}) = n^2, n = 1, 2, 3, \cdots$.
    }
  \label{fig:chi2_vs_k1k2_case4}
 \end{center}
\end{figure}

\begin{figure}[htbp]
 \begin{center}
  \includegraphics[width=15.5cm]{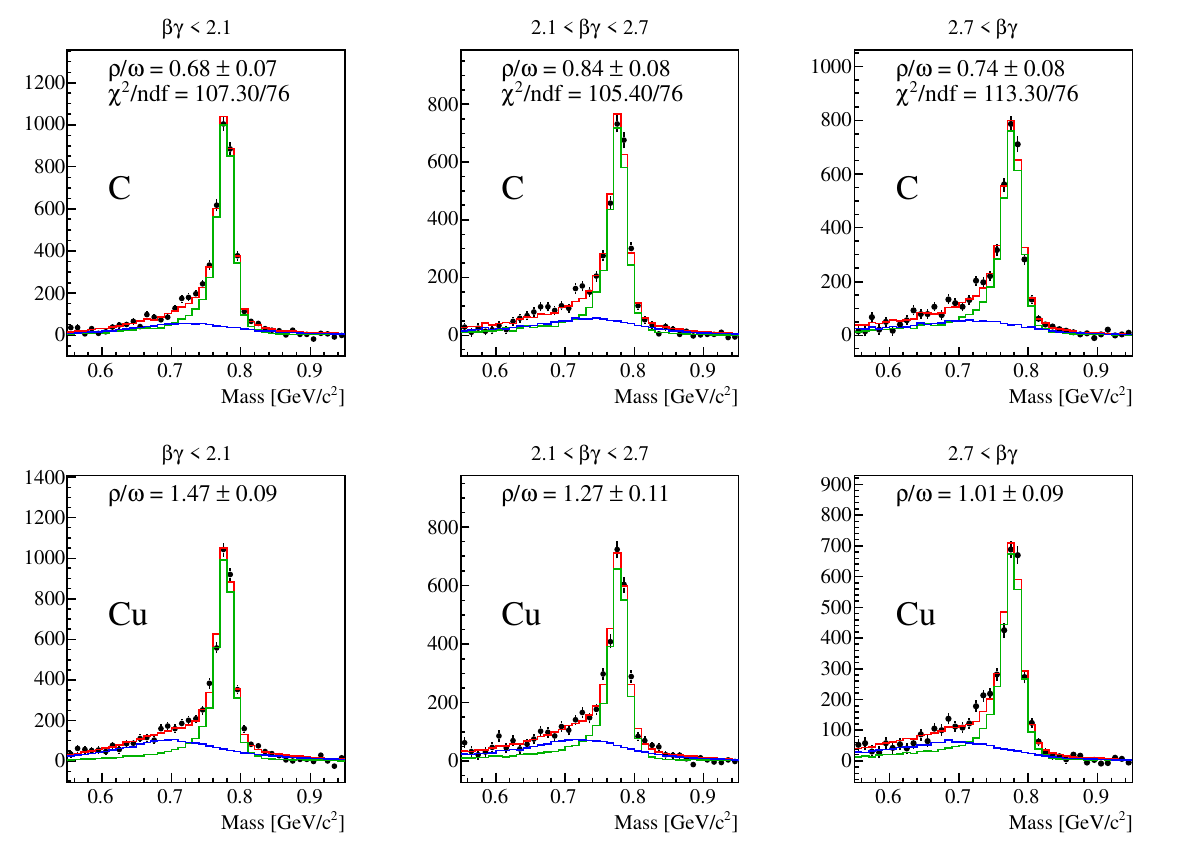}
    \caption{Fitting results near minimum $\chi^2$ values assuming no modification of $\omega$ as in the CLAS experiment.
    The combinatorial background has been subtracted.
    Red lines depict the fitting results obtained as the sum of $\omega\rightarrow e^+e^-$ (green) and $\rho\rightarrow e^+e^-$ (blue).
    }
  \label{fig:model_fit_result_case4}
 \end{center}
\end{figure}

To clarify the origin of the difference, 
we also performed model calculations assuming 
solely $\rho$ meson modifications, since the CLAS experiment analysis ignored $\omega$ meson modifications.
The asymmetric distribution of $\rho$ meson is assumed in the calculation to follow the method used by the CLAS experiment.
The obtained results are summarized in Table~\ref{tab:clas_model_results}
and the corresponding fit results are shown in Fig~\ref{fig:chi2_vs_k1k2_case4} and Fig~\ref{fig:model_fit_result_case4}.
Even in this case, the calculation still shows significant mass shifts of 
$\rho$ mesons, though worse $\chi^2$ values are obtained compared to the result with both $\rho$ and $\omega$ modifications.

This analysis is not the same in a strict sense as the CLAS analysis;
in CLAS, the fit is performed using only the $\rho$ contribution after subtracting the $\omega\rightarrow e^+e^-$ contribution, but in this analysis, the $\omega$ contribution is also included in the fit.
In the
case of our data compared to CLAS, the $\rho$ amplitude is smaller than the $\omega$ amplitude.
Therefore, the accuracy of the subtraction depends strongly on the accuracy of the determination of $\omega$ amplitude.
In order to improve the accuracy of the $\omega$ amplitude, we fitted with not only deformed $\rho$ but also unmodified $\omega$.
Nevertheless, we also tried the fitting after subtraction of unmodified $\omega$ and the results agreed within the errors.

\begin{table}[hbpt]
   \caption{
   Obtained values of $k_1, k_2$ near the $\chi^2$ minimum, $\rho/\omega$ ratios, and $\chi^2$ values assuming no modification of $\omega$ as in the CLAS experiment.
   The listed error values are statistical errors. The obtained $k_2$ values are consistently zero, with the presented numbers indicating the upper limits of the 99\% confidence level.
   }
   \label{tab:clas_model_results}
   \begin{center}
      \begin{tabular}{cccccc}
         \hline
        & $k_1$ & $k_2$ & $\rho/\omega$~(C) & $\rho/\omega$~(Cu) & $\chi^2_\mathrm{min}$/ndf \\ \hline\hline
           $\beta\gamma < 2.1$       & $0.11^{+0.05}_{-0.05}$ & $<0.4$  &  $0.68 \pm 0.07$  &  $1.47 \pm 0.09$  &  $107.3/76$ \\
          $2.1 < \beta\gamma < 2.7$ & $0.07^{+0.08}_{-0.03}$ & $<0.6$  &  $0.84 \pm 0.08$  &  $1.27 \pm 0.11$  &  $105.4/76$ \\
                 $2.7 < \beta\gamma$       & $0.09^{+0.05}_{-0.07}$ & $<0.7$  &  $0.74 \pm 0.08$  &  $1.01 \pm 0.09$  &  $113.3/76$ \\
         \hline
         \hline
      \end{tabular}
   \end{center}
\end{table}


One potential reason for the observed disparities between the two experimental outcomes could be differences in reaction mechanisms.
The production processes of vector mesons may differ between $pA$ and $\gamma A$ reactions.
For instance, vector mesons could be generated coherently by the $\gamma$ beam. The LEPS experiment measured nucleus dependence of $\phi$ meson production cross sections in $\gamma A$ reactions and selected an incoherent process for their analysis~\cite{LEPS}.
Thus, without relevant selection criteria, nuclear matter effects on vector mesons could vary between these reactions.
Another plausible reason is the difference in background normalization determination.
In our analysis, absolute normalization methods for combinatorial backgrounds could not be applied since like-sign pairs were not measured. 
However, we successfully confirmed the robustness of our combinatorial background estimation within the scope of our analysis;
i.e., manually adjusting the amount of background within a reasonable range produced consistent results as 
described in Section \ref{Systematic errors of the modification parameters}. 
For a test, we fixed the amount of combinatorial background to half of the fitted result to 
examine unexpected effects of background evaluations. 
In this case, the resulting $\rho/\omega$ ratios of 
C and Cu were approximately 
3 and 6, respectively. 
They also contradicted 
the result (1.0$\pm$0.2) of a former 12 GeV/$c$ $pp$ experiment~\cite{bib:pp_12GeV}.
In addition, the optimal $k_1$ parameter in this calculation showed a
significant change in the pole mass position.
This indicated that significant changes to the background amount are unlikely to reproduce our data.
Thus, the discrepancy between the two experimental results may not stem from the differences in background evaluation methods.

\subsubsection{Comparison with theoretical calculations}
Our updated results are compared with theoretical calculations for the
mass modifications of the $\rho$ and
$\omega$ mesons. Calculations based on the QCD sum rule 
show that the mass decreasing of $\rho$ and $\omega$ mesons is $16 \pm 6$\% 
at normal nuclear matter density~\cite{bib:HL92,bib:HL95} which is consistent with our results, as shown in Fig. \ref{fig:k1_vs_momentum}.
Momentum dependence is also calculated based on the QCD sum rule~\cite{bib:shlee98},
but the momentum range is limited to less than 1~GeV/$c$.
Additionally, in our data analysis, we have not been able to separate longitudinal and transverse components.

Several calculations showed a huge increase of a $\omega$ meson width~\cite{Ramos13, CR14}.
The obtained $\omega$ meson width in the model calculations 
is $120 \sim 200$~MeV/$c^2$. Our evaluations don't show 
such huge increases. 

Mass spectra of $\rho$ and $\omega$ mesons were also calculated by using several theoretical models. 
One model was based on an 
effective Lagrangian which combined chiral SU(3) dynamics with 
vector meson dominance~\cite{KKW97}. Although the model didn't show a peak mass decrease, the calculated results had asymmetric modifications of $\rho$ meson mass spectra and an increase in the yield on the low mass side. 
Another calculation included effects of nucleon resonances~\cite{PM02} 
and the calculation showed similar asymmetric modifications of $\rho$ mesons mass spectra. 
These results support our analysis method as  
adopting an asymmetric mass shape.  
To have a conclusion,
more detailed comparisons are required in the mass spectrum level. 

\section{Conclusion}
Study on the mass spectra of vector mesons in finite-density QCD matter offers crucial insights into its properties.
To date, numerous experimental studies have focused on exploring the in-medium properties
of vector mesons using nuclear targets and low-energy heavy-ion 
collisions. 
However, despite significant modifications of the vector meson mass spectra, definitive conclusions remain unclear.

The KEK-PS E325 experiment measured the mass spectra of vector mesons in nuclei through 12~GeV $p + A$ reactions, focusing on the $e^+e^-$ decay mode of mesons. 
The obtained mass spectra exhibited an excess yield in the 
low-mass region 
of the $\omega$ meson peak.  
Model calculations suggested that this excess could be attributed to a mass reduction of $9.2 \%$ for $\rho$ and $\omega$ mesons.
Analyzing $\gamma A$ collisions, the CLAS-g7 experiment
measured the invariant mass spectra of
$e^+e^-$ pairs.
The result of the experiment revealed significant mass broadening of the $\rho$ meson for Fe-Ti target, by a factor of 1.45.
Conversely, the KEK experiment reported a finite mass shift without any broadening of $\rho$ and $\omega$ mesons.
These discrepancies prompted us to perform an updated analysis of the E325 data.

In the current updated analysis, the $\beta\gamma$ dependence of mass spectra
in the $\rho$ mass region was first established
by refining the analysis details.
The kinematic acceptance of the final sample was increased, in which
the differences in acceptance across target positions was minimized. 
Refinements in simulation conditions well accounted for effects of position detection chamber misalignment and improved the reproducibility of
the peak positions and the width of the $\omega$ meson. 

Further, model calculations incorporated several updates. 
Form factors of the Dalitz decays of $\omega$ and $\eta$ mesons were
updated, and evaluations of the radiation tails of resonance decays 
were improved based on recent theoretical advancements. 
An asymmetric mass distribution was adopted,
similar to that used in the CLAS experiment, in addition to the original relativistic Breit--Wigner distribution.

Compared to known mass spectra and experimental effects, our spectra exhibited a significant excess in the mass range from 0.65 to 0.86~GeV/$c^2$ across all $\beta \gamma$ 
kinematic regions. 
Updated model calculations evaluated the mass modifications of $\rho$ and $\omega$ mesons. In this case, mass spectra 
modifications were parameterized using two parameters: a shift in the pole mass ($k_1$) and a broadening of the width  ($k_2$). 
Accordingly, for $\beta\gamma < 2.1$, $2.1 < \beta\gamma < 2.7$, and $2.7 < \beta\gamma$,
the obtained values of the $k_1$ parameter were 
$0.12 ^{+0.03}_{-0.03}\mathrm{(stat.)} ^{+0.01}_{-0.03}\mathrm{(sys.)}$,
$0.12 ^{+0.04}_{-0.06}\mathrm{(stat.)} ^{+0.01}_{-0.09}\mathrm{(sys.)}$, and
$0.10 ^{+0.03}_{-0.05}\mathrm{(stat.)} ^{+0.02}_{-0.02}\mathrm{(sys.)}$, respectively.
The obtained values of the $k_2$ parameter exhibited no significant broadening. 
These obtained values are consistent with the previous findings~\cite{bib:naruki_prl}.

\section*{Acknowledgment}
We would like to sincerely thank all the staff members of KEK, 
including the PS accelerator division, online group, 
electronics group and computing research center, 
PS floor group and particularly the PS beam channel group for their 
helpful support. We would express our gratitude to the staff members
of RIKEN-CCJ, RIKEN RSCC and RIKEN Hokusai,
for their support on the data analysis.
This study was partly supported by
JSPS Research Fellowship for Young Scientists,
RIKEN Special Postdoctoral Researchers Program, 
and MEXT/JSPS KAKENHI Grant Numbers
JP06640391, JP07640396, JP08404013, JP12440064,
JP15340089, JP20H05647 and JP23H05440.

\appendix

\section{Mass shape formulae}                                                     \label{section-App-massform}                       
\begin{figure}[htbp]
 \begin{center}
  \includegraphics[width=10cm]{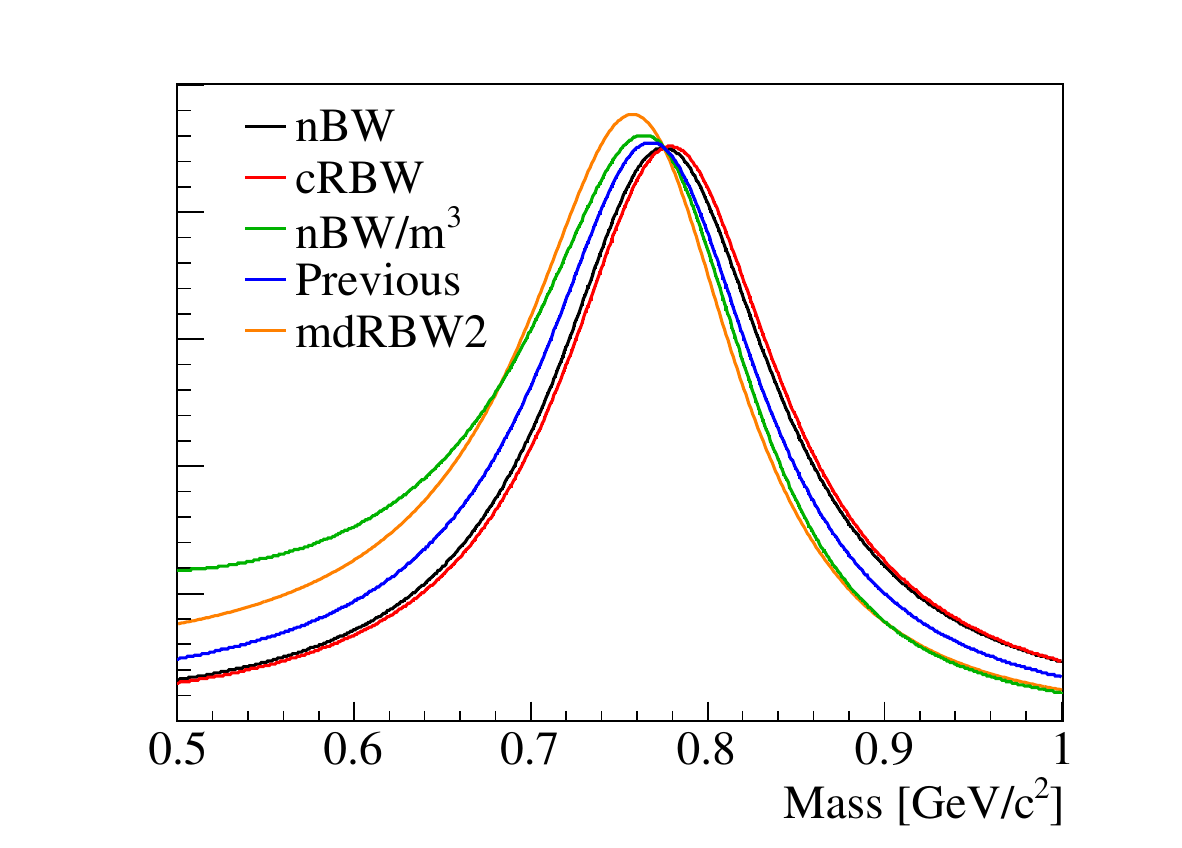}
    \caption{Mass distributions of $\rho$ meson ($m_0 = 0.775$~GeV/$c^2$ and $\Gtot = 0.151$~GeV/$c^2$) modeled  using several functions.
    The black line depicts the distribution derived from the non-relativistic Breit--Wigner (nBW) function, used in the analysis case (i),
    while the red line depicts the distribution derived from its relativistic counterpart (cRBW). 
    Furthermore, 
    the orange line 
    depicts a relativistic BW using the mass-dependent width (mdRBW2).
    The green line illustrates the asymmetric mass shape adopted 
    in the case (ii) in our analysis (nBW/$m^3$). 
    The blue line depicts the formula (\ref{form-naruki}), which is the mass shape adopted in our previous study~\cite{bib:naruki_prl}.
    Notably, all functions were scaled to match at the pole mass $m_0$.
    }
  \label{fig:rbw_compare2}
 \end{center}
\end{figure}

Non-relativistic Breit--Wigner formula (nBW) is already shown as 
the formula (\ref{form-nBW}),
\begin{eqnarray}
   \mathrm{nBW}(m) = \frac{\Gtot}{2\pi}\frac{1}{(m-m_0)^2+\Gtot^2/4},             
\end{eqnarray}
and 
constant-width relativistic Breit-Wigner (cRBW) shape~\cite{bib:giacosa_bw_2021}
is
\begin{eqnarray}
   \mathrm{cRBW}(m) = \frac{2}{\pi} \frac{mm_0\Gee}{(m^2-m_0^2)^2+m_0^2\Gtot^2},  
\end{eqnarray}
where $m$ denotes the invariant mass, $m_0$ represents the pole mass, $\Gtot$ indicates the
total decay width of the meson, and $\Gee$ represents the partial decay width
of the dielectron channel.
As depicted in Fig.~\ref{fig:rbw_compare2},
even for relatively wide resonances such as those of $\rho$ mesons,
the differences between the two distributions, cRBW and nBW, is negligible.

Figure~\ref{fig:rbw_compare2} also depicts other mass shapes.

To include the mass-dependent widths, cRBW is modified as
\begin{eqnarray}
   \mathrm{mdRBW}(m) = \frac{2}{\pi} \frac{m^2\Gee(m)}{(m^2-m_0^2)^2+m^2\Gtot(m)^2},  
\end{eqnarray}
and take the mass-dependent widths as 
$\Gtot(m)=\frac{m}{m_0} \Gtot$
and
$\Gee(m)=\frac{m_0^3}{m^3} \Gee$,
according to a reference~\cite{Fang88}.
Then the shape is written as 
\begin{eqnarray}
   \mathrm{mdRBW2}(m) = \frac{2}{\pi} \frac{(m_0^3/m)\Gee}{(m^2-m_0^2)^2+(m^4/m_0^2)\Gtot^2}
\end{eqnarray}
which is plotted as mdRBW2 in Fig.~\ref{fig:rbw_compare2},
which has an asymmetric shape as nBW/m$^3$, 
but lower than that in the low mass region.

The mass shape discussed in the CLAS analysis~\cite{bib:clas_g7} is
\begin{eqnarray}
   A(m) = \frac{2}{\pi} \frac{m^2\Gamma_\mathrm{tot}(m)}{(m^2-m_0^2)^2+m^2\Gamma_\mathrm{tot}^2(m)}.
\end{eqnarray}
They claimed this formula is well approximated by Breit-Wigner/m$^3$, and
it used in their analysis.

The mass shape employed in our previous study~\cite{bib:naruki_prl} is
\begin{eqnarray}
   f(m) \propto \frac{m_0^2 \Gtot \Gee}{(m^2-m_0^2)^2+m^2\Gtot^2}, 
   \label{form-naruki} 
\end{eqnarray}
which is deduced from
\begin{equation}
\frac{d\sigma}{dm} \propto \frac{m^2\ \Gtot(m)\ \Gee(m)}{(m^2-m_0^2)^2+m_0^2\ \Gtot(m)^2} .
\label{eq-Naruki-RBW0}
\end{equation}
where taking the mass-dependent widths as
$\Gtot(m)=\frac{m}{m_0} \Gtot$
and
$\Gee(m)=\frac{m_0^3}{m^3} \Gee$, as same as mentioned above.
The form is 
equivalent to the shape used by the CERES experiment~\cite{bib:CERES:2005},
\begin{equation}
\frac{d\sigma}{dm} \propto \frac{(1-4m_\pi^2/m^2)}{(m^2-m_0^2)^2+m^2\ \Gtot^2}(2\pi mT)^{(3/2)} \exp(-m/T) .
\label{eq-CERES}
\end{equation}
by removing the Boltzmann term and $(1-4m_\pi^2/m^2)$, 
considering that the matter is cold and the initial channel is not dominated 
by $\pi\pi \rightarrow \rho$ in our experiment.
It also has an asymmetric shape, between nBW and nBW/m$^3$.

Thus we	used as	two extreme cases, nBW and nBW/m$^3$, symmetric and asymmetric.

\vspace{0.2cm}
\noindent


\let\doi\relax


\appendix

\end{document}